\documentclass[aip,pop,reprint,amsmath,amssymb,floatfix]{revtex4-1}

\pdfoutput=1

\usepackage{graphicx}
\usepackage{subfigure}
\usepackage{xcolor}
\usepackage{cancel}
\usepackage{cleveref}

\let\oldhat\hat
\renewcommand{\vec}[1]{\mathbf{#1}}
\renewcommand{\hat}[1]{\oldhat{\mathbf{#1}}}

\begin{document}

\title{Spatio-temporal behavior of magnetohydrodynamic fluctuations with cross-helicity and background magnetic field}

\author{R. Lugones}\email{rlugones@df.uba.ar}
\affiliation{Departamento de F\'\i sica, Facultad de Ciencias Exactas y Naturales,
  Universidad de Buenos Aires and IFIBA, CONICET, Ciudad
  Universitaria, 1428 Buenos Aires, Argentina.}
  
\author{P. Dmitruk}
\affiliation{Departamento de F\'\i sica, Facultad de Ciencias Exactas y Naturales,
  Universidad de Buenos Aires and IFIBA, CONICET, Ciudad
  Universitaria, 1428 Buenos Aires, Argentina.}
  
\author{P.D. Mininni}
\affiliation{Departamento de F\'\i sica, Facultad de Ciencias Exactas y Naturales,
  Universidad de Buenos Aires and IFIBA, CONICET, Ciudad
  Universitaria, 1428 Buenos Aires, Argentina.}

\author{A. Pouquet}
\affiliation{Laboratory for Atmospheric and Space Physics, University of Colorado,
  Boulder, CO 80309, USA, and National Center for Atmospheric Research, P.O. Box 3000,
  Boulder, CO 80307, USA.}

\author{W.H. Matthaeus}
\affiliation{Bartol Research Institute and Department of Physics and Astronomy,
University of Delaware, Newark, DE 19716, USA.}

\begin{abstract}
  We study the spatio-temporal behavior of the Els\"asser variables
  describing magnetic and velocity field fluctuations, using direct
  numerical simulations of three-dimensional magnetohydrodynamic
  turbulence. We consider cases with relatively small, intermediate,
  and large values of a mean background magnetic field, and with null,
  small, and high cross-helicity (correlations between the velocity
  and the magnetic field). Wavenumber-dependent time correlation
  functions are computed for the different simulations. From these
  correlation functions, the decorrelation time is computed and
  compared with different theoretical characteristic times: the local
  non-linear time, the random-sweeping time, and the Alfv\'enic
  time. It is found that decorrelation times are dominated by sweeping
  effects for low values of the mean magnetic field and for low values
  of the cross-helicity, while for large values of the background
  field or of the cross-helicity and for wave vectors sufficiently
  aligned with the guide field, decorrelation times are controlled by
  Alfv\'enic effects. Finally, we observe counter-propagation of
  Alfv\'enic fluctuations due to reflections produced by
  inhomogeneities in the total magnetic field. This effect becomes
  more prominent in flows with large cross-helicity, strongly
  modifying the propagation of waves in turbulent magnetohydrodynamic
  flows.
\end{abstract}

\maketitle

\section{Introduction}\label{sec_Intro}

Turbulent fluctuations are essentially broadband, both in spatial and
temporal scales, involving non-linear couplings among a wide range of
scales \cite{frisch_turbulence_1995}. In incompressible
magnetohydrodynamics (MHD) \cite{pouquet_strong_1976,
  zhou_magnetohydrodynamic_2004} these couplings are based on
interactions of triads of modes \cite{zhou_non-gaussian_1993, 
  alexakis_turbulent_2007, teaca_energy_2009, aluie_2010, 
  mininni_scale_2011} which can be of different types, such as (local 
in wavenumber space) nonlinear distortions of eddies, or (non-local 
in wavenumber space) sweeping of small eddies by larger ones
\cite{kraichnan_structure_1959,
  tennekes_eulerian_1975, chen_sweeping_1989, nelkin_time_1990,
  matthaeus_eulerian_2010, servidio_time_2011,
  carbone_anisotropy_2011}. Of course, these non-linear couplings also
involve interactions with waves in the flow, which are ubiquitous in
MHD flows as well as in plasma turbulence.

The incompressible MHD equations sustain Alfv\'en waves, which in the
presence of a background magnetic field $\vec{B}_0'$ are described by
a linear dispersion relation of frequency
$\omega={\bf k} \cdot {\bf V}_\textrm{A}$ for the wavevector
${\bf k}$, with Alfv\'en velocity
${\bf V}_\textrm{A}={\bf B}_0'/\sqrt{4\pi \rho}$ and with mass density
$\rho$. It is well known that these waves, when considered in
isolation, are also exact solutions of the non-linear MHD (ideal)
equations. Simultaneous presence of counter-propagating fluctuations
however activate nonlinear interactions among modes, producing
dispersion, and in consequence the waves are no longer exact solutions
of the system \cite{dobrowolny_1980_HydromagneticTurbulence}. As the
background magnetic field controls the propagation velocity (i.e., the
Alfv\'en velocity), the non-linear interaction is influenced by the
Alfv\'en crossing time of counter-propagating wave packets. There is
therefore a competition between non-linear interactions (i.e.,
turbulence) and wave propagation \cite{dmitruk_waves_2009, rappazzo_coronal_2007}.

The strength of counter-propagating fluctuations can be measured by
the cross-helicity, a quantity which is a quadratic invariant of the
ideal MHD equations (see Sec.~\ref{sec_EqNumSim} for specifics). This
quantity is also of relevance for the solar wind and for space
plasmas, as large-scale flows with cross-helicity (in the presence of
a guide field) are often found in the interplanetary medium. A
spatio-temporal analysis of field fluctuations
\cite{servidio_time_2011, clark_di_leoni_spatio-temporal_2015} can
thus be performed to quantitatively study the importance of these
different effects, and to distinguish which is the dominant timescale
among the different ones depending on the controlling parameters of
the system. This kind of analysis was performed in the past for MHD
flows without cross-helicity \cite{meyrand_weak_2015,
  lugones_2016_spatiotemporal, meyrand_direct_2016}, observing
different behaviors depending on whether the turbulence is weak or
strong. The prevailing conclusion, for strong turbulence, is that the
time decorrelation of Fourier modes in the inertial range is typically
dominated by the sweeping decorrelation due to large scale flows
\cite{servidio_time_2011, chen_sweeping_1989,
  lugones_2016_spatiotemporal}.  However, the effect of changing the
strength of counter-propagating fluctuations in the spatio-temporal
behavior of the flow, and in its decorrelation time, was not
considered before.

In the present paper we perform a spatio-temporal analysis of MHD
turbulence, controlling simultaneously and separately the intensity of
the background magnetic field and the amount of cross-helicity in the
flow, extending our previous study \cite{lugones_2016_spatiotemporal}
of incompressible MHD with a background magnetic field and no
cross-helicity. We present several numerical solutions of the
incompressible MHD equations in a turbulent steady state, and analyze
each timescale in the system using wavenumber-dependent time
correlation functions, and spatio-temporal spectra of the Els\"asser
variables. The spatio-temporal study of the Els\"asser variables
allows us to disentangle the two possible polarizations of the
Alfv\'en waves, as well as their direction of propagation, and to
quantify any imbalance between the two polarizations. We find that
decorrelation times are dominated by sweeping effects for low values
of the mean magnetic field and for low values of the cross-helicity,
while for large values of the background field or of the
cross-helicity decorrelation times are controlled by the Alfv\'enic
times. Moreover, for large values of the cross-helicity we also
observe counter-propagation of Alfv\'enic fluctuations (i.e., an
inversion in the direction of propagation of one polarization of
Alfv\'en waves), resulting from reflections in inhomogeneities of the
total magnetic field produced by the turbulence. Under some
conditions, this can result in the propagation of both polarizations
of the Alfv\'en waves in the same direction. This effect strongly
affects non-linear interactions.

The structure of the paper is as follows. In Sec.~\ref{sec_EqNumSim}
we introduce the equations and the numerical methods employed, as well
as a description of the spatio-temporal spectrum and of the
correlation functions. Results are presented in
Sec.~\ref{sec_results}. Finally, discussions and conclusions are
presented in Sec.~\ref{sec_Conclusions}.

\section{Equations and numerical simulations}\label{sec_EqNumSim}

\subsection{The MHD equations and the Els\"asser fields}\label{sec_eq}
The incompressible MHD equations (momentum and induction equations) in
dimensionless units as solved in the numerical simulations are
\begin{equation}\label{eq:MHD_v}
  \frac {\partial {\bf v}}{\partial t} +
  {\bf v }\cdot \nabla {\bf v} = -\frac{1}{\rho}\nabla p +
  {\bf j} \times {\bf B} + \frac{1}{R} \nabla^2{\bf v} + \vec{F}_v,
\end{equation}
\begin{equation}\label{eq:MHD_b}
  \frac{\partial {\bf b}}{\partial t} = \nabla \times ({\bf v} 
  \times {\bf B}) + \frac{1}{R_m} \nabla^2 {\bf b} + \vec{F}_b,
\end{equation}
where ${\bf v}$ is the plasma bulk velocity, ${\bf B} = {\bf b} + {\bf
  B}_0$ is the total magnetic field (in units of an Alfv\'enic speed,
and obtained from the total magnetic field ${\bf B}'$ in Gaussian
units after dividing by $\sqrt{4\pi\rho}$, where $\rho$ is the plasma
density), and $\vec{F}_v$ and $\vec{F}_b$ are forcing terms to be
discussed in more detail below. The total magnetic field has a
fluctuating part ${\bf b}$, and a mean DC field ${\bf
  B}_0=B_0\hat{x}$. Finally, ${\bf j} = \nabla \times {\bf b}$ is the
current density and $p$ is the plasma pressure. The units are based on
a characteristic speed $v_0$, which for MHD is chosen to be the
typical Alfv\'en speed of the magnetic field fluctuations, $v_0 =
\sqrt{\langle b^2 \rangle /(4\pi\rho)}$, where $\langle . \rangle$
denotes a spatial average. The dimensionless parameters appearing in
the equations are the kinetic and magnetic Reynolds numbers, $R=v_0
L/\nu$ and $R_m = v_0 L /\mu$ respectively, with $\nu$ the kinematic
viscosity, $\mu$ the magnetic diffusivity, and $L$ the characteristic
length scale (the simulation box side length is defined as $2\pi
L$). The unit time is $t_0 = L/v_0$, which for MHD becomes the
Alfv\'en crossing time based on magnetic field fluctuations. The
Els\"asser fields are then defined as
\begin{equation}\label{eq:MHD_zdef}
\vec{z}^\pm = \vec{v} \pm \vec{b} .
\end{equation}

In terms of the Els\"asser fields, the MHD equations can be written
\cite{servidio_time_2011} as
\begin{equation}
\partial_t \vec{z}^\pm = \pm \vec{V}_\textrm{A} \cdot \nabla
\vec{z}^\pm - \vec{z}^\mp \cdot \nabla \vec{z}^\pm - \nabla{P} +
\frac{1}{R} \nabla^2 \vec{z}^\pm ,
\label{eq:Elsasser}
\end{equation}
with $P$ the total pressure divided by the plasma density, and with
the assumption that $R=R_m$. In the r.h.s.~of Eq.~(\ref{eq:Elsasser}) 
we explicitly separated the convective term into a linear part 
describing Alfv\'enic propagation with $\vec{V}_\textrm{A} = \vec{B}_0$ 
the Alfv\'en velocity based on the background magnetic field (with
$\vec{B}_0$ the field in units of velocity), and the non-linear part
describing the interaction among counter-propagating wave-like 
fluctuations. It is evident from these equations that both Els\"asser 
fields must be present to activate the non-linear interactions.

The ideal invariants (i.e., with zero viscosity and resistivity) of
incompressible MHD theory can be written in terms of the Els\"asser
fields. The total energy $E$ (kinetic plus magnetic) in terms of these
variables is
\begin{equation}
E = \frac{1}{2}\int{\left(\left|\vec{v}\right|^2 +
    \left|\vec{b}\right|^2 \right)\,dV} =
    \frac{1}{4}\int{\left(\left|\vec{z}^+\right|^2 +
    \left|\vec{z}^-\right|^2 \right)\,dV},
\label{eq:ener}
\end{equation}
while the cross-helicity $H_c$ is
\begin{equation}
H_c = \int{\vec{v}\cdot\vec{b} \, dV} =
    \frac{1}{4}\int{\left(\left|\vec{z}^+\right|^2 
    - \left|\vec{z}^-\right|^2 \right)\,dV} .
\label{eq:cross}
\end{equation}
The ratio $\sigma_c = H_c/E$ measures the amount of counter-propagating
fluctuations in the system. A value $\sigma_c = \pm 1$ corresponds to
only one type of fluctuations $\vec{z}^\pm$, while $\sigma_c=0$
represents equipartition between both fields.

As later in the analysis we will be interested in the effect of flow inhomogeneities in the propagation of these fluctuations, following 
the works of Matthaeus {\it et al.} \cite{matthaeus1994transport} 
and Zhou {\it et al.} \cite{zhou1990remarks} we can linearize the 
ideal MHD equations considering the presence of an inhomogeneous 
background magnetic field and/or an inhomogeneous background flow. From 
these works, the general MHD equations (including density fluctuations) 
can be written as
\begin{equation}\label{eq:MHD_zpzm}
  \partial_t \vec{z}^\pm
  + \left( L^\pm_\vec{x} + L^\pm \right) \vec{z}^\pm
  + M^\pm_{ik} \vec{z}^\mp_k
  = 0,
\end{equation}
The linear operators $L^\pm_\vec{x}$, $L^\pm$, and $M^\pm_{ik}$
involve gradients acting on both the large- and the small-scale
fields, and are given by
\begin{equation}\label{eq:MHD_Lx}
  L^\pm_\vec{x} = \left( \vec{U} \mp \vec{V}_\textrm{A} \right) \cdot \nabla ,
\end{equation}
\begin{equation}\label{eq:MHD_L}
  L^\pm = \frac{1}{2} \nabla \cdot \left( \frac{\vec{U}}{2} \pm \vec{V}_\textrm{A} 
  \right) ,
\end{equation}
and
\begin{equation}\label{eq:MHD_Mik}
  M^\pm_{ik} = \nabla_k U_i \pm \frac{1}{\sqrt{4\pi\rho}} \nabla_k B_i'
  - \frac{1}{2} \delta_{ik} \nabla\cdot \left( \frac{\vec{U}}{2} \pm
  \vec{V}_\textrm{A} \right) ,
\end{equation}
where $\vec{U}$ is a background flow. Here, both $\vec{U}$ and
$\vec{V}_\textrm{A}$ can include large-scale inhomogeneities
(including, for $\vec{V}_\textrm{A}$, inhomogeneities associated to
density fluctuations). The mixing terms (those involving the
$M_{ik}^\pm$ operators) allow the possibility of creating counter
propagating fluctuations out of a single-sign propagating fluctuation,
by means of reflections due to inhomogeneities in any of the
background fields \cite{velli_1993_propagation}. In this sense, even
if the system starts from an initial condition with only one sign of
propagating fluctuations, the reflections by inhomogeneous background
fields will create an amount of counter propagating fluctuations which
will turn on non-linearities, producing dispersion and turbulence
\cite{matthaeus_1999_coronal, dmitruk_2001_coronal}. But this effect
can also result, in flows with both polarizations of Alfv\'enic
excitations, in the counter-propagation of one of the excitations, as
will be shown from numerical data in Sec.~\ref{sec_results}.

\subsection{Wavenumber-frequency spectrum and correlation functions}\label{sec_Wfspectrum_and_Gamma}

Using scaling arguments, different characteristic times in the system
can be estimated. The local eddy turnover time or isotropic non-linear
timescale can be defined as $\tau_{nl} \sim 1/\left[ k v(k) \right]$,
where $v(k)$ is the amplitude of the velocity fluctuations at scale
$\sim 1/k$. Considering a Kolmogorov-like scaling
$v(k) \sim v_{rms} \left(kL\right)^{-1/3}$, the nonlinear time in the
inertial range can be written as
\begin{equation}
\tau_{nl} = C_{nl} \left [
   v_{rms} L^{-1/3} \left(\sqrt{k^2_\perp +
   k^2_\parallel}\right)^{2/3}\right ]^{-1},
\label{eq:taunl}
\end{equation}
where $C_{nl}$ is a dimensionless constant of order one, and
$k_\parallel$ and $k_\perp$ denote the wavenumbers parallel and
perpendicular to the background magnetic field. Here,
$v_{rms} = \left\langle |{\bf v}|^2 \right\rangle ^{1/2}$ is a global
quantity, dominated by contributions from the large scales
\cite{zhou_magnetohydrodynamic_2004, matthaeus_anisotropic_2009}.

Another time decorrelation effect is governed by the sweeping
characteristic time, which at the scale $\sim 1/k$ can be expressed as
\begin{equation}
\tau_{sw} = C_{sw} \left( v_{rms}\sqrt{k^2_\perp + k^2_\parallel}
    \right)^{-1} .
\label{eq:tausw}
\end{equation}
This time corresponds to the advection of small-scale structures by
the large-scale flow. Finally, a characteristic Alfv\'en time can be
defined as
\begin{equation}
\tau_A= C_A \left( v_A k_\parallel \right)^{-1} .
\label{eq:taua}
\end{equation}
In the last two expressions, $C_{sw}$ and $C_A$ also are dimensionless
constants of order unity.

These are not all the times scales that could be present in MHD turbulence, 
but the ones most relevant for the discussions in the following sections. 
As an example, another time scale worth mentioning is the decorrelation 
time of triple moments when there is no equipartition between magnetic 
and kinetic energies, e.g., in the dynamo context 
\cite{baerenzung_2008_spectral}.

To disentangle these time scales in the flow, and to identify which is 
the most relevant time scale at a given spatial scale, two tools can be 
used: the statistical properties of the correlation function in space
and time, and the wavenumber-frequency spectrum. We start by
introducing the former. The statistics of the Els\"asser fields can be
characterized by the spatio-temporal two-point autocorrelation
function \cite{servidio_time_2011},
 \begin{equation}
 R^\pm({\bf r},\tau) = \left\langle {\bf z}^\pm( {\bf x},t) \cdot
   {\bf z} ^\pm( {\bf x} + {\bf r},t+\tau) \right\rangle \Big/ 
   \left\langle \left|{{\bf z}^\pm}\right|^2 \right\rangle.
 \label{eq:Rzij}
 \end{equation}
 The Fourier transform in $\vec{r}$ leads to a time-lagged spectral
 density which can be further factorized as
 $S({\bf k},\tau) = S({\bf k})\Gamma({\bf k},\tau)$. The function
 $\Gamma({\bf k},\tau)$ is the scale-dependent correlation function
 \cite{heisenberg_zur_1948,
   comte-bellot_simple_1971,orszag_numerical_1972} which describes the
 time decorrelation of each spatial mode ${\bf k}$, that is, the loss
 of memory of fluctuations with characteristics lengths of order
 $k_x^{-1}$, $k_y^{-1}$, and $k_z^{-1}$.

\begin{figure}
\centering
\includegraphics[width=1\columnwidth]{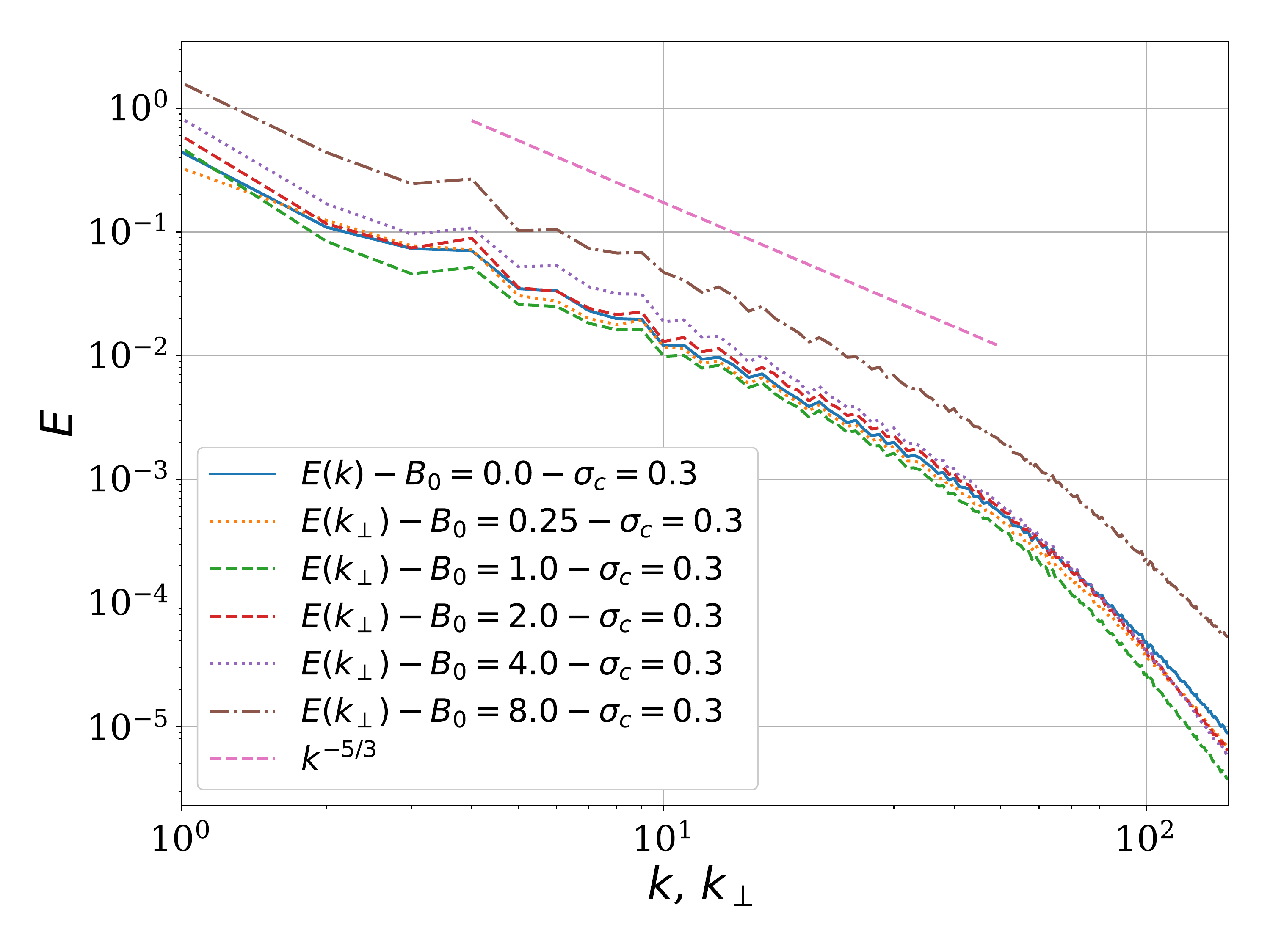}
\caption{Reduced perpendicular energy spectra $E(k_\perp)$ for
  simulations with $B_0=0$, $0.25$, $1$, $2$, $4$, and $8$. All curves
  correspond to the case $\sigma_c = 0.3$, but the cases with
  $\sigma_c = 0$ and $0.9$ show the same behavior. Kolmogorov scaling,
  $\sim k_\perp^{-5/3}$, is shown as reference.}
\label{fig1:E}
\end{figure}

When there is a preferential direction in the flow (as in the present
case of MHD turbulence with a guide magnetic field), it is useful to
assume axial symmetry in Fourier space and to write
$\Gamma({\bf k},\tau) = \Gamma(k_\perp,k_\parallel,\tau)$. As this
function is three dimensional, it is also useful to study
$\Gamma(k_\perp,k_\parallel,\tau)$ with one of the arguments fixed;
for instance, fixing a value of $k_\perp$ and analyzing
$\Gamma(k_\perp,k_\parallel,\tau)$ as a function of $k_\parallel$ and
$\tau$ gives us information on fluctuations that vary only in the
parallel direction, and allow us to distinguish between decorrelation
arising from Alfv\'enic non-linear interactions or sweeping.

The Fourier transform in the time lag of the scale-dependent
correlation function results in the wavenumber-frequency spectrum
$E^\pm(\vec{k},\omega)$ for each of the Els\"asser fields. 
This property follows directly from the 1, that 
states that the Fourier transform of a signal auto-correlation is the 
power spectrum of the same signal (see Refs. 
\cite{clark_di_leoni_quantification_2014, 
clark_di_leoni_spatio-temporal_2015}, and pp.~35-36 from Ref. 
\cite{batchelor_theory_1953} for further details). The spectra
$E^\pm(\vec{k},\omega)$ allow identification of modes satisfying a
generalized dispersion relation of the system, and provide a direct
measurement of how much energy is in those modes, and of how much
energy is in other modes. For the two separate Els\"asser fields, from
Eqs.~(\ref{eq:ener}) and (\ref{eq:cross}) it is easy to see that
\begin{equation}
  E = E^+ + E^- , \,\,\,\, H_c = E^+ - E^- ,
\end{equation}
where $E^\pm = \int |\vec{z}^\pm|^2/4 \, dV$. Thus, for the
wavenumber-frequency spectra of the Els\"asser fields, the two
following relations hold
\begin{eqnarray}
  E^+(\vec{k},\omega) &=& [E(\vec{k},\omega) + H_c(\vec{k},\omega)]/2, \\
  E^-(\vec{k},\omega) &=& [E(\vec{k},\omega) - H_c(\vec{k},\omega)]/2.
\end{eqnarray}
Therefore, computation of the wavenumber-frequency spectra of the
energy and of the cross-helicity allows unique determination of the
wavenumber-frequency spectra of the Els\"asser fields, and vice-versa.

\begin{figure*}
  \centering
  \subfigure[$B_0=0$]{\includegraphics[width=0.48\textwidth]{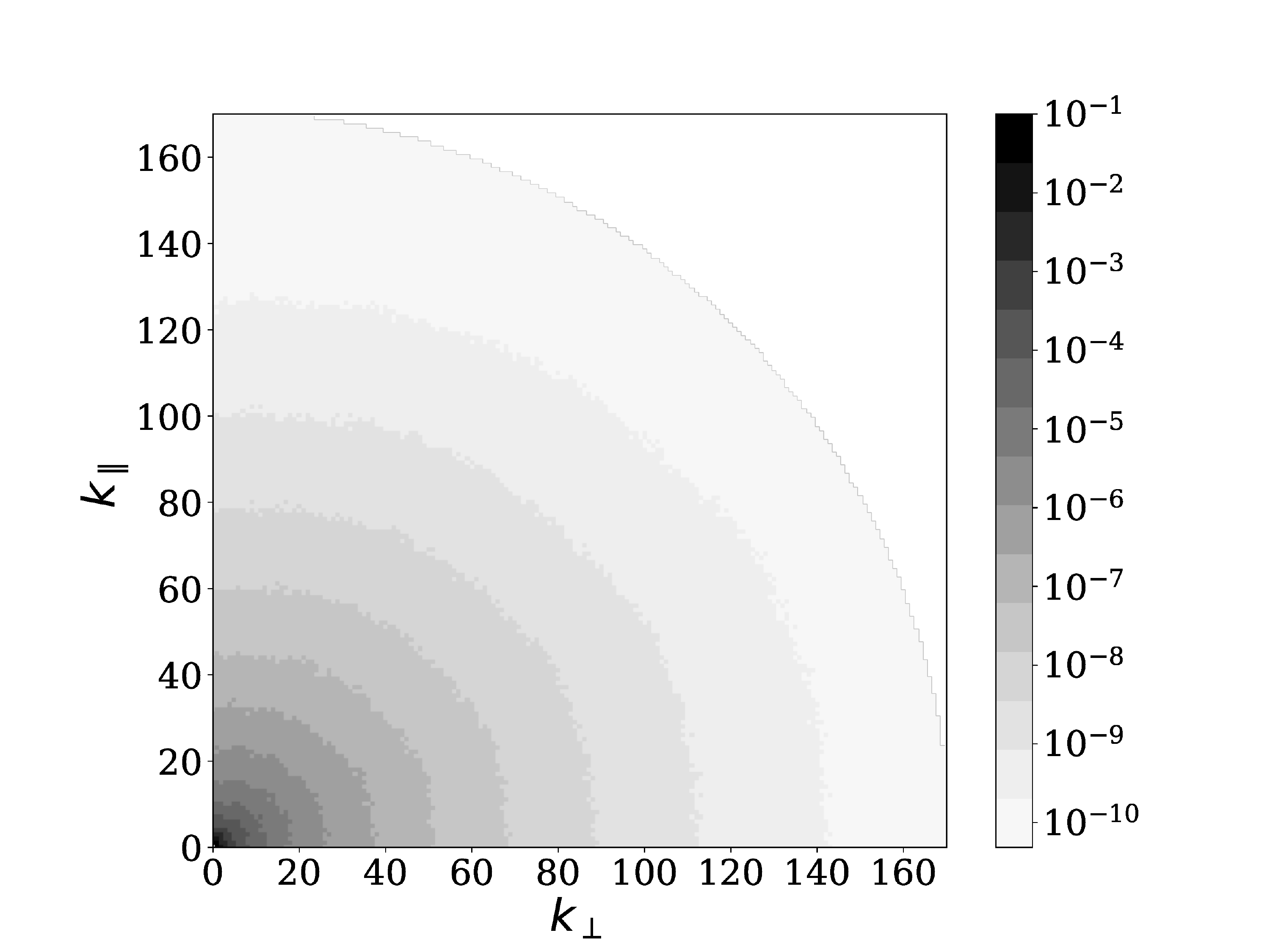}}
  \subfigure[$B_0=1$]{\includegraphics[width=0.48\textwidth]{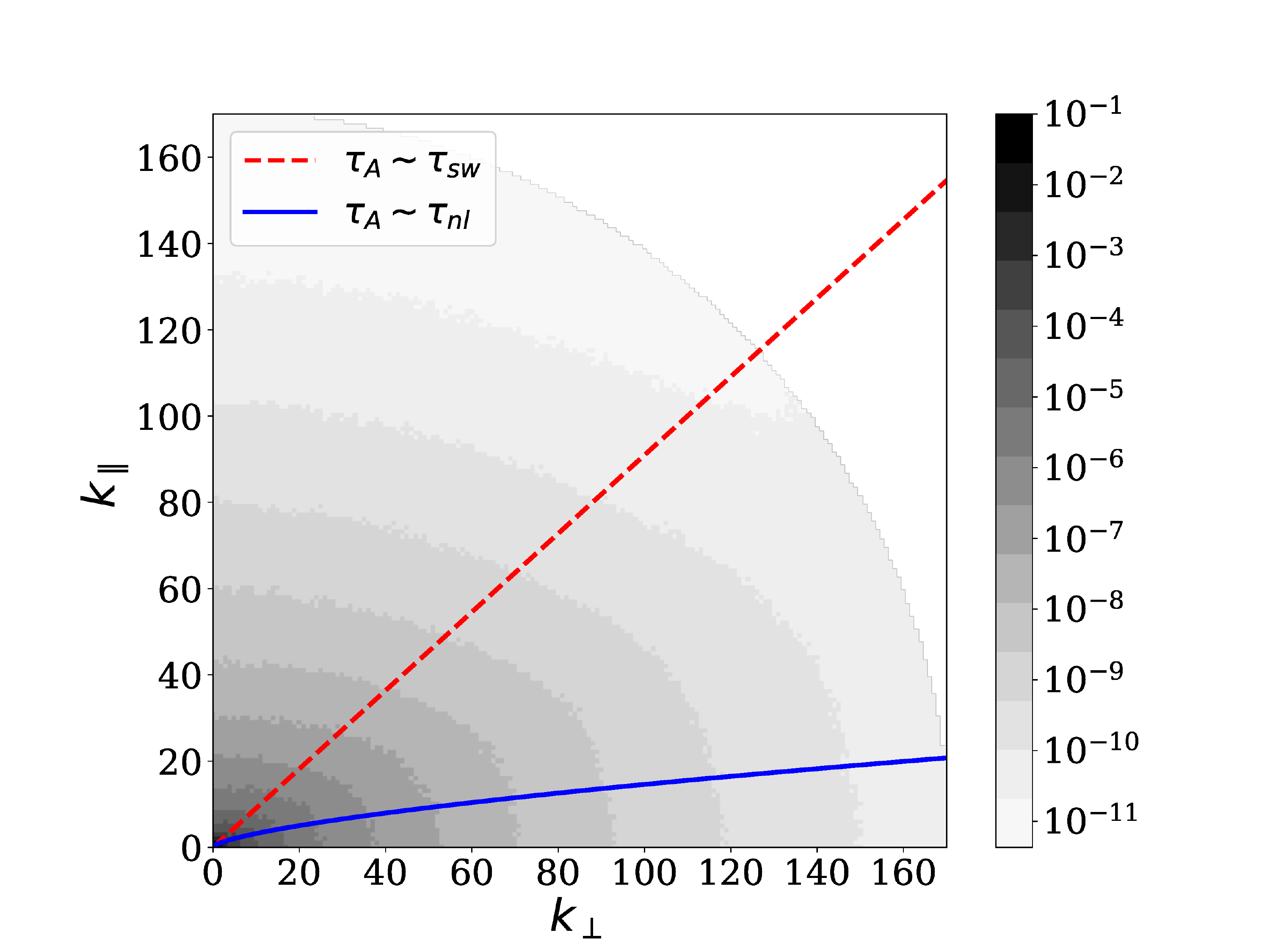}}
  \subfigure[$B_0=4$]{\includegraphics[width=0.48\textwidth]{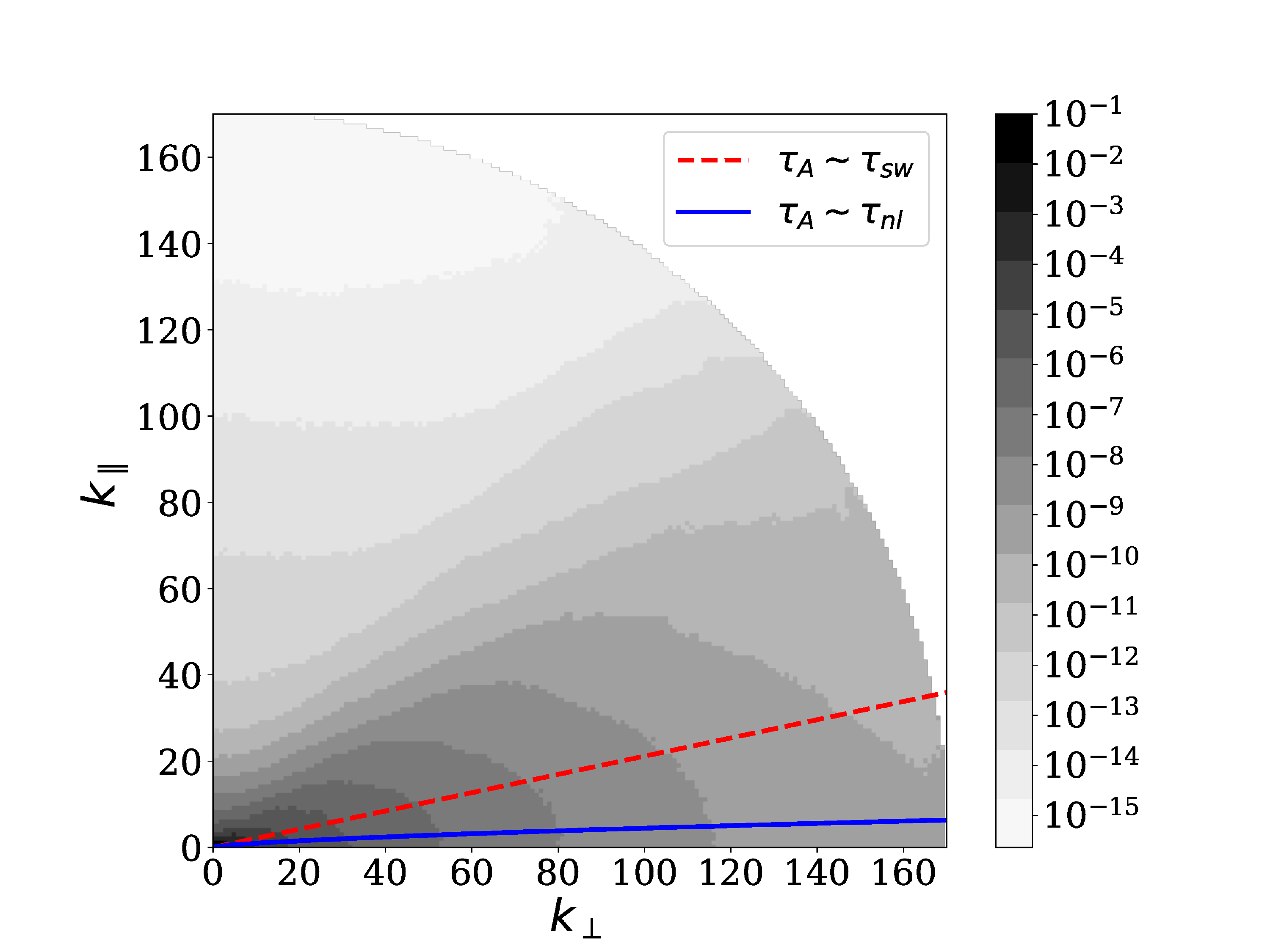}}
  \subfigure[$B_0=8$]{\includegraphics[width=0.48\textwidth]{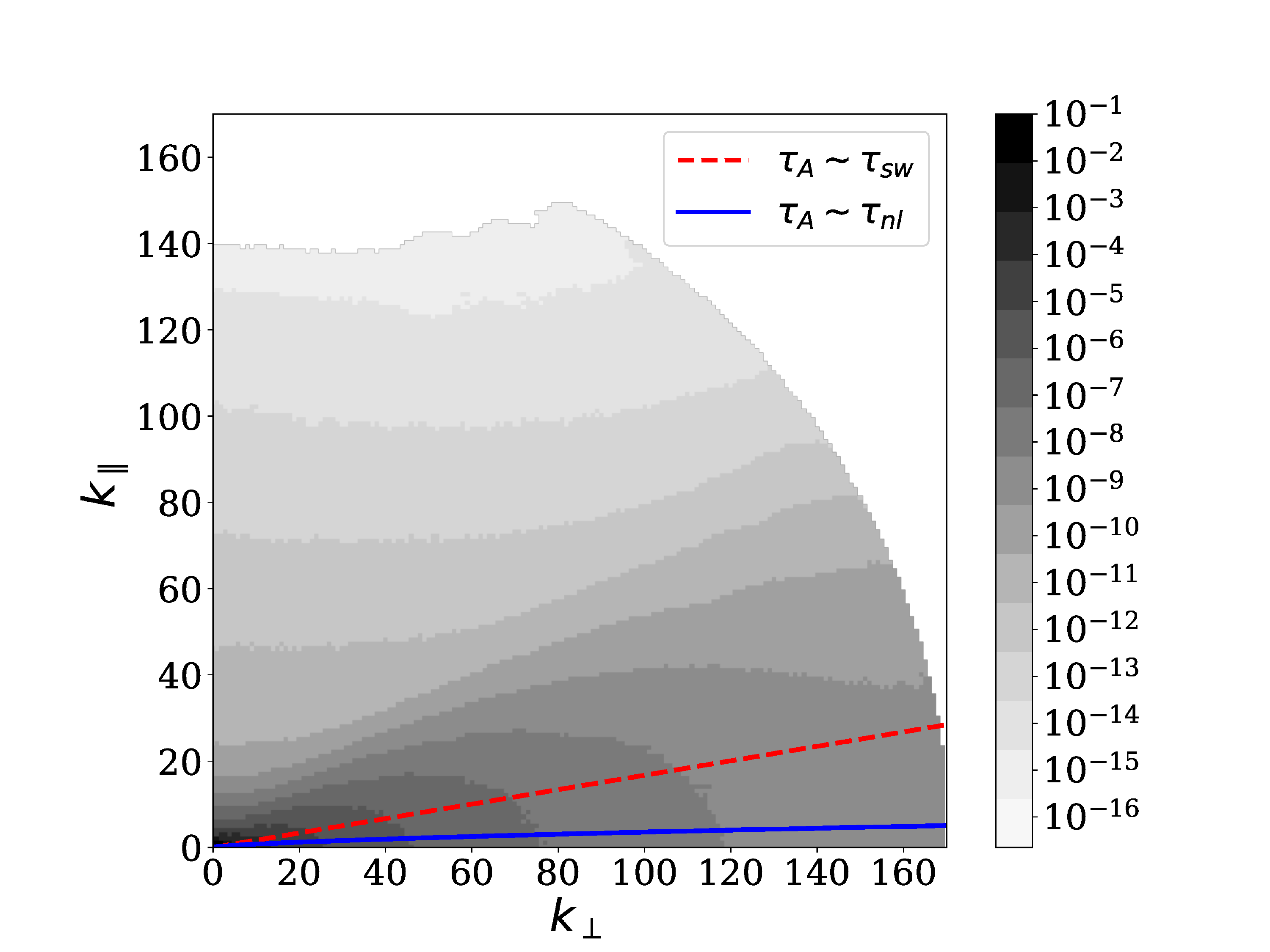}}
  \caption{Isocontours of the axisymmetric energy spectrum
    $e(k_\perp,k_\parallel)$ for $B_0=0$, $1$, $4$ and $8$, and for
    $\sigma_c = 0.3$. In all cases, dark means larger energy density (in
    logarithmic scale). The lines indicate the modes for which the
    sweeping time (red dashed line) or the local non-linear time
    (solid blue line) become equal to the Alfv\'en time. For large
    $B_0$ the flow becomes more anisotropic, and isocontours change
    shape as they cross these lines. Note also the increase in the
    energy in modes that have the Alfv\'en time as the fastest time
    (i.e., of modes below the solid blue curve) as $B_0$ increases.}
  \label{fig2:isocontourns}
\end{figure*}

\subsection{Numerical simulations}\label{sec_NumSim}

To solve numerically the incompressible MHD Eqs.~(\ref{eq:MHD_v})
and (\ref{eq:MHD_b}) we employ a parallel pseudo-spectral code
\cite{gomez_parallel_2005, gomez_mhd_2005, hybrid2011}. We consider a
spatial resolution of $N^3 = 512^3$ grid points, with a second-order
Runge-Kutta time integration scheme. Spatial resolution is moderate as
we need to store a large amount of data in space and time to compute
the correlation functions and spectra defined in
Sec.~\ref{sec_Wfspectrum_and_Gamma}.  Values considered for the
intensity of the external magnetic field are $B_0 = 0, 0.25$, $1$,
$2$, $4$ and $8$ (in units of the initial r.m.s.~magnetic fluctuations
value). We assume periodic boundary conditions in a cube of side $2\pi
L$ (with $L$ the initial correlation length of the fluctuations,
defined as the unit length). Aliasing is removed by the two-thirds
rule truncation method.

\begin{figure*}
  \centering 
  \subfigure[$B_0 = 0$, $\vec{z}^-$, $\sigma_c = 0.3$]{\includegraphics[width=0.45\textwidth]{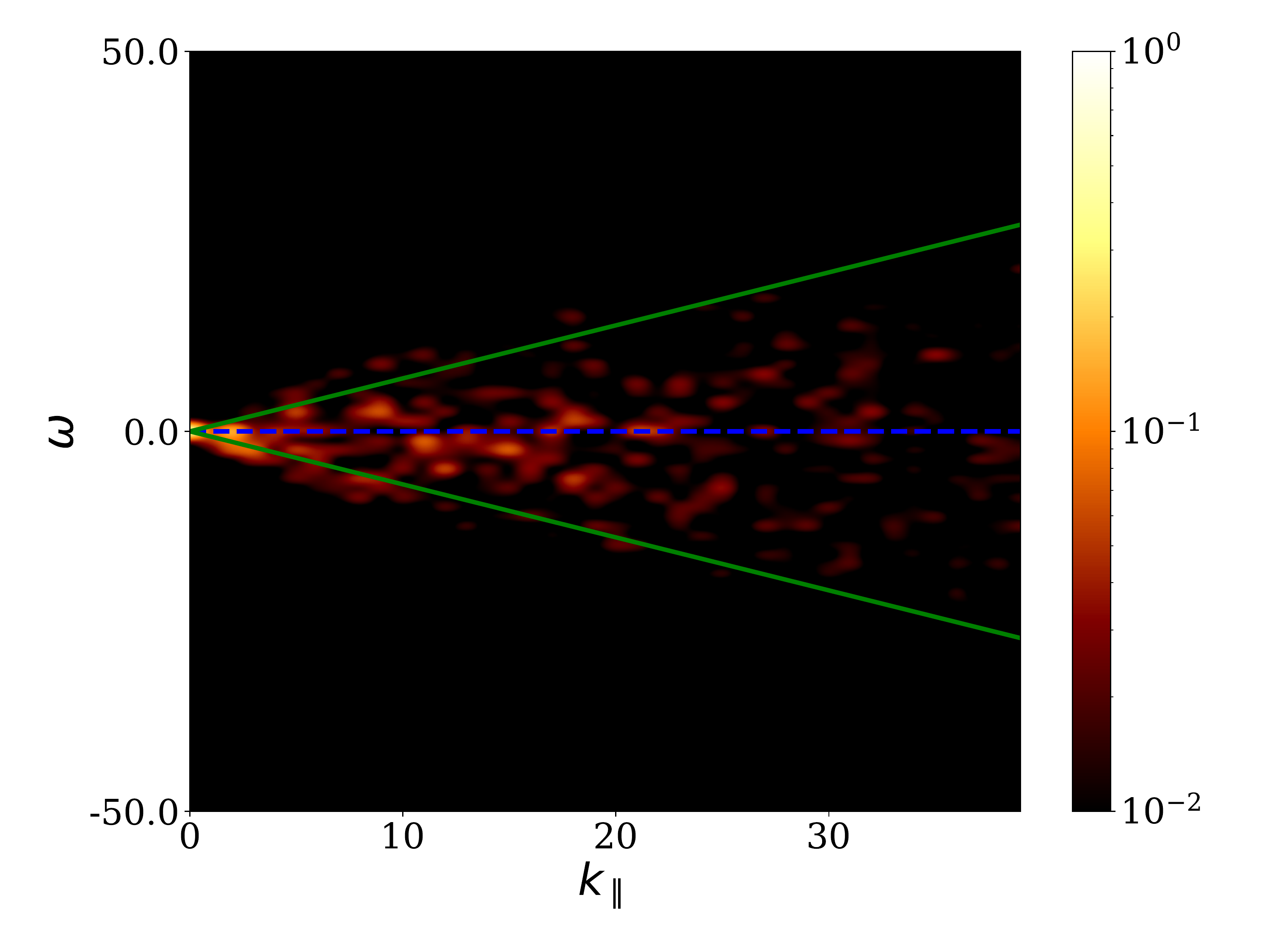}}
  \subfigure[$B_0 = 0$, $\vec{z}^+$, $\sigma_c = 0.3$]{\includegraphics[width=0.45\textwidth]{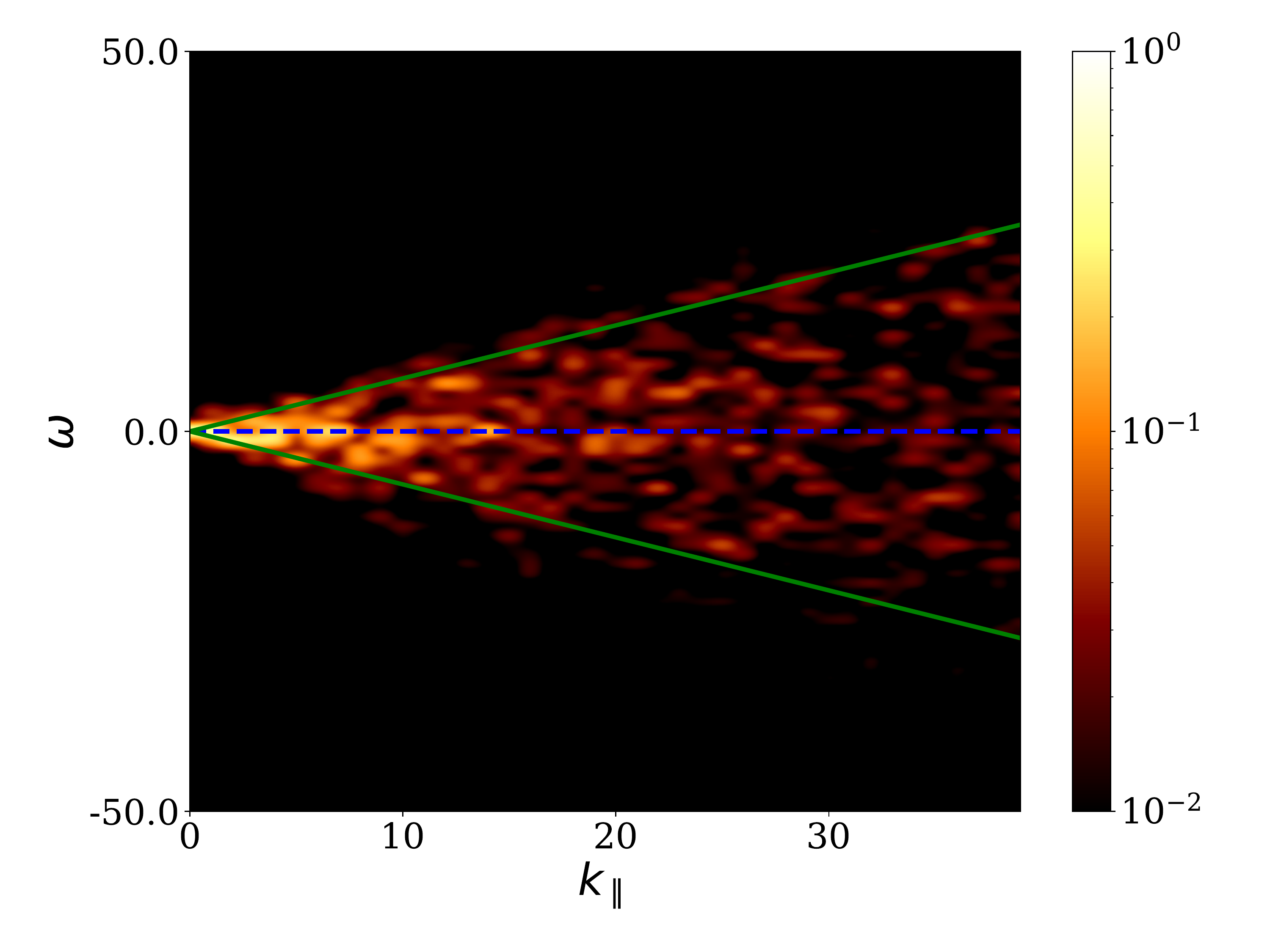}}
  \subfigure[$B_0 = 0$, $\vec{z}^-$, $\sigma_c = 0.9$]{\includegraphics[width=0.45\textwidth]{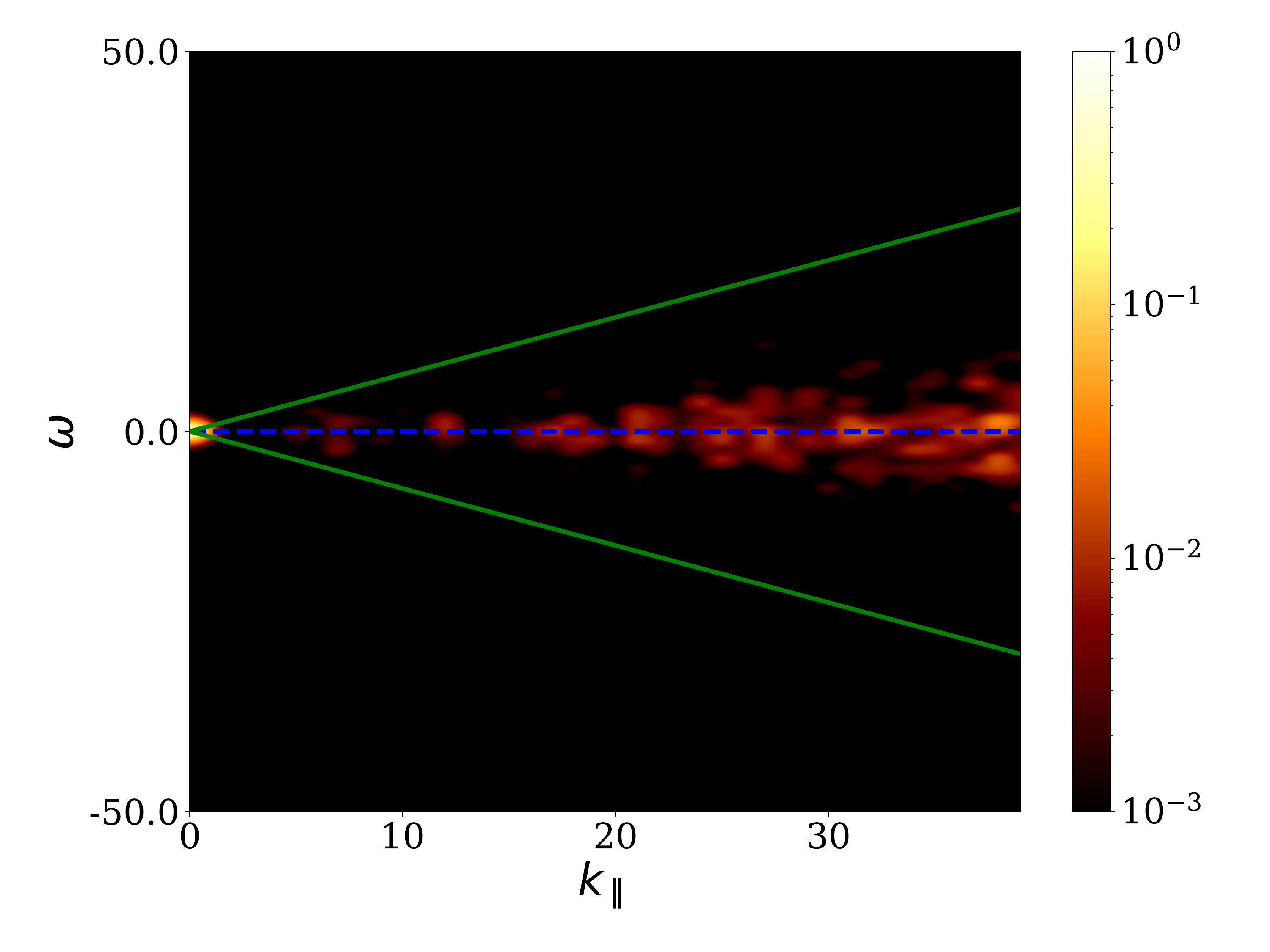}}
  \subfigure[$B_0 = 0$, $\vec{z}^+$, $\sigma_c = 0.9$]{\includegraphics[width=0.45\textwidth]{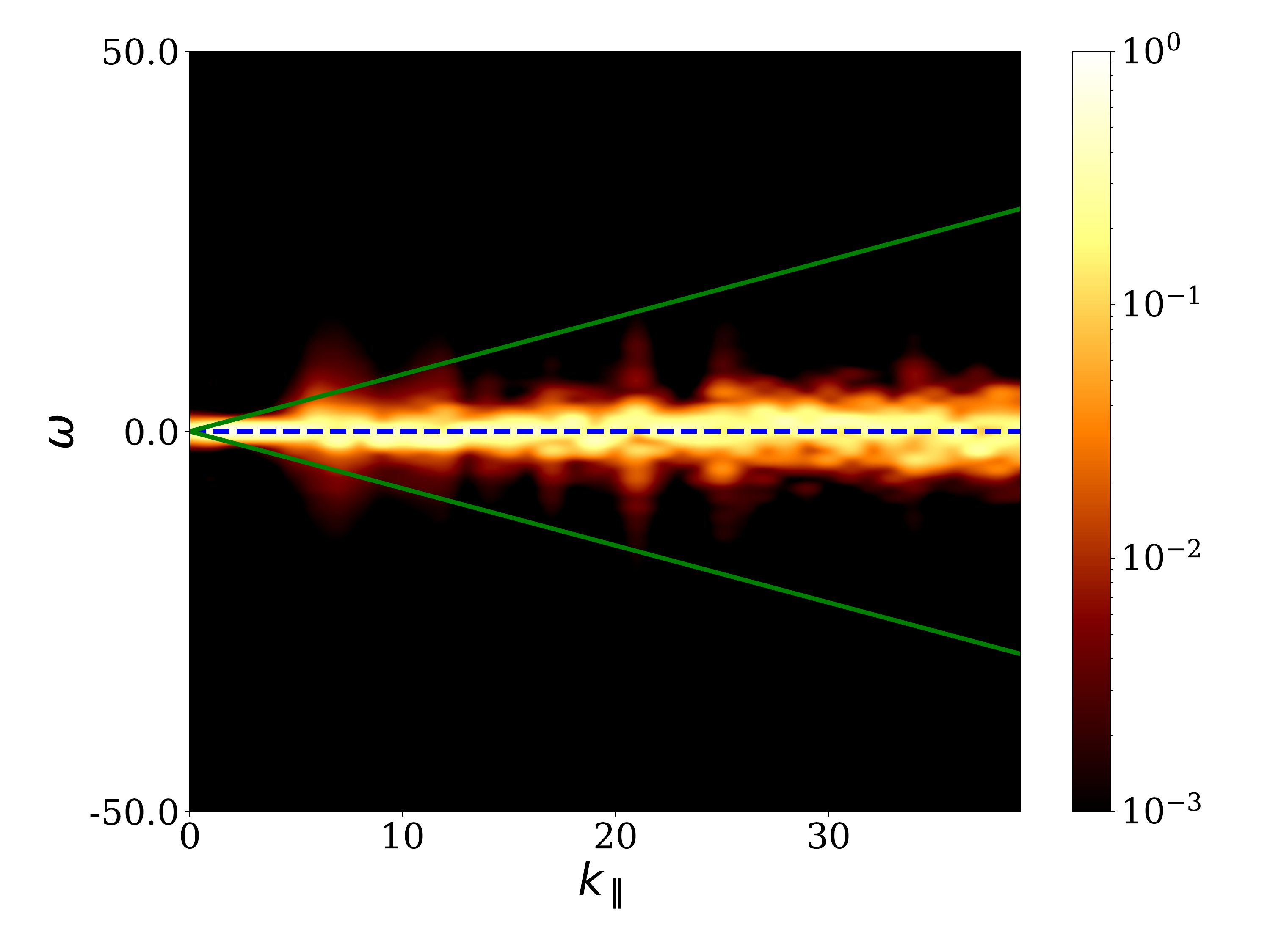}}
  \caption{Normalized wave vector and frequency spectra $E^\pm({\bf
      k}, \omega)/E^+({\bf k})$ of $\vec{z}^-$ (left) and $\vec{z}^+$
    (right), for the isotropic simulations ($B_0=0$) with $\sigma_c =
    0.3$ [top, panels (a) and (b)] and $\sigma_c = 0.9$ [bottom,
      panels (c) and (d)], as a function of $k_\parallel$ and for
    fixed $k_\perp=0$. Lighter regions indicate larger energy
    density. The spectra correspond to the power in the time and space
    Fourier transform of the fields, such that accumulation of energy
    in modes near the dispersion relation (or in all modes below the
    sweeping curve) points to a dominance of a physical effect (i.e.,
    of its associated frequency) in the dynamics of a given scale
    $\sim 1/k_\parallel$. As a reference, the sweeping time relation
    giving by Eq.~(\ref{eq:tausw}) is indicated by solid (green)
    lines. A broad excitation of modes is observed for all modes with
    $\omega \leq 1/\tau_{sw}$ (sweeping) in panels (a) and (b), and
    for $\omega \approx 0$ in panels (c) and (d).}
  \label{fig3:B0_spectrum_Hc}
\end{figure*}

The initial condition in all simulations consists of nonzero
amplitudes for the ${\bf v}({\bf k})$ and ${\bf b}({\bf k})$ fields,
equipartitioned in all the wavenumbers within shells $1.1 \leq k \leq
4$ (in units of $2\pi L/\lambda$, with $\lambda$ the
wavelength). Random phases are chosen for all Fourier modes in both
fields. To keep the system in a turbulent steady state we apply a
driving consisting of forcing terms $\vec{F}_b$ and $\vec{F}_v$ for
$\vec{b}$ and $\vec{v}$ respectively, in Eqs.~(\ref{eq:MHD_v}) and
(\ref{eq:MHD_b}). $\vec{F}_b$ and $\vec{F}_v$ are band limited to a
fixed set of Fourier modes in the band $0.9\leq k \leq 1.8$. The
driving has a random and a time-coherent component, and the
correlation time of the forcing is $\tau_f \approx 1$ (of the order of
the unit time $t_0$), which is larger than all the characteristic
times defined in the previous section. To change the level of
cross-helicity in the flow, correlations were introduced between the
mechanical and electromotive drivings, resulting at late times
(depending on the level of cross-correlation between the drivers) in a
normalized cross-helicity of $\sigma_c=0$, $0.3$, or $0.9$ (these
values correspond to the time average in the turbulent steady state;
in practice, in each simulation the instantaneous cross-helicity
fluctuates in time around the reported mean values).

\begin{table}
\begin{tabular}{|l||c||c||c||c||c||c|}
\hline
        & $B_0=0$ & $B_0 = 0.25$ & $B_0 = 1$ & $B_0 = 2$ & $B_0 = 4$ & $B_0 = 8$ \\ \hline\hline
              & $0$ & $0$ & $0$ & $0$ & $0$ & $0$ \\ \cline{2-7} 
$\sigma_c \approx $ & $0.3$ & $0.3$ & $0.3$ & $0.3$ & $0.3$ & $0.3$ \\ \cline{2-7} 
              & $0.9$ & $0.9$ & $0.9$ & $0.9$ & $0.9$ & $0.9$ \\ \hline
\end{tabular}
\caption{List of numerical simulations performed, with guide field
$\vec{B} = B_0 \hat{x}$ and normalized cross-helicity $\sigma_c$.}
\label{tab:listSim}
\end{table}

Note the different values of $B_0$ and of $\sigma_c$ explored result
in a total of 18 simulations (see \cref{tab:listSim}). All simulations
were continued until the system reached a turbulent steady state, and
then continued further to perform the spatio-temporal analysis on the
evolution of the Els\"asser fields presented in the next section. We
will first characterize the spatial behavior of the flows (specially
considering the degree of anisotropy as the intensity of the
background flow is increased), to then study the behavior of the
Els\"asser fluctuations using the spatio-temporal information.

\section{Results}\label{sec_results}

\begin{figure*}
  \centering
  \subfigure[$B_0 = 0.25$, $\vec{z}^-$, $\sigma_c = 0$]{\includegraphics[width=0.45\textwidth]{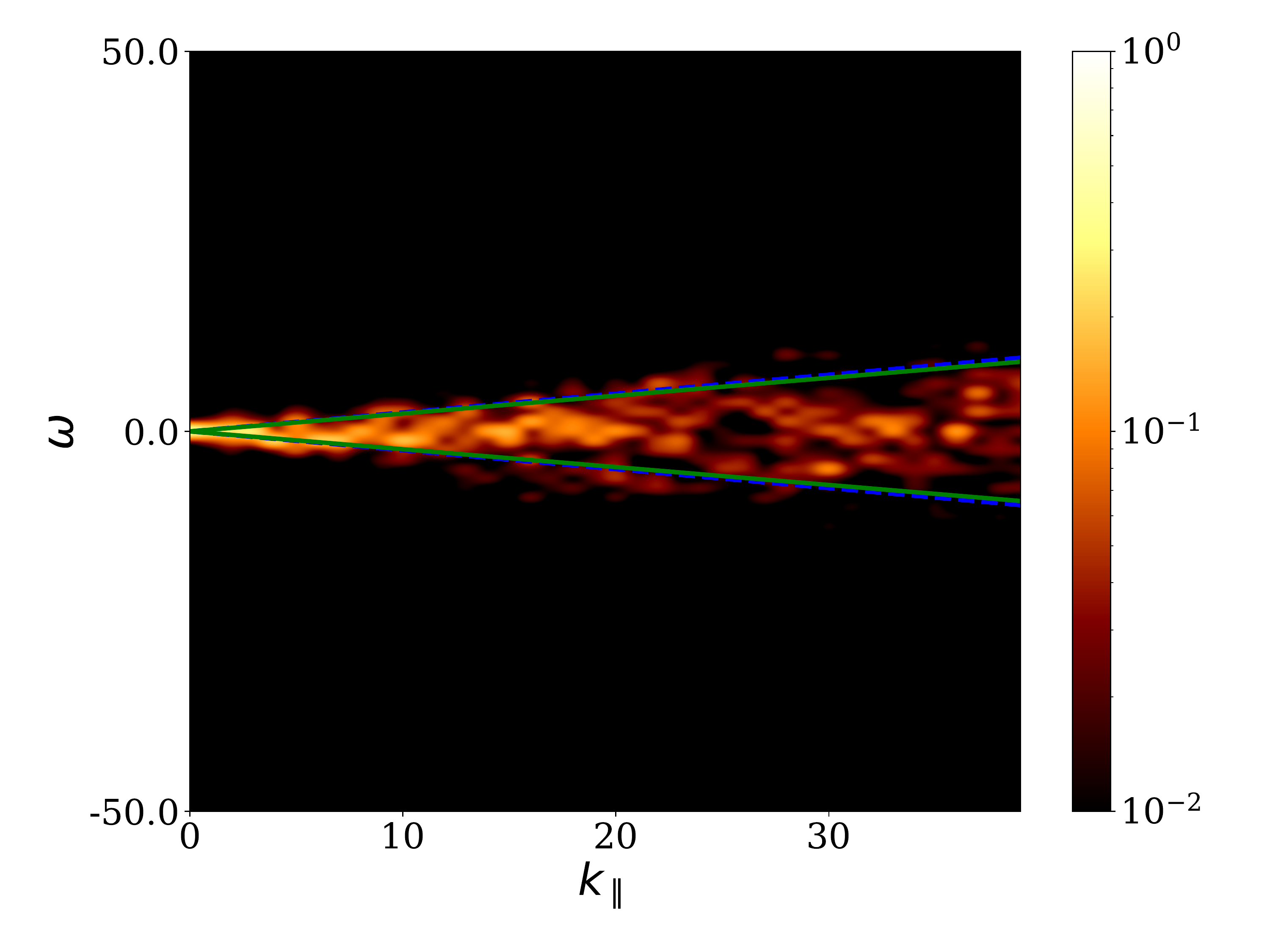}}
  \subfigure[$B_0 = 0.25$, $\vec{z}^+$, $\sigma_c = 0$]{\includegraphics[width=0.45\textwidth]{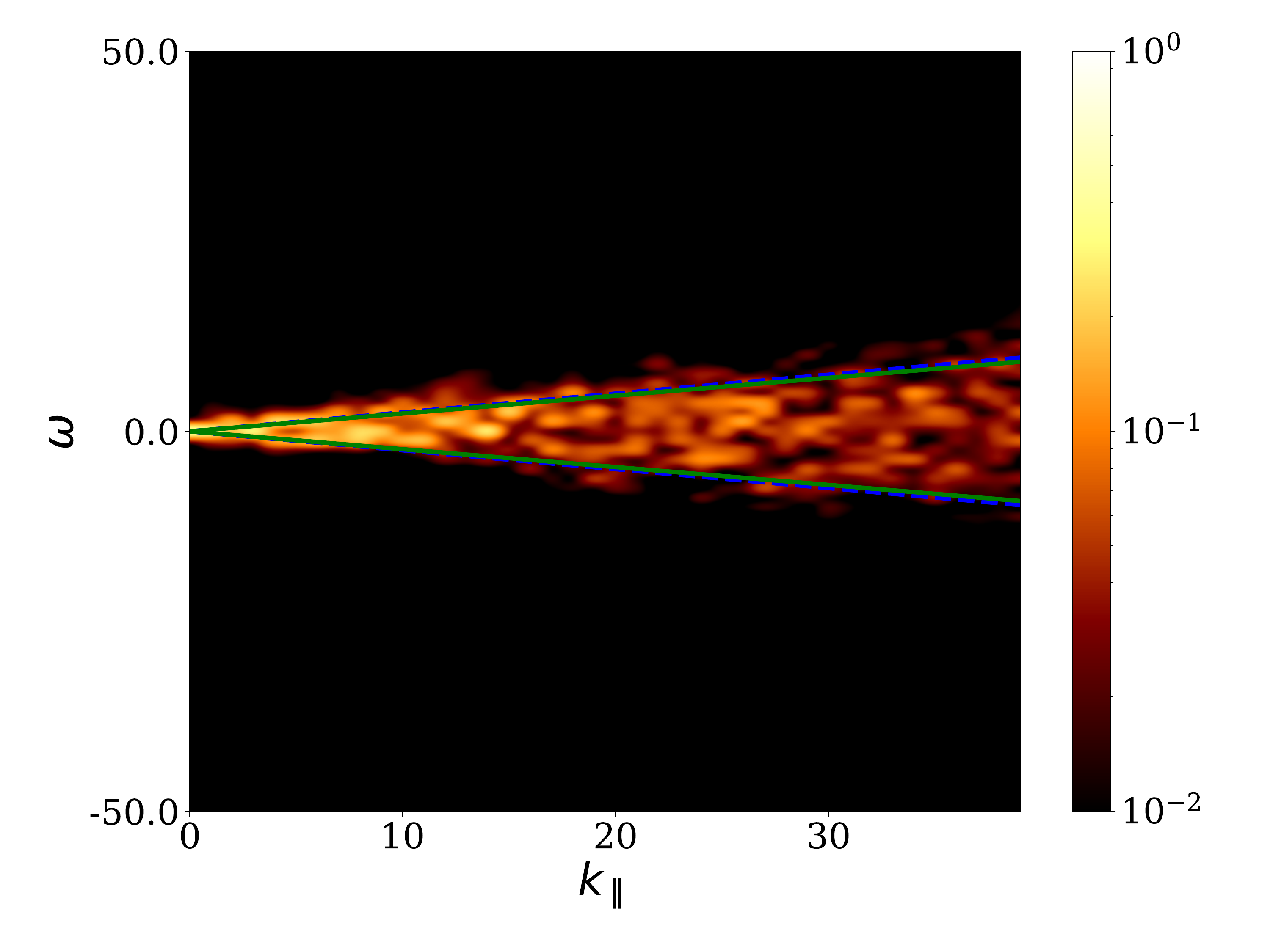}}
  \subfigure[$B_0 = 0.25$, $\vec{z}^-$, $\sigma_c = 0.3$]{\includegraphics[width=0.45\textwidth]{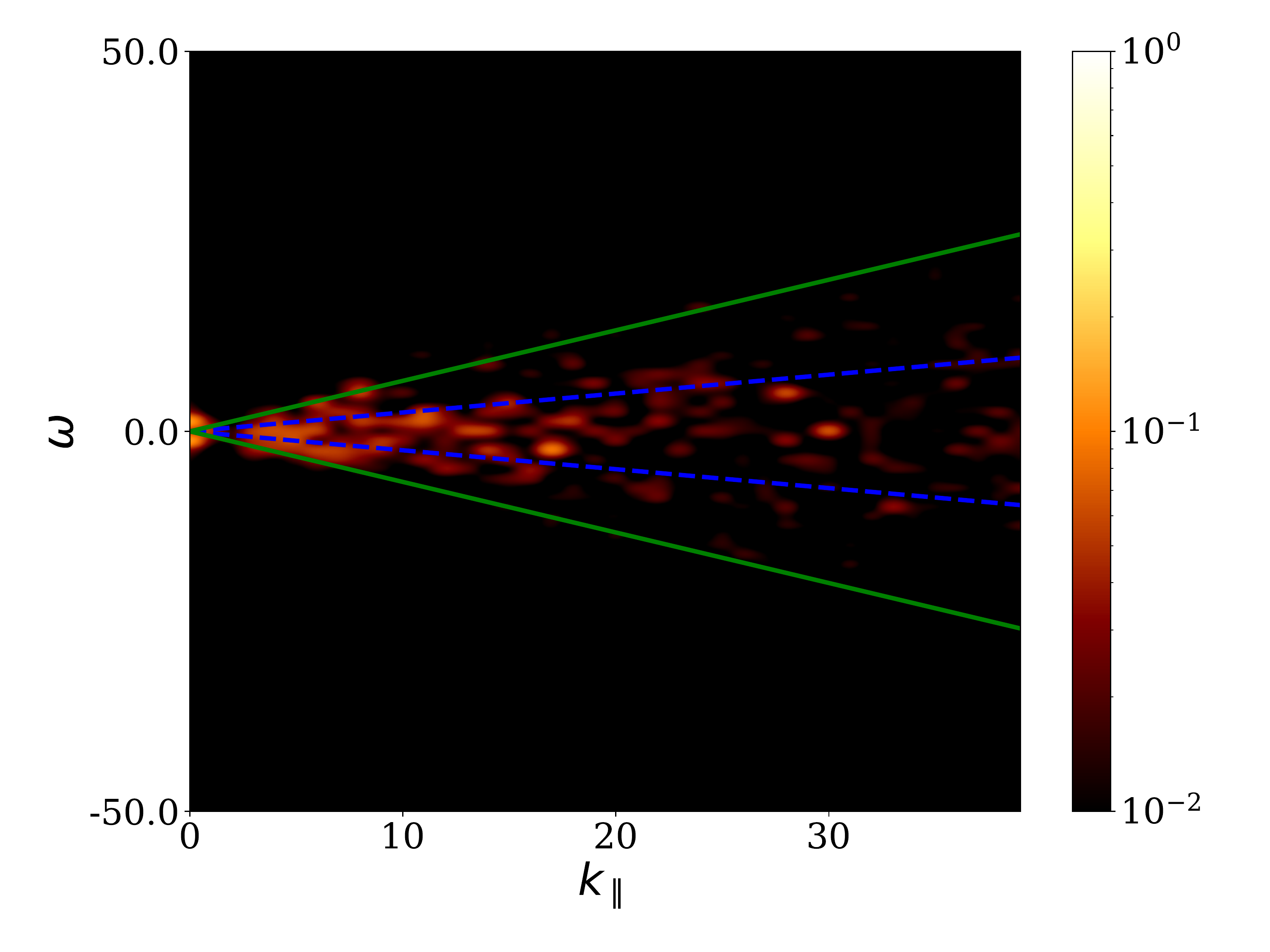}}
  \subfigure[$B_0 = 0.25$, $\vec{z}^+$, $\sigma_c = 0.3$]{\includegraphics[width=0.45\textwidth]{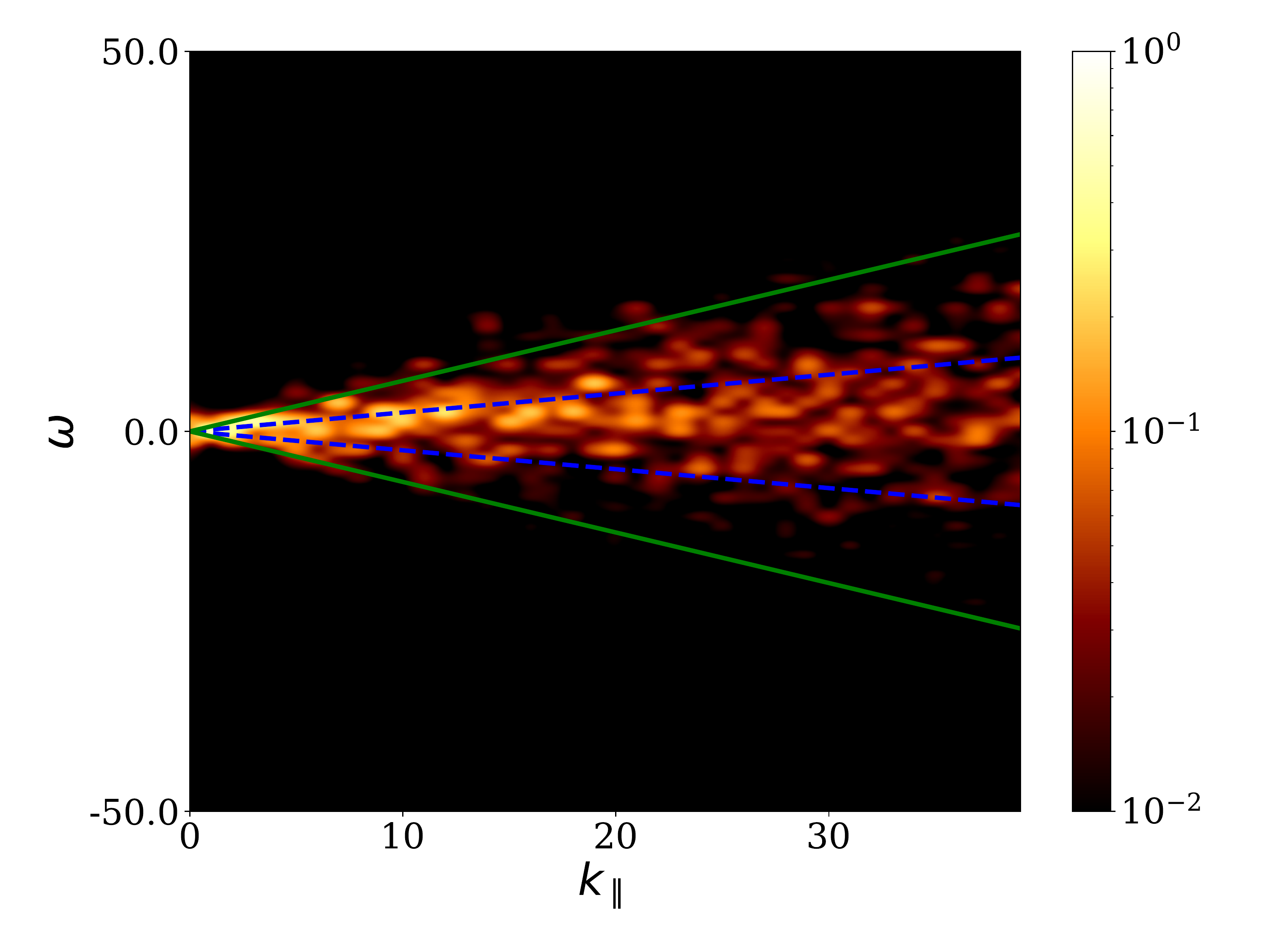}}
  \subfigure[$B_0 = 0.25$, $\vec{z}^-$, $\sigma_c = 0.9$]{\includegraphics[width=0.45\textwidth]{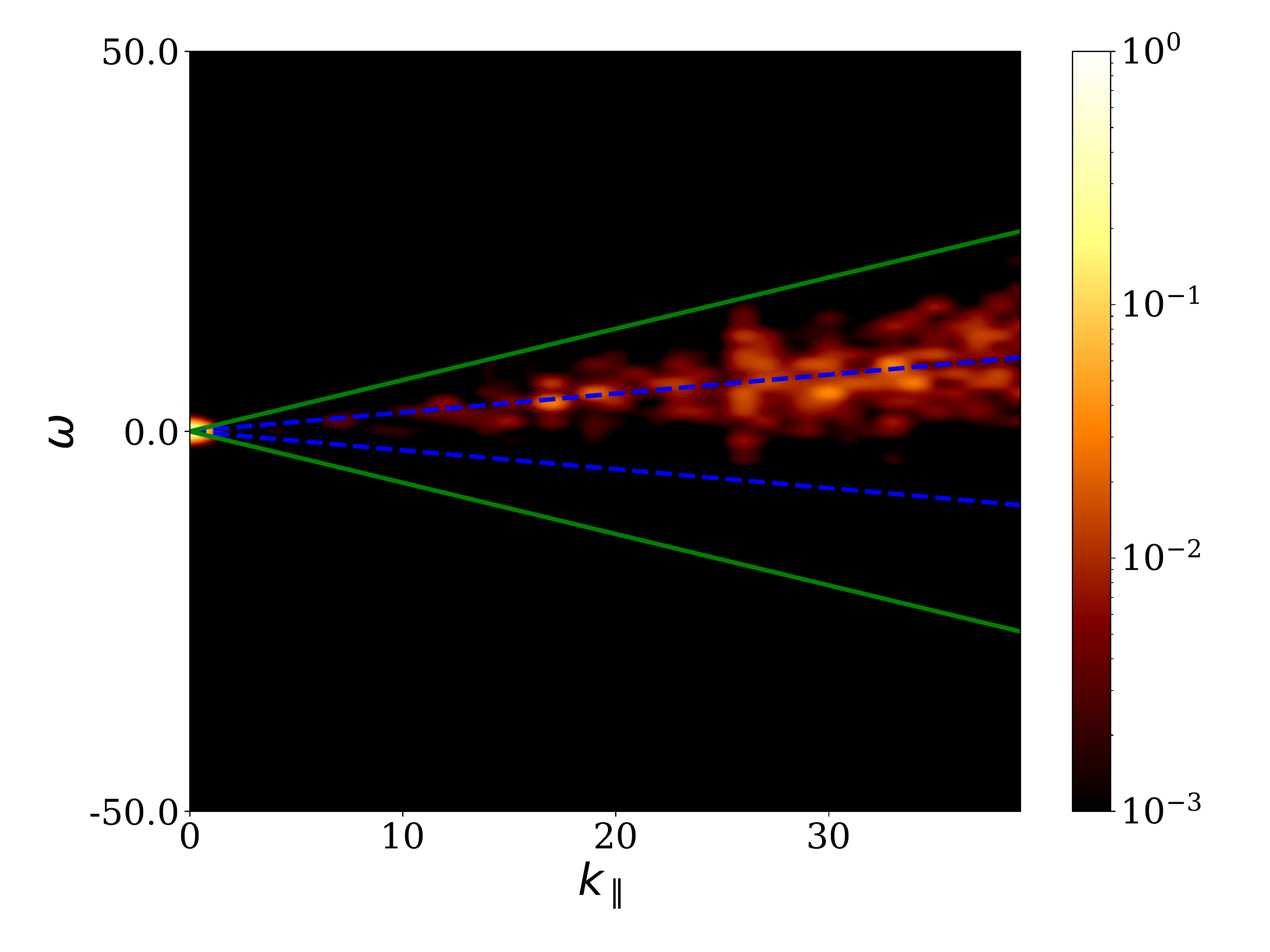}}
  \subfigure[$B_0 = 0.25$, $\vec{z}^+$, $\sigma_c = 0.9$]{\includegraphics[width=0.45\textwidth]{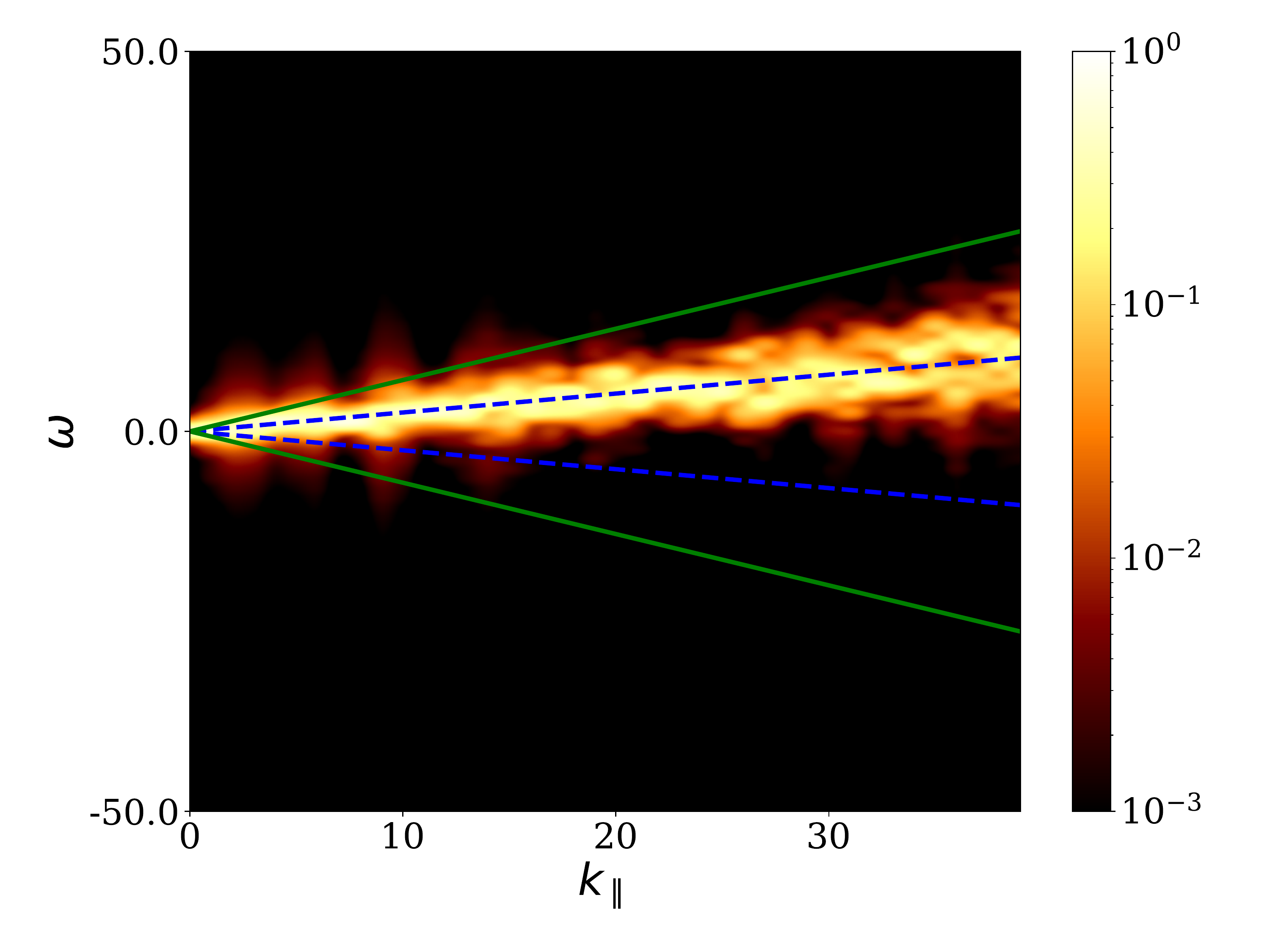}}
    \caption{Normalized spectra $E^\pm({\bf k}, \omega)/E^+({\bf k})$
      of $\vec{z}^-$ (left) and $\vec{z}^+$ (right), for the runs with
      $B_0=0.25$, for modes with $k_\perp=0$, and thus as a function
      of $k_\parallel$ and $\omega$. Panels (a) and (b) correspond to
      $\sigma_c = 0$, (c) and (d) to $\sigma_c = 0.3$, and (e) and (f)
      to $\sigma_c = 0.9$.  The sweeping time relation, given by
      Eq.~(\ref{eq:tausw}), is indicated by solid (green) lines, and
      the dashed (blue) lines indicate the dispersion relation of
      Alfv\'en waves.  Lighter regions indicate larger energy density,
      and the accumulation of energy in modes near the dispersion
      relation (or in all modes below the sweeping curve) points to a
      dominance of a physical effect (i.e., of its associated
      frequency) in the dynamics of a given scale $\sim
      1/k_\parallel$. For low normalized cross-helicity $\sigma_c$
      sweeping is the dominant effect, while for large $\sigma_c$
      energy accumulates near the dispersion relation of the waves,
      albeit for both $\vec{z}^+$ and $\vec{z}^-$ with the same sign
      of the frequency $\omega$.}
  \label{fig3:B025_spectrum_Hc}
\end{figure*}

\begin{figure*}
  \centering
  \subfigure[$B_0 = 1.0$, $\vec{z}^-$, $\sigma_c = 0$]{\includegraphics[width=0.45\textwidth]{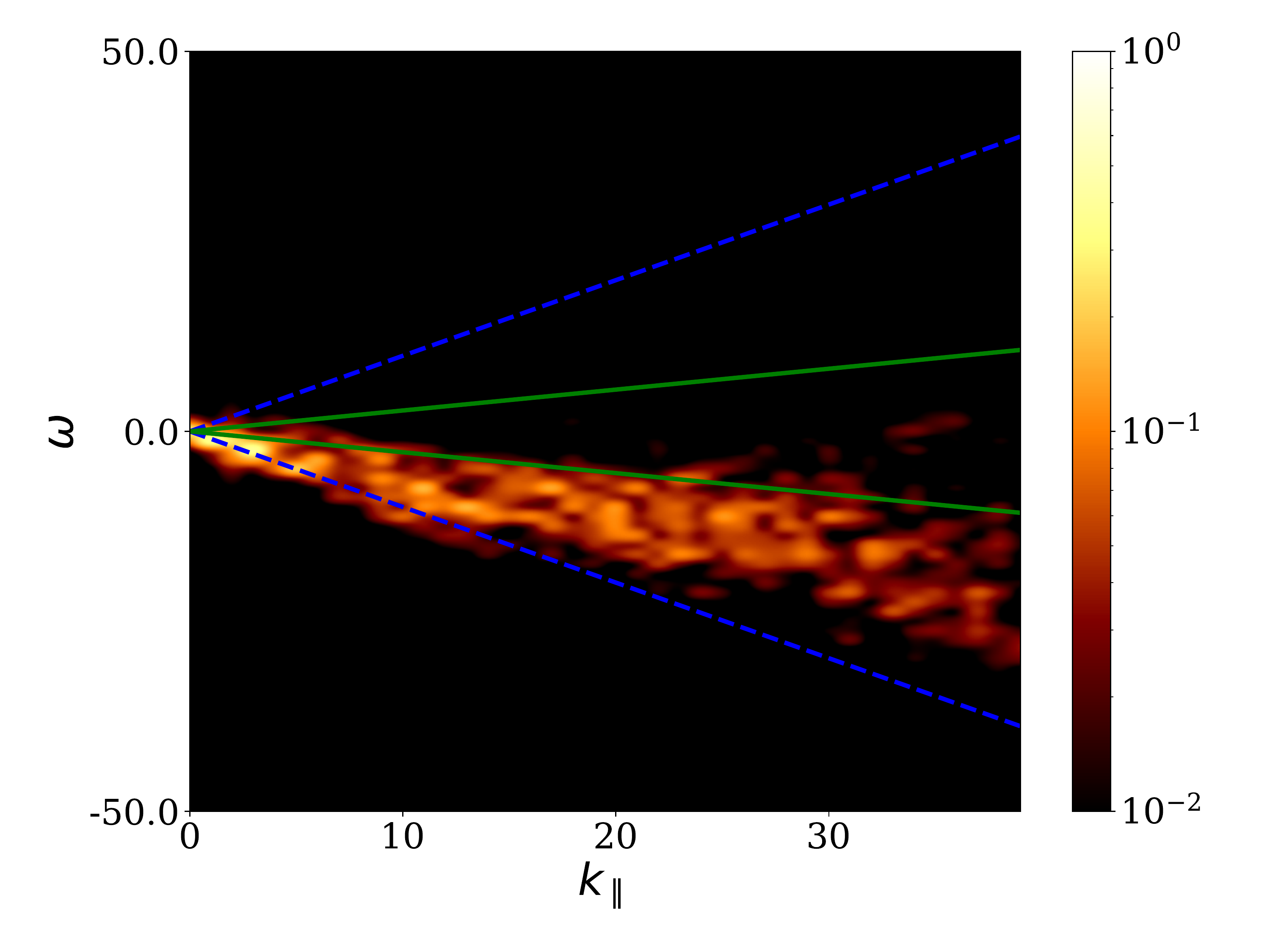}}
  \subfigure[$B_0 = 1.0$, $\vec{z}^+$, $\sigma_c = 0$]{\includegraphics[width=0.45\textwidth]{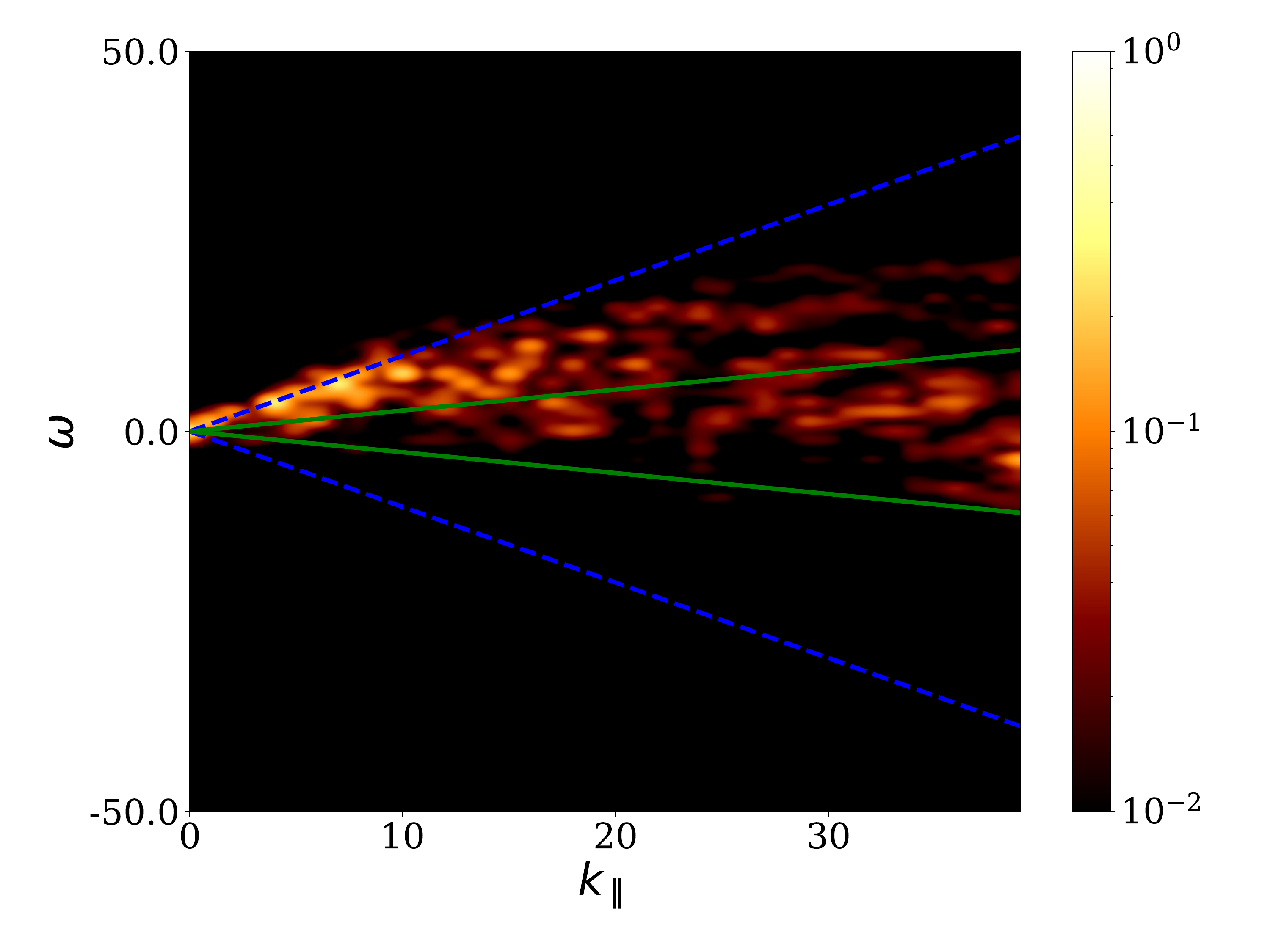}}
  \subfigure[$B_0 = 1.0$, $\vec{z}^-$, $\sigma_c = 0.3$]{\includegraphics[width=0.45\textwidth]{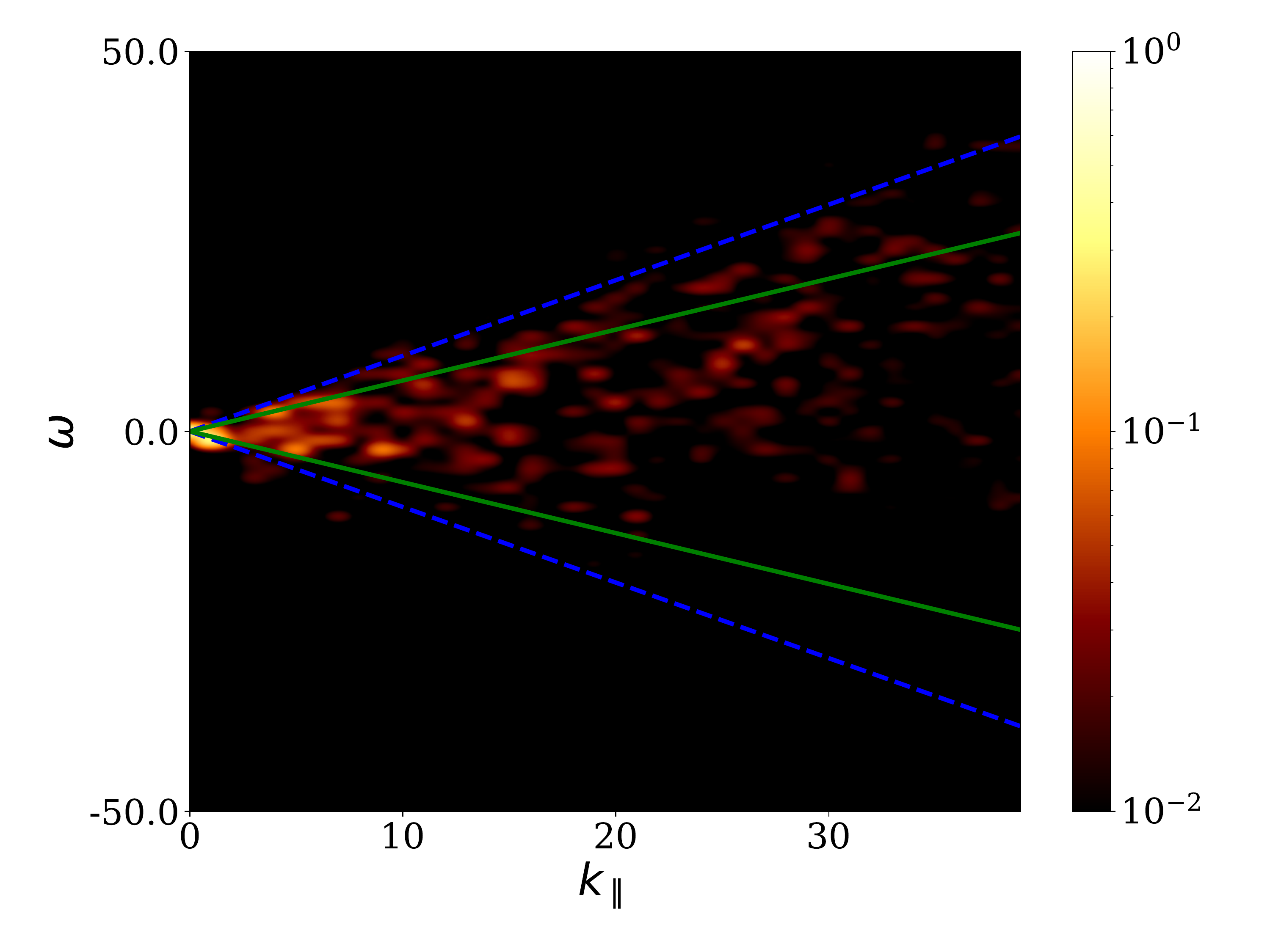}}
  \subfigure[$B_0 = 1.0$, $\vec{z}^+$, $\sigma_c = 0.3$]{\includegraphics[width=0.45\textwidth]{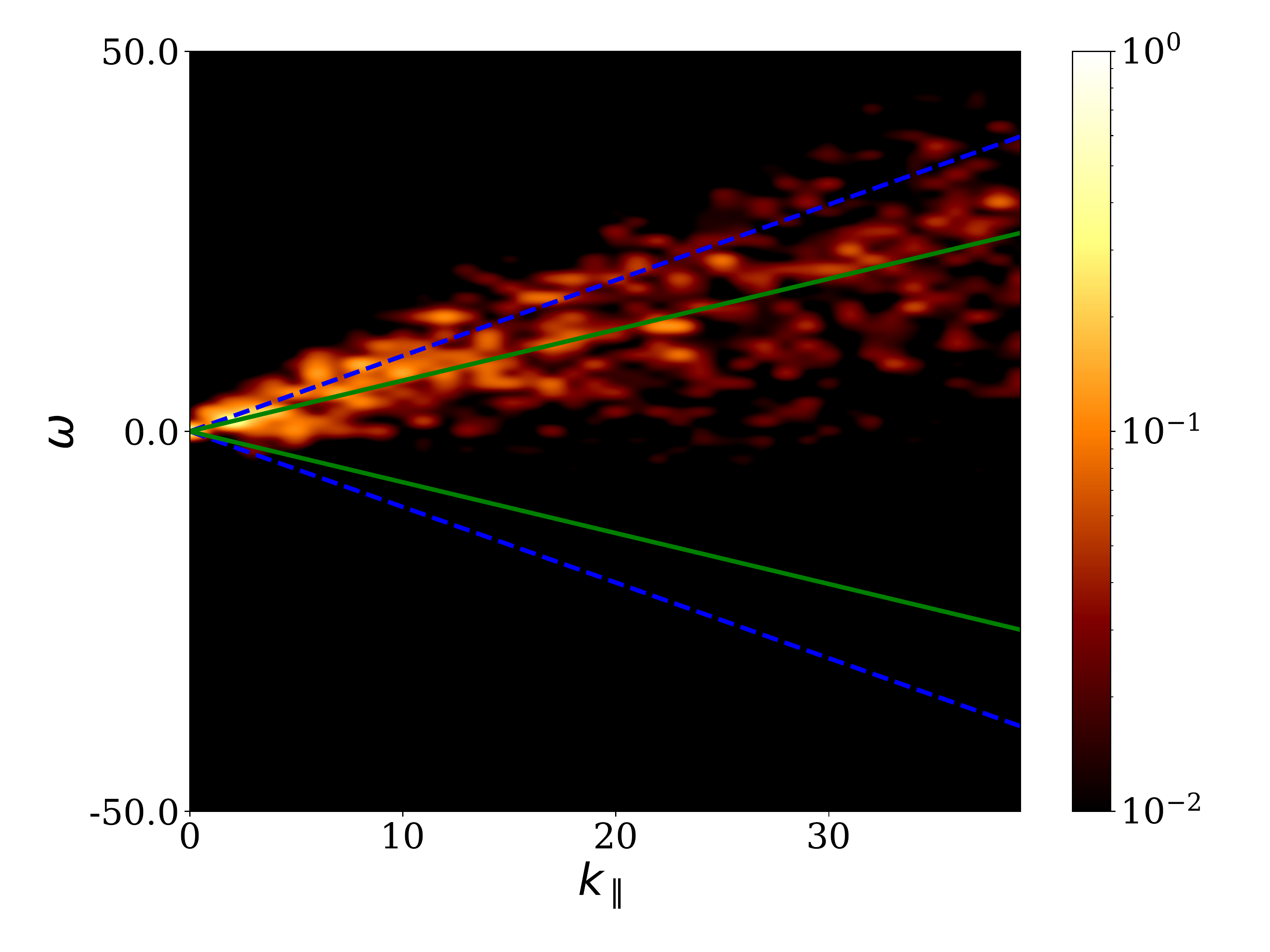}}
  \subfigure[$B_0 = 1.0$, $\vec{z}^-$, $\sigma_c = 0.9$]{\includegraphics[width=0.45\textwidth]{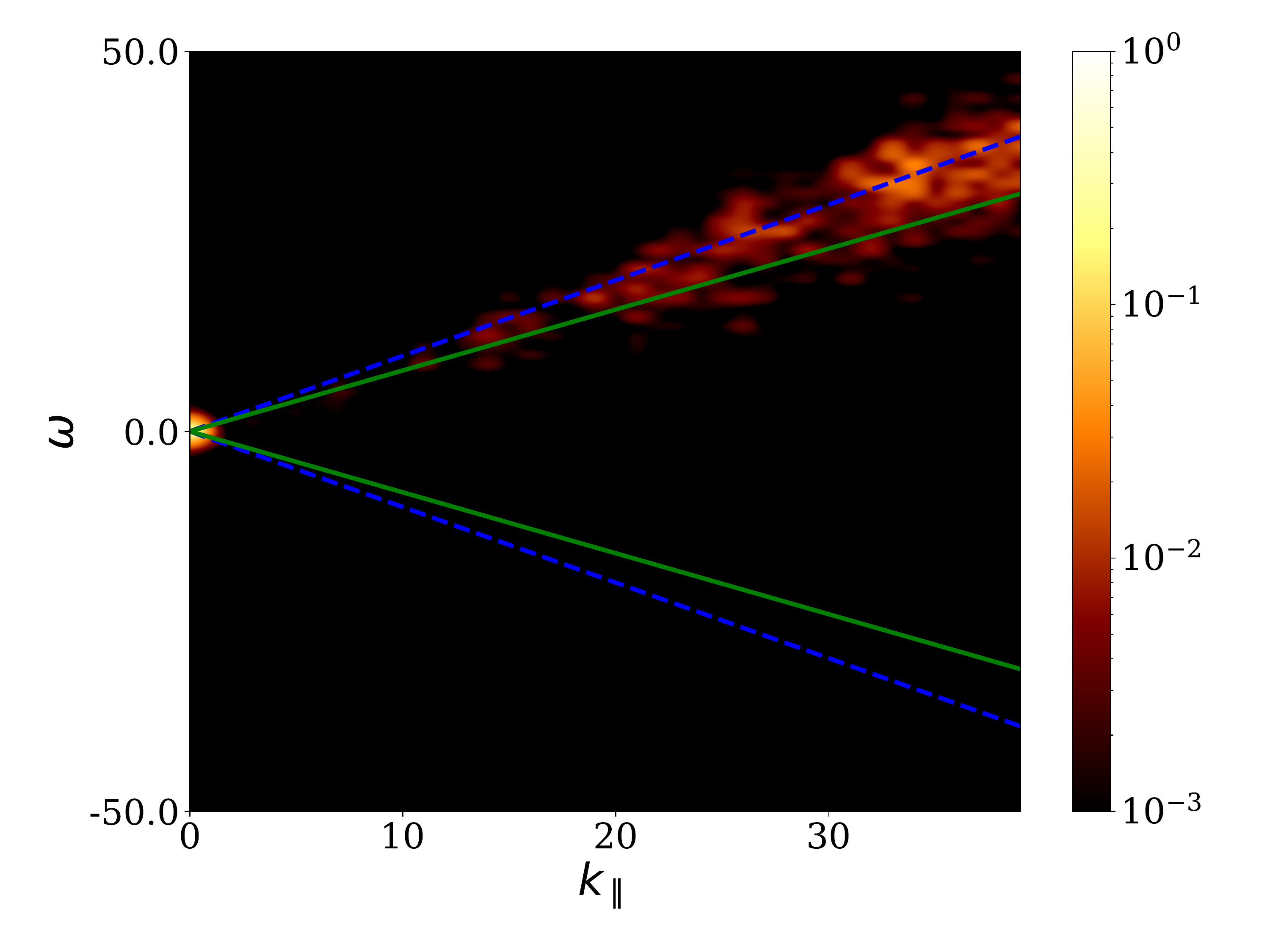}}
  \subfigure[$B_0 = 1.0$, $\vec{z}^+$, $\sigma_c = 0.9$]{\includegraphics[width=0.45\textwidth]{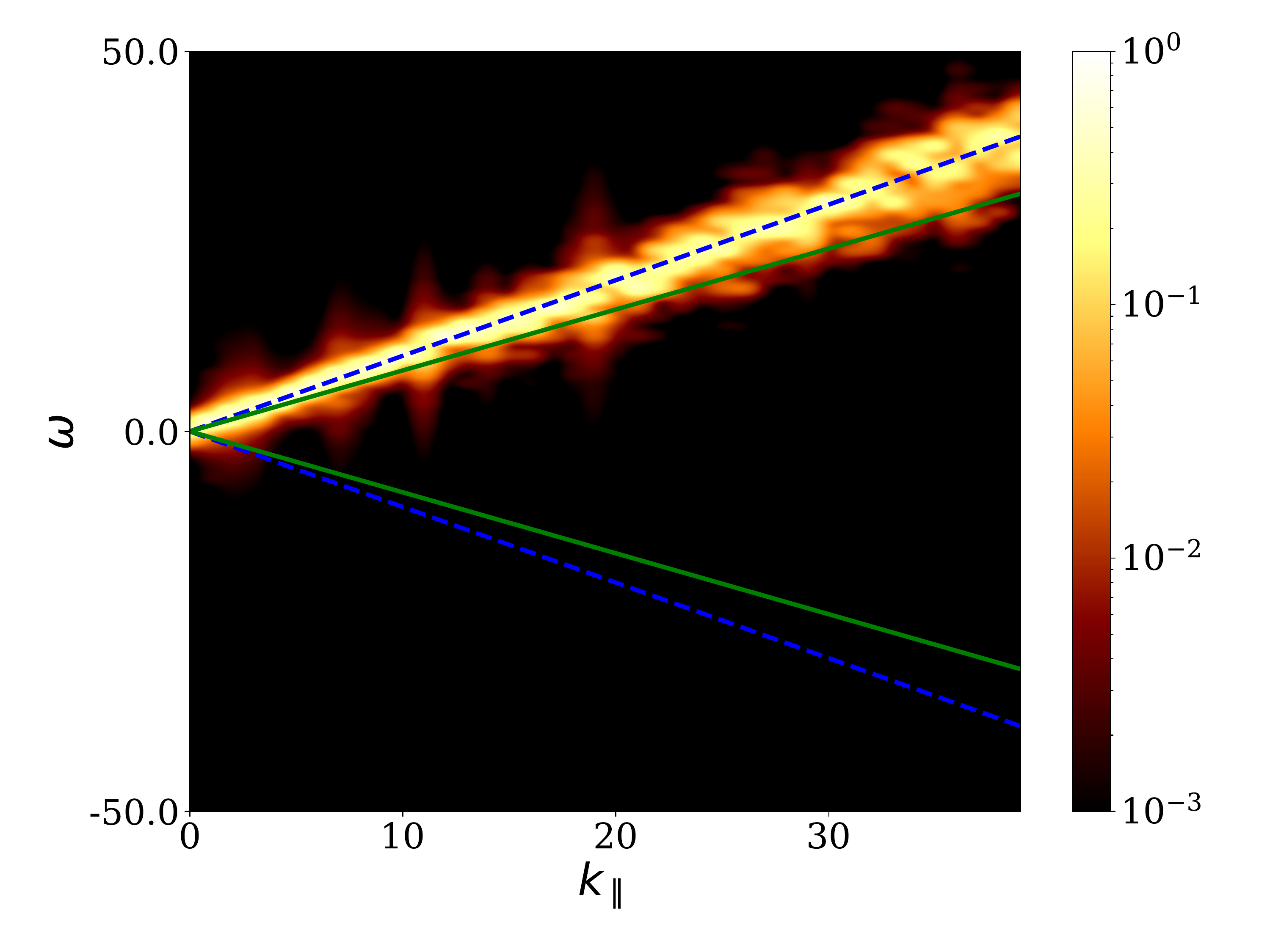}}
    \caption{Normalized spectra $E^\pm({\bf k}, \omega)/E^+({\bf k})$
      of $\vec{z}^-$ (left) and $\vec{z}^+$ (right), for the runs with
      $B_0=1$, for modes with $k_\perp=0$, and thus as a function of
      $k_\parallel$ and $\omega$. Panels (a) and (b) correspond to
      $\sigma_c = 0$, (c) and (d) to $\sigma_c = 0.3$, and (e) and (f)
      to $\sigma_c = 0.9$. The sweeping time relation, given by
      Eq.~(\ref{eq:tausw}), is indicated by solid (green) lines, and
      the dashed (blue) lines indicate the dispersion relation of
      Alfv\'en waves. Lighter regions indicate larger energy
      density. In this case power for $\sigma_c = 0$ is concentrated
      in a region near the wave dispersion relations $\omega^\pm
      \approx \pm \vec{V}_\textrm{A}\cdot\vec{k}$ up to $k_\parallel
      \approx 10$. For $\sigma_c = 0.9$, both fields $\vec{z}^+$ and
      $\vec{z}^-$ follow the same dispersion relation $\omega \approx
      + \vec{V}_\textrm{A}\cdot\vec{k}$, and Alfv\'enic excitations
      dominate over all scales.}
  \label{fig3:B1_spectrum_Hc}
\end{figure*}

\begin{figure*}
  \centering
  \subfigure[$B_0 = 8.0$, $\vec{z}^-$, $\sigma_c = 0$]{\includegraphics[width=0.45\textwidth]{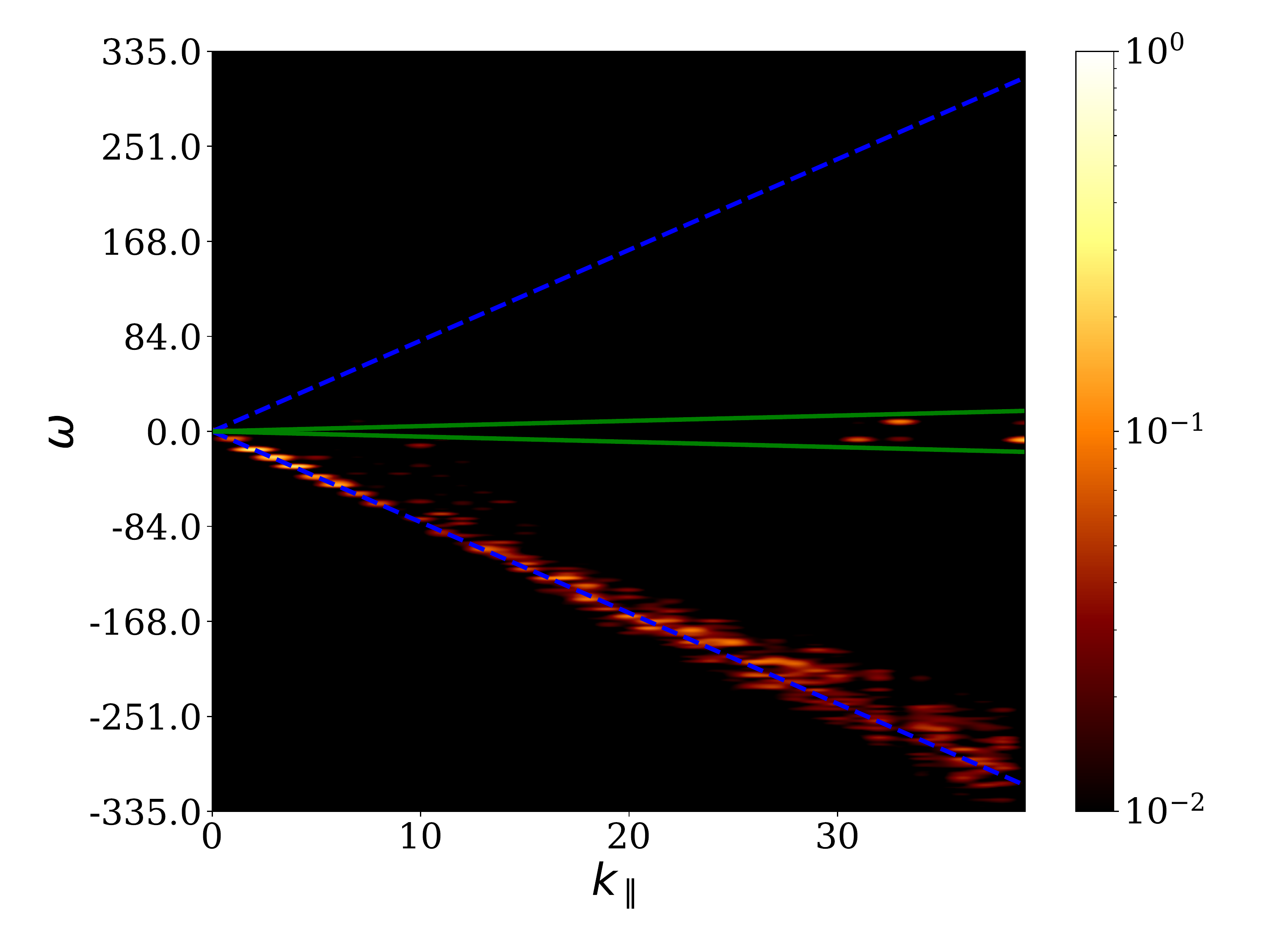}}
  \subfigure[$B_0 = 8.0$, $\vec{z}^+$, $\sigma_c = 0$]{\includegraphics[width=0.45\textwidth]{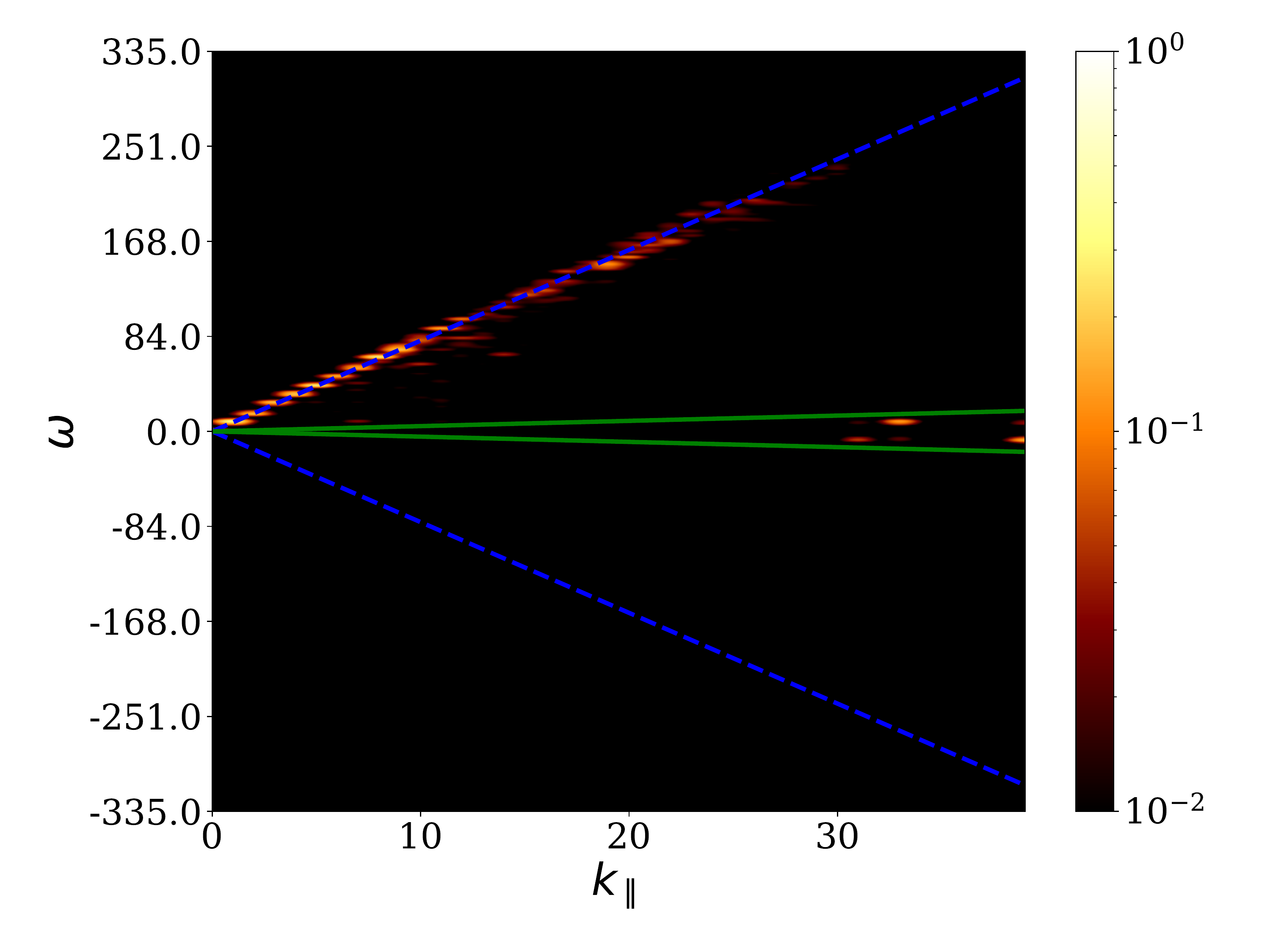}}
  \subfigure[$B_0 = 8.0$, $\vec{z}^-$, $\sigma_c = 0.3$]{\includegraphics[width=0.45\textwidth]{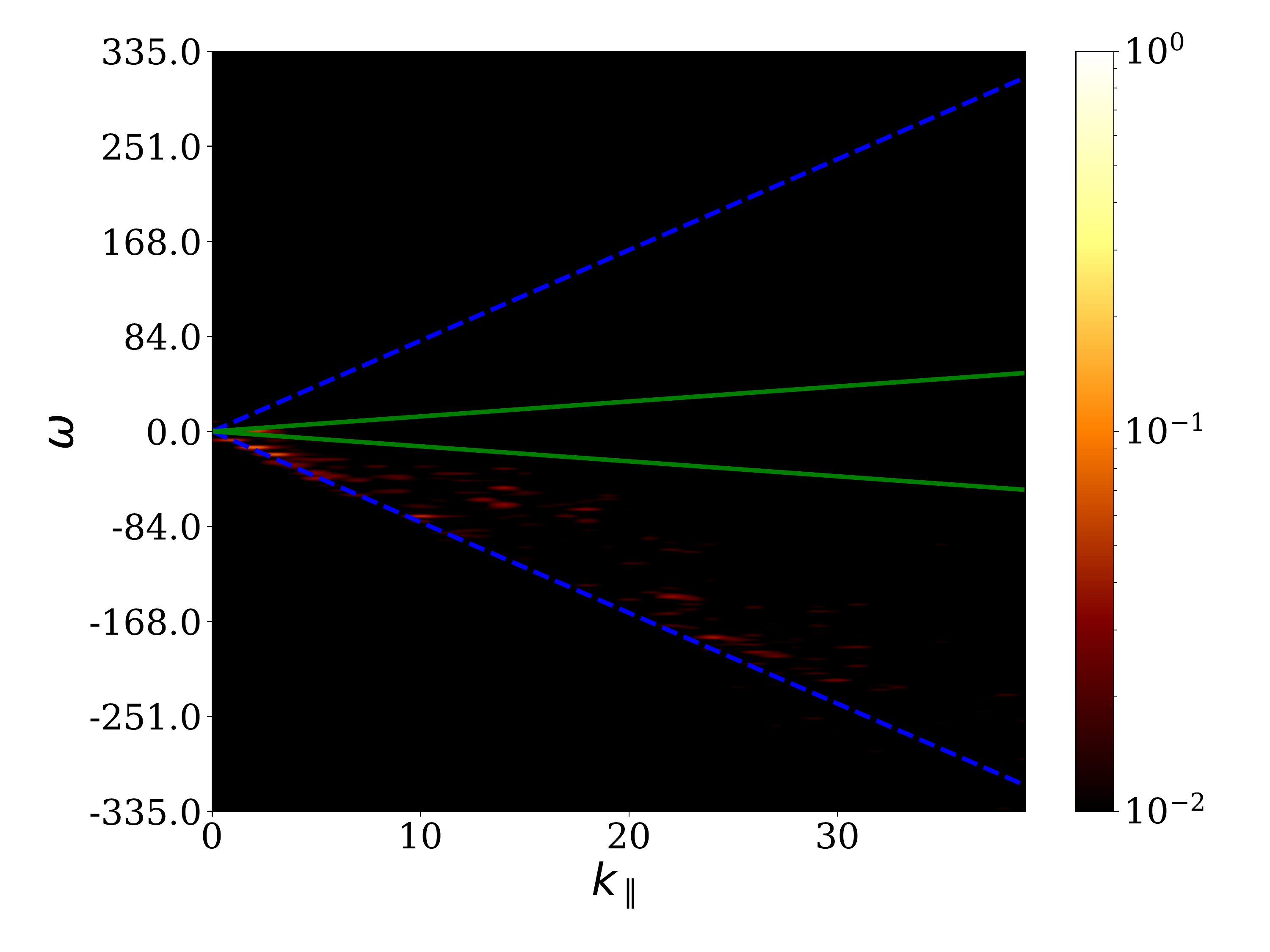}}
  \subfigure[$B_0 = 8.0$, $\vec{z}^+$, $\sigma_c = 0.3$]{\includegraphics[width=0.45\textwidth]{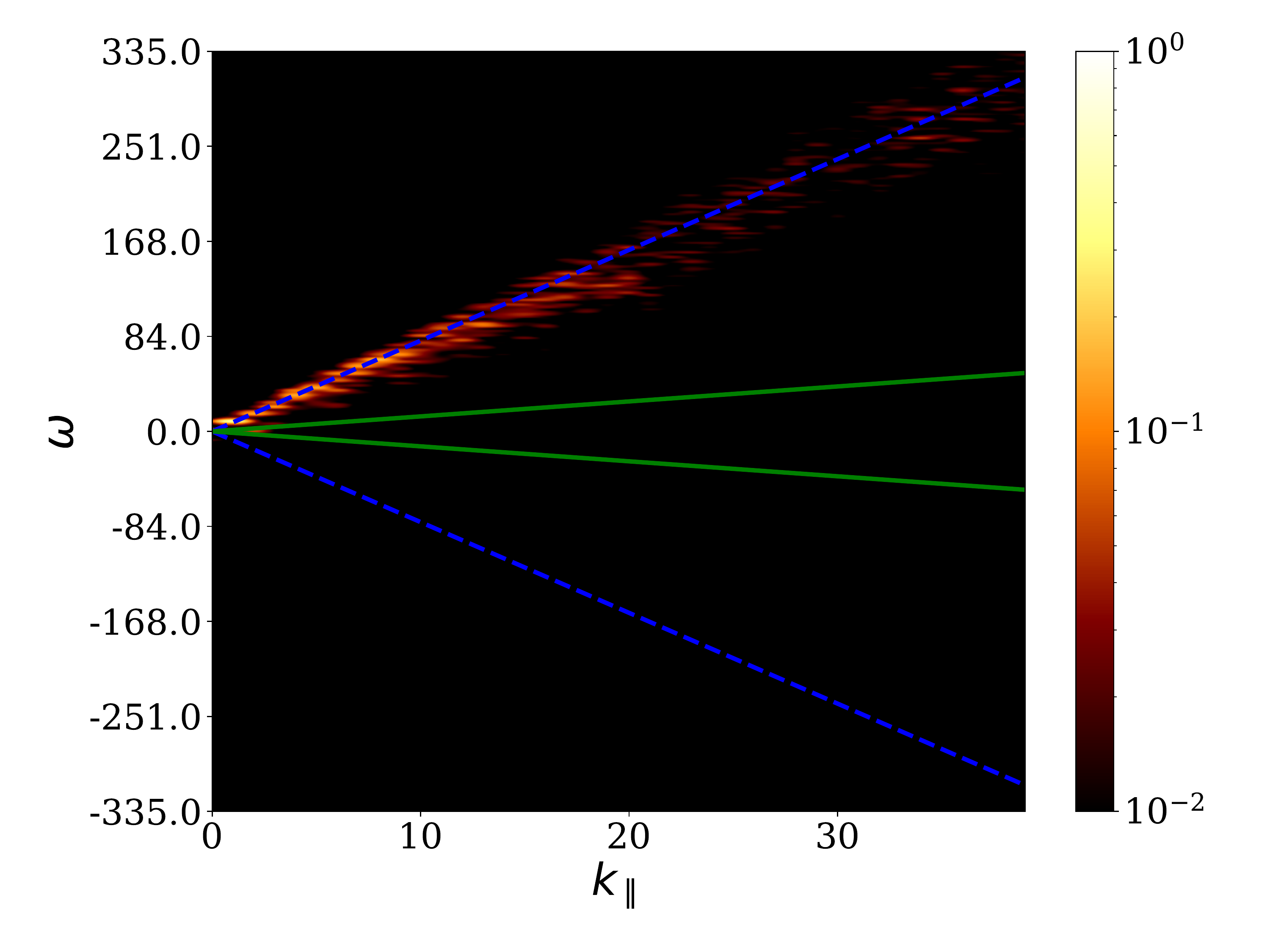}}
  \subfigure[$B_0 = 8.0$, $\vec{z}^-$, $\sigma_c = 0.9$]{\includegraphics[width=0.45\textwidth]{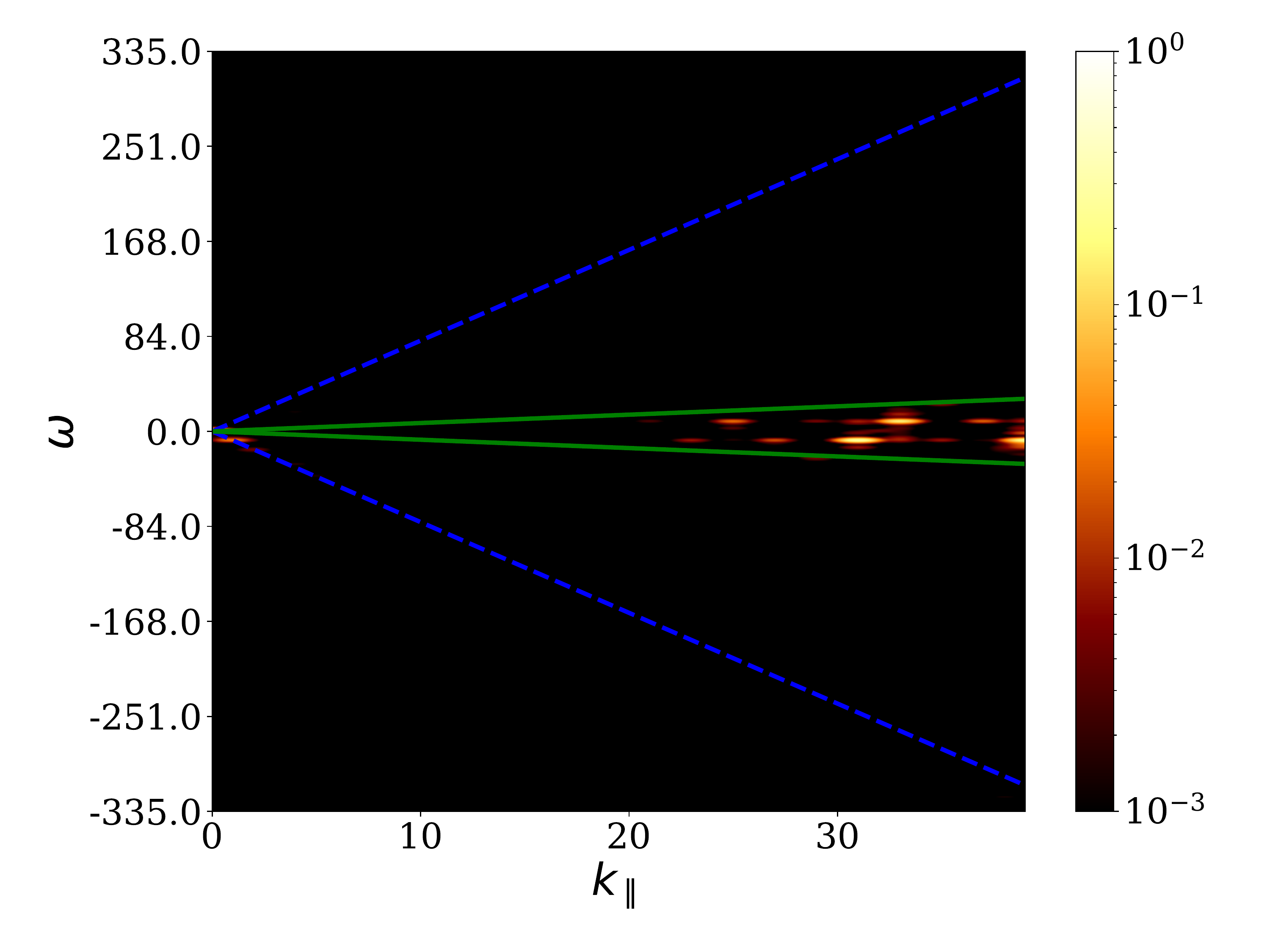}}
  \subfigure[$B_0 = 8.0$, $\vec{z}^+$, $\sigma_c = 0.9$]{\includegraphics[width=0.45\textwidth]{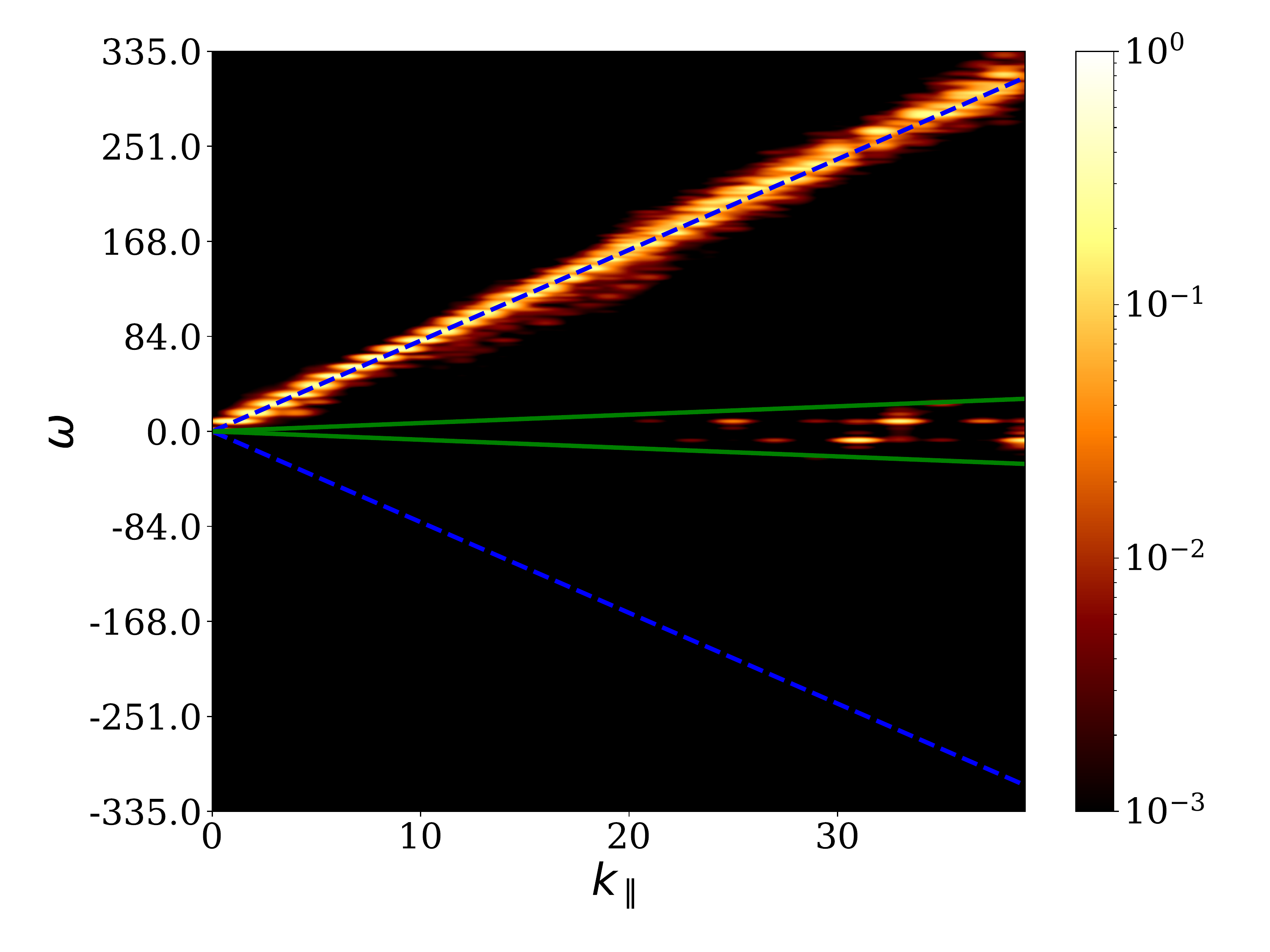}}
    \caption{Normalized spectra $E^\pm({\bf k}, \omega)/E^+({\bf k})$
      of $\vec{z}^-$ (left) and $\vec{z}^+$ (right), for the runs with
      $B_0=8$, for modes with $k_\perp=0$, and thus as a function of
      $k_\parallel$ and $\omega$. Panels (a) and (b) correspond to
      $\sigma_c = 0$, (c) and (d) to $\sigma_c = 0.3$, and (e) and (f)
      to $\sigma_c = 0.9$. The sweeping time relation, given by
      Eq.~(\ref{eq:tausw}), is indicated by solid (green) lines, and
      the dashed (blue) lines indicate the dispersion relation of
      Alfv\'en waves. Lighter regions indicate larger energy
      density. In all cases power is concentrated in a narrow region
      near the wave dispersion relations $\omega^\pm \approx \pm
      \vec{V}_\textrm{A}\cdot\vec{k}$ or near $\omega \approx 0$, for
      all the wavenumbers studied, and there is no evidence of
      counter-propagation of waves.}
  \label{fig3:B8_spectrum_Hc}
\end{figure*}

\subsection{Wavenumber spectra}

After the system reached the turbulent steady state, we analyzed the
results during $10$ large-scale unit times, after verifying that this
time span was enough to ensure convergence of spatio-temporal spectra
and correlation functions.

We start discussing the spatial spectral, to characterize the
turbulence and to quantify its anisotropy as the intensity of the
guide field is varied, for different values of the cross-helicity.
But first we need to define some quantities, as we are dealing with
anisotropic flows. In principle we could study spectra in terms of 
the wave vector ${\bf k}$, but this results in a three-dimensional
spectral density. Using the preferential direction associated to 
the guide field, reduced spectra can be defined that simplify
substantially the data analysis.

The axisymmetric energy spectrum $e(k_\perp, k_\parallel, t)$ provides
information on the anisotropy of the turbulence relative to the the
guide field \cite{mininni_isotropization_2012}. It is defined as
\begin{equation}\label{eq:axisymmetric}
\begin{split}  e(k_\perp, k_\parallel, t) = \sum_{\substack{k_\perp \leq |\vec{k}\times\hat{x}| < k_\perp+1 \\ k_\parallel \leq k_x < k_\parallel +1}} |\hat{u}(\vec{k},t)|^2 +|\hat{b}(\vec{k},t)|^2 = \\ = \int \left(|\hat{u}(\vec{k},t)|^2 +|\hat{b}(\vec{k},t)|^2\right) |\vec{k}| \sin \theta_k~d\phi_k.
\end{split}
\end{equation}
The first equality corresponds to the way the spectra is computed 
in the simulations (as Fourier modes are discrete), while the 
second corresponds to the theoretical definition in the continuum 
case. Since the guide field is $\vec{B_0} = B_0 \hat{x}$, in both 
cases the wave vector components $k_\parallel = k_x$ and 
$k_\perp = \sqrt{k_y^2+k_z^2}$, and the polar angles in Fourier 
space $\theta_k$ and $\phi_k$, are relative to the $x$-axis. That is, 
in Eq.~(\ref{eq:axisymmetric}), 
$\theta_k = \arctan(k_\perp/k_\parallel)$ is the co-latitude in
Fourier space with respect to the $x$-axis, and $\phi_k$ is the
longitude with respect to the $y$-axis. Note that below we treat the 
discrete and continuum expressions of Fourier spectra as equivalent,
bearing in mind that in all cases integrals should be replaced by sums
when required for the numerics.

Using the axisymmetric spectrum, one can define the time averaged
isotropic energy spectrum $E(k)$ as
\begin{equation}
  E\left(k \right) = \frac{1}{T}\int\int e(|\vec{k_\perp}|,
  k_\parallel, t) |{\bf k}|~d\theta_k~dt,
\end{equation}
and the reduced perpendicular energy spectrum $E(k_\perp)$
\cite{mininni_isotropization_2012} as
\begin{equation}\label{eq:reducedspectrum}
  E\left(k_\perp\right) = \frac{1}{T}\int\int e(|\vec{k_\perp}|,
  k_\parallel, t) \, dk_\parallel~dt,
\end{equation}
where in the latter case we integrate over parallel wave numbers to 
obtain a spectrum that depends only on $k_\perp$, and in both cases
we average in time over a (sufficiently long) time $T$.

The reduced perpendicular
energy spectra $E(k_\perp)$ are shown in Fig.~\ref{fig1:E} for the
simulations with $B_0 = 0.25$, $1$, $2$, $4$, and $8$ with normalized
cross-helicity $\sigma_c=0.3$. In this figure we also show the 
isotropic energy spectrum $E(k)$ for the simulation with $B_0 = 0$, 
with $\sigma_c=0.3$.  The simulations with $\sigma_c=0$ and 
$\sigma_c=0.9$ display a similar behavior. A Kolmogorov power law is 
also indicated in the figure as reference.  As can be seen, despite 
the moderate spatial resolution of the runs, the observed spatial 
spectra are compatible with Kolmogorov scaling $\sim k_\perp^{-5/3}$, 
and the simulations are well resolved displaying a dissipative range 
for large wavenumbers (for example, the Kolmogorov dissipation 
wavenumbers $k_\nu$ are $k_\nu \approx 91$, $152$, and $122$ for the 
simulations with $B_0 = 1$ and $\sigma_c = 0$, $0.3$, and $0.9$ 
respectively).

We can see the spectral behavior (and of the anisotropy of the flows)
in more detail in Fig.~\ref{fig2:isocontourns}. There, we show
isocontours of the axisymmetric energy spectrum $e(k_\perp,
k_\parallel)$ (i.e., the energy density as a function of perpendicular
and parallel wavenumbers) for $B_0=0$, 1, 4, and 8, and in all cases
for flows with $\sigma_c = 0.3$. As a reference we also indicate the
curves (in Fourier space) where the Alfv\'en time is equal to either
the sweeping time, or the non-linear time. In other words, these
curves separate regions in which (from theoretical arguments) the
fastest time scale can be expected to be either $\tau_A$ (above the
dashed red curve) or $\tau_{nl}$ (below the solid blue curve). The
sweeping time can be relevant for all modes below the dashed red
curve.

\begin{figure}
  \centering
  \subfigure[$\vec{z}^-$, $B_0=1$, $\sigma_c=0.3$, $k_\parallel=10$]{\includegraphics[width=0.95\columnwidth]{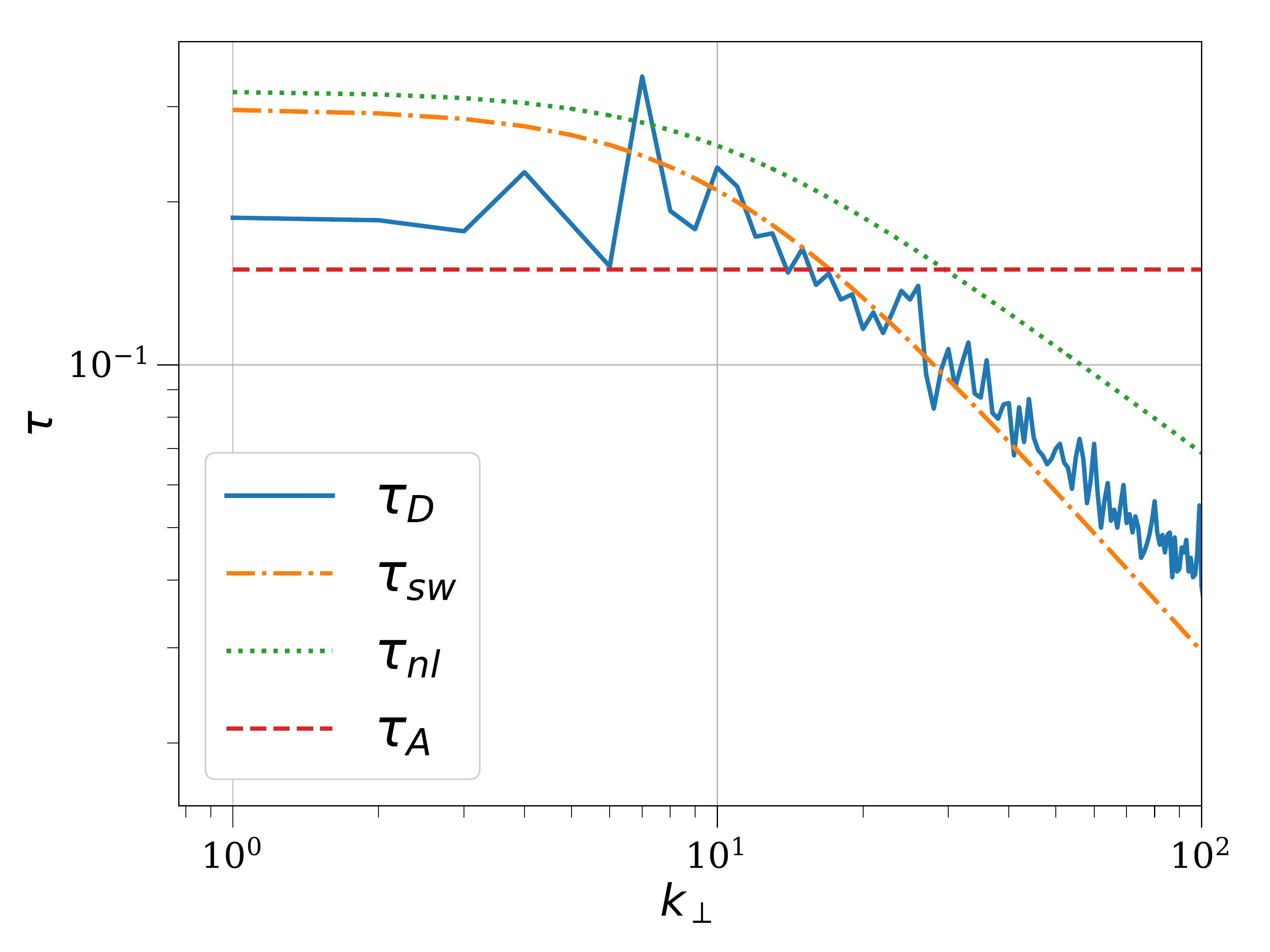}}
  \subfigure[$\vec{z}^+$, $B_0=1$, $\sigma_c=0.3$, $k_\parallel=10$]{\includegraphics[width=0.95\columnwidth]{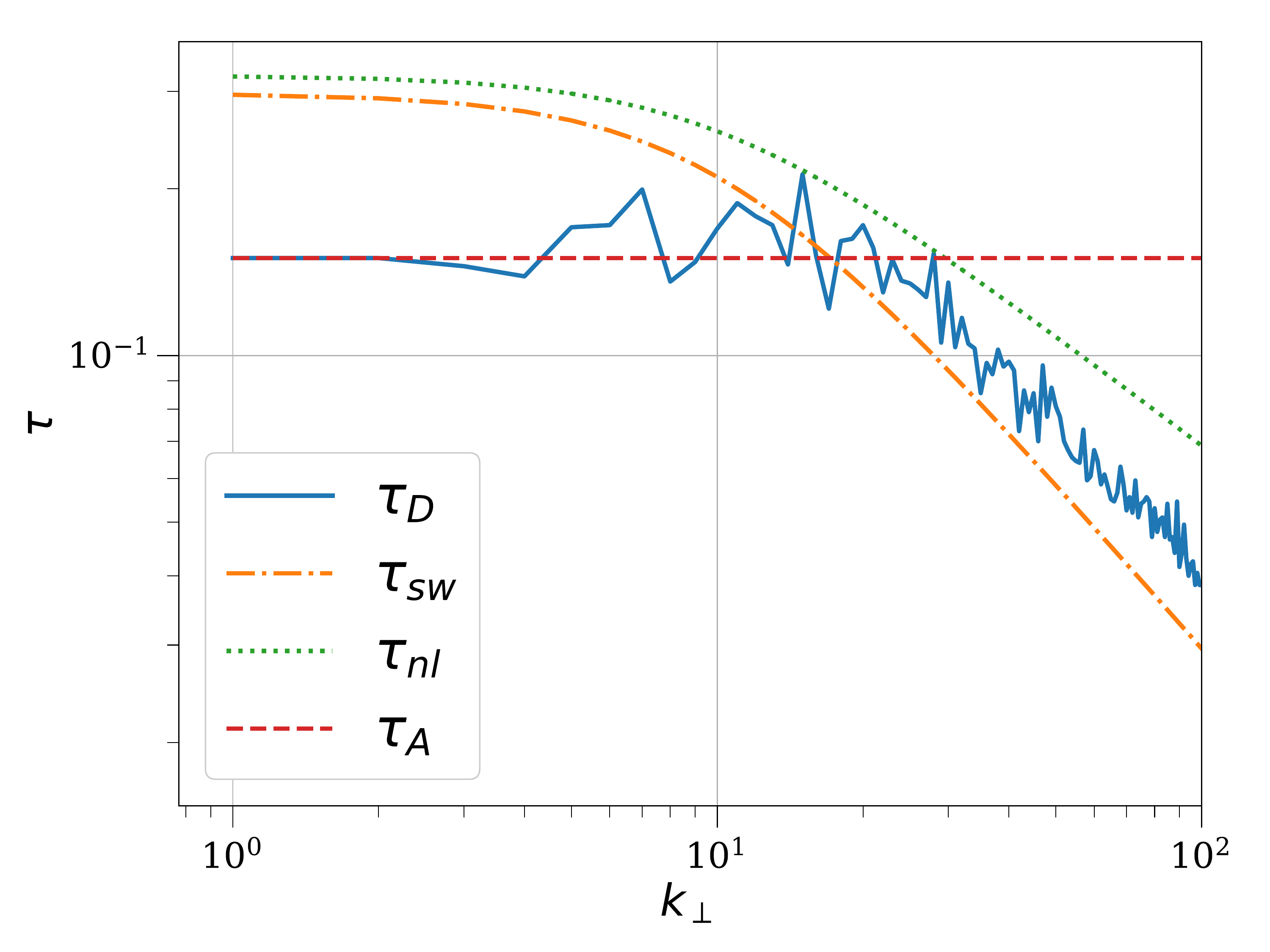}}
  \caption{Decorrelation times $\tau_D$ for the run with $B_0=1$ and
    $\sigma_c=0.3$, for $k_\parallel=10$ constant and as a function of
    $k_\perp$. Panel (a) corresponds to $\vec{z}^-$ and panel (b) to
    $\vec{z}^+$. The theoretical prediction for the sweeping time
    $\tau_{sw}$, the non-linear time $\tau_{nl}$, and the Alfv\'en
    time $\tau_A$ are indicated as references.}
  \label{fig5:z+_vs_z-}
\end{figure}

Note that for $B_0 \neq 0$ the energy is not distributed isotropically
in the axisymmetric spectra in Fig.~\ref{fig2:isocontourns}. Energy
tends to accumulate in modes with small $k_\parallel$ as $B_0$ is
increased, and for $B_0=4$ and 8, a substantial fraction of the energy
accumulates in the vicinity of the curves satisfying $\tau_A \approx
\tau_{sw}$ and $\tau_A \approx \tau_{nl}$.

\subsection{Wavenumber-frequency spectra}\label{sec:wk}

We calculate the energy spectrum $E(\vec{k}, \omega)$ from the relation
\begin{equation}
  E(\vec{k}, \omega) = \frac{1}{2} \left|\hat{u}(\vec{k},
  \omega)\right|^2 + \frac{1}{2} \left|\hat{b}(\vec{k},
  \omega)\right|^2
\end{equation}
where $\hat{u}(\vec{k}, \omega)$ and $\hat{b}(\vec{k}, \omega)$ are
the Fourier transforms in time and in space of the velocity and the
magnetic fields, respectively.  The main results of the present study
are summarized in Figs.~\ref{fig3:B0_spectrum_Hc} to
\ref{fig3:B8_spectrum_Hc}, which quantify the spatio-temporal behavior
of the Els\"asser fields separately. These figures show the normalized
wavevector and frequency spectra of the $\vec{z}^+$ and $\vec{z}^-$
variables, for simulations with different values of the background
mean field $B_0$ and normalized cross-helicity $\sigma_c$. As the
spectra are multidimensional, in all cases we show slices of the
spectrum for $k_\perp=0$ and as a function of $k_\parallel$ and
$\omega$ (other slices, with other values of $k_\perp$, display the
same behavior for the waves reported below).

Figure \ref{fig3:B0_spectrum_Hc} shows these spatio-temporal spectra
for simulations with $B_0=0$. In this case, the dispersion relation
for Alfv\'enic fluctuations becomes $\omega=0$, and Alfv\'en waves are
indistinguishable (in this spectrum) from slow modes such as turbulent
eddies. The sweeping relation, for eddies with velocity $v_{rms}$,
becomes $\omega=\pm v_{rms} k$, and in practice, as all turbulent
eddies with this velocity (or a smaller velocity) can randomly sweep
small-scale structures in the flow, the relation for random sweeping
becomes $|\omega| \leq v_{rms} k$. Both relations are indicated
respectively by dashed and solid lines in
Fig.~\ref{fig3:B0_spectrum_Hc}.

Accumulation of energy in the spectra in
Fig.~\ref{fig3:B0_spectrum_Hc} can be seen for all modes in the region
enclosed by the sweeping relation, evidencing the presence of
broadband (strong) turbulence rather than of wave turbulence or linear
wave propagation. Moreover, for large values of the normalized
cross-helicity ($\sigma_c = 0.9$), energy accumulates instead in modes
with $\omega\approx 0$, and more energy can be observed in $\vec{z}^+$
modes when compared to the $\vec{z}^-$ modes. From these spectra we
can conclude that for $B_0=0$ and $\sigma_c=0$ the dominant timescale
is that of the sweeping, while for large values of $\sigma_c$ either
the non-linear timescale or the Alfv\'en time become dominant.

Figure \ref{fig3:B025_spectrum_Hc} shows the spatio-temporal spectra
for simulations with $B_0=0.25$. The case with $\sigma_c=0$ shows
again a broad range of fluctuations in the range of frequencies
enclosed by the sweeping relation. As the value of $\sigma_c$ is
increased the $\vec{z}^+$ fluctuations become dominant, a situation
which is more evident in the case with $\sigma_c=0.9$. Also, as
$\sigma_c$ is increased, energy in $\vec{z}^+$ fluctuations leaves the
funnel defined by the sweeping relation, and concentrates in the
vicinity of the dispersion relation of Alfv\'en waves $\omega^+ =
+\vec{V}_\textrm{A} \cdot \vec{k}$ (see the case with $\sigma_c=0.9$
in Fig.~\ref{fig3:B025_spectrum_Hc}). Note that the choice of signs
for waves described by $\vec{z}^\pm={\vec z}_0^\pm e^{i(\vec{k} \cdot
  \vec{x}+\omega^\pm t)}$ follows from the fact that the Fourier
transforms used in space and in time follow the same sign convention,
and where ${\vec z}_0^\pm$ are the amplitudes of the waves.  This way,
the sign of $\omega^+$ implies $\vec{z}^+$ fluctuations propagate
anti-parallel to the guide field, as expected. However, in an apparent
contradiction, the waves with the opposite polarization, i.e., the
$\vec{z}^-$ fluctuations, also populate (albeit with smaller
amplitude) the same upper branch of the Alfv\'enic wave dispersion
relation. As the $\vec{z}^-$ fluctuations satisfy another dispersion
relation ($\omega^- = -\vec{V}_\textrm{A} \cdot \vec{k}$), in the
linear regime these fluctuations should populate instead the lower
branch of the dispersion relation shown in
Fig.~\ref{fig3:B025_spectrum_Hc}. This behavior indicates that
$\vec{z}^-$ fluctuations also propagate in real space in the direction
anti-parallel to the guide field (i.e., with negative velocity),
instead of parallel to this field (i.e., with positive velocity) as
would be expected. Such a behaviour was predicted by Hollweg
\cite{hollweg_1990_wkb} for the solar wind and caused by, e.g.,
reflections of waves in density fluctuations in the interplanetary
medium, using a WKB expansion. In our case, the flow is incompressible
and density is uniform in space and constant in time.

\begin{figure*}
  \centering
  \subfigure[$\vec{z}^+$, $B_0=0.25$, $\sigma_c=0.3$, $k_\parallel=15$]{\includegraphics[width=0.9\columnwidth]{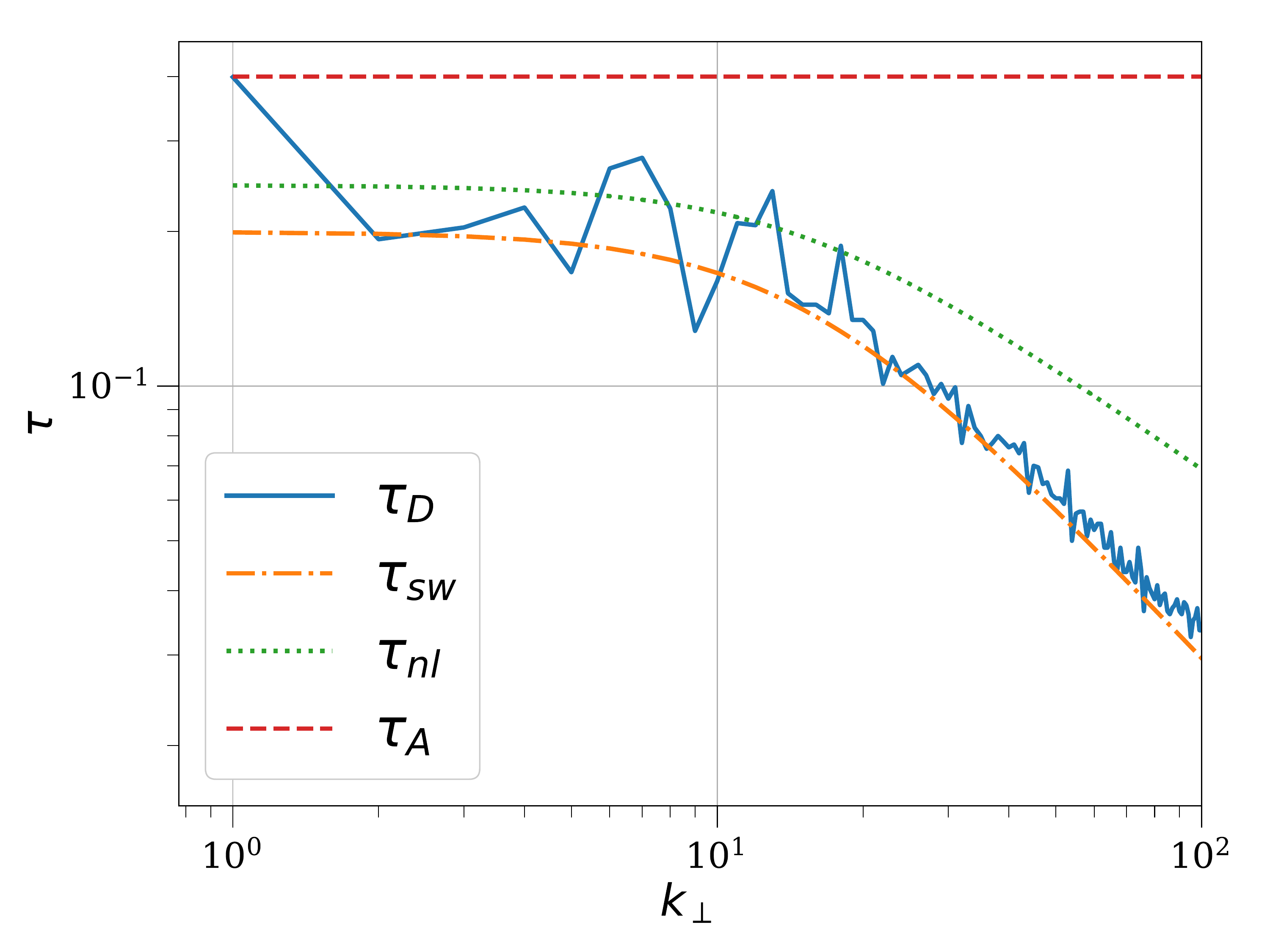}}
  \subfigure[$\vec{z}^+$, $B_0=1$, $\sigma_c=0.3$, $k_\parallel=15$]{\includegraphics[width=0.9\columnwidth]{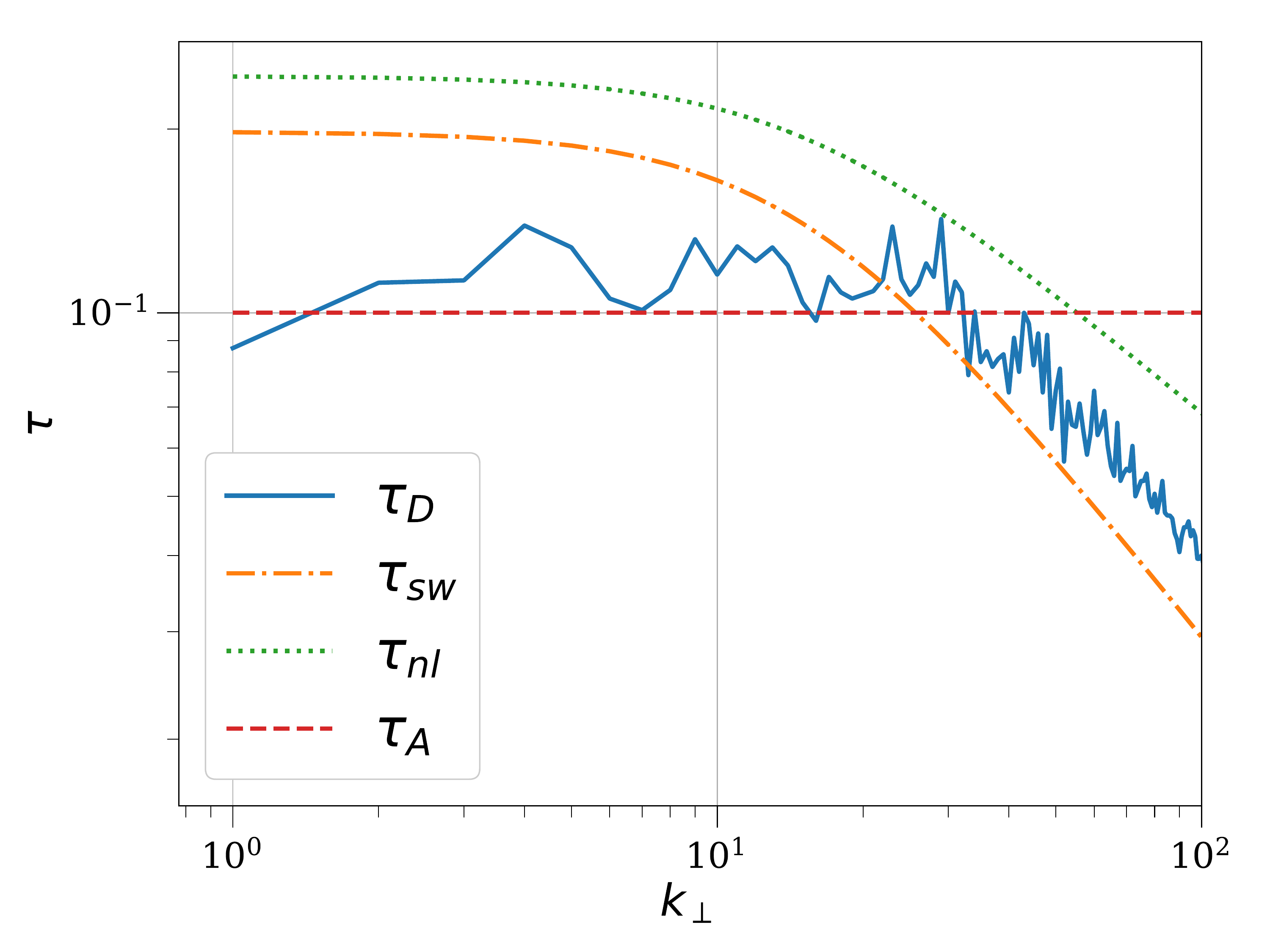}}
  \subfigure[$\vec{z}^+$, $B_0=4$, $\sigma_c=0.3$, $k_\parallel=15$]{\includegraphics[width=0.9\columnwidth]{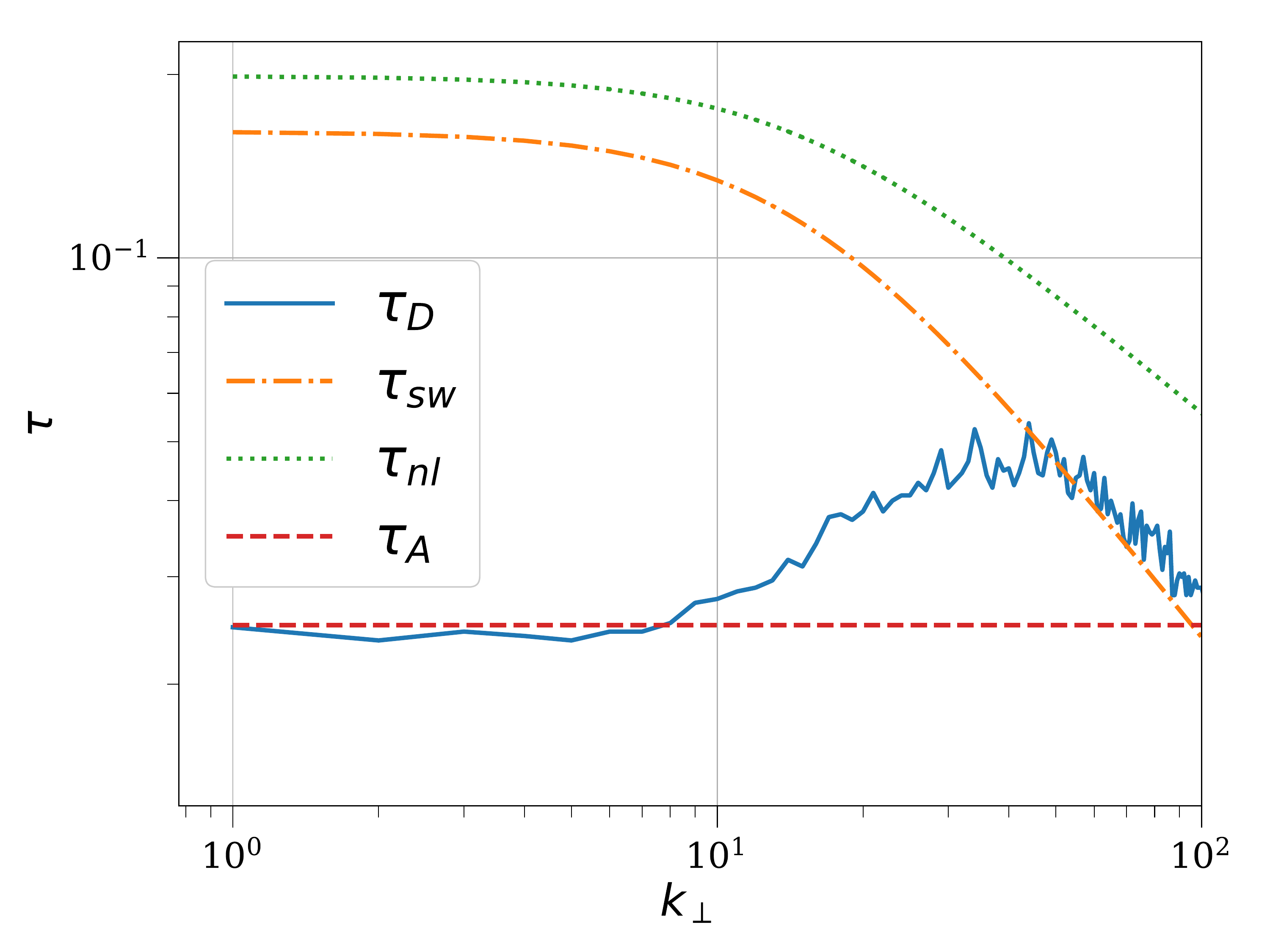}}
  \subfigure[$\vec{z}^+$, $B_0=8$, $\sigma_c=0.3$, $k_\parallel=15$]{\includegraphics[width=0.9\columnwidth]{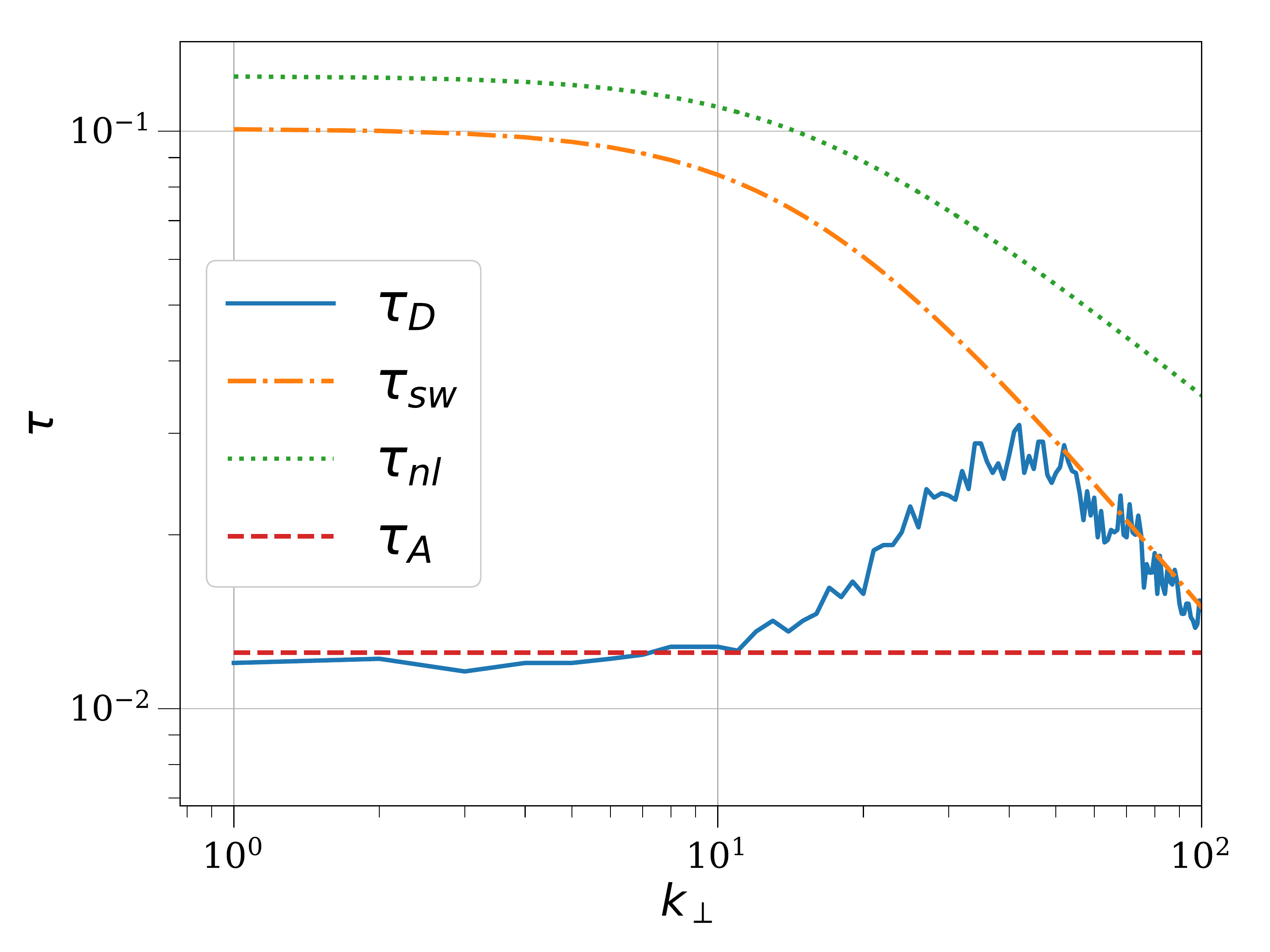}}
    \caption{Decorrelation times $\tau_D$ for the $\vec{z}^+$ field in
      simulations with $\sigma_c = 0.3$ and (a) $B_0=0.25$, (b) $1$,
      (c) $4$, and (d) $8$, for $k_\parallel = 15$ and as a function
      of $k_\perp$. The theoretical prediction for the sweeping time
      $\tau_{sw}$, the non-linear time $\tau_{nl}$, and the Alfv\'en
      time $\tau_A$ are indicated as references.}
  \label{fig5:tD_vs_B0}
\end{figure*}

\begin{figure*}
  \centering
  \subfigure[$\vec{z}^+$, $B_0=0.25$, $\sigma_c=0.3$, $k_\perp=15$]{\includegraphics[width=0.9\columnwidth]{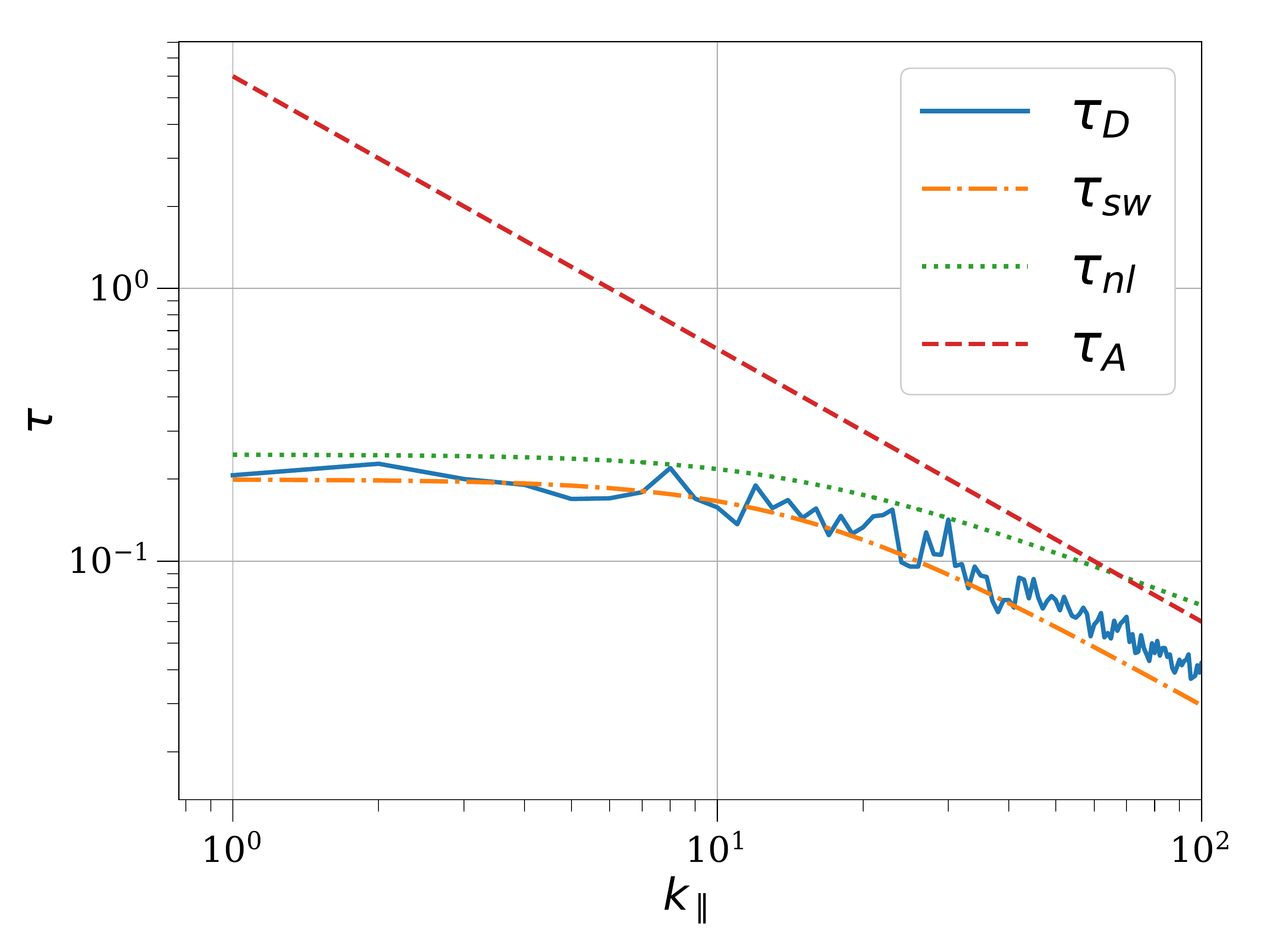}}
  \subfigure[$\vec{z}^+$, $B_0=1$, $\sigma_c=0.3$, $k_\perp=15$]{\includegraphics[width=0.9\columnwidth]{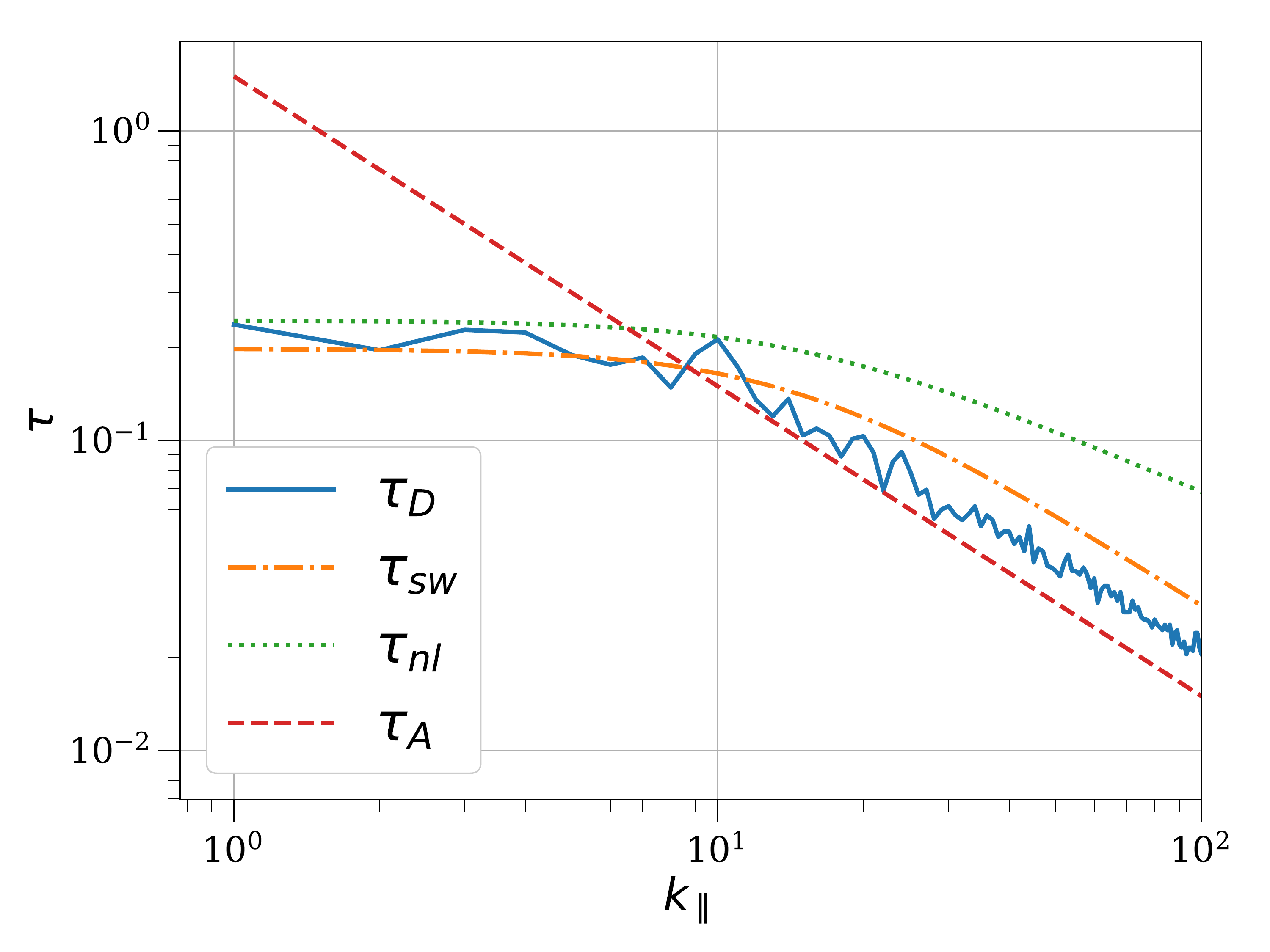}}
  \subfigure[$\vec{z}^+$, $B_0=4$, $\sigma_c=0.3$, $k_\perp=15$]{\includegraphics[width=0.9\columnwidth]{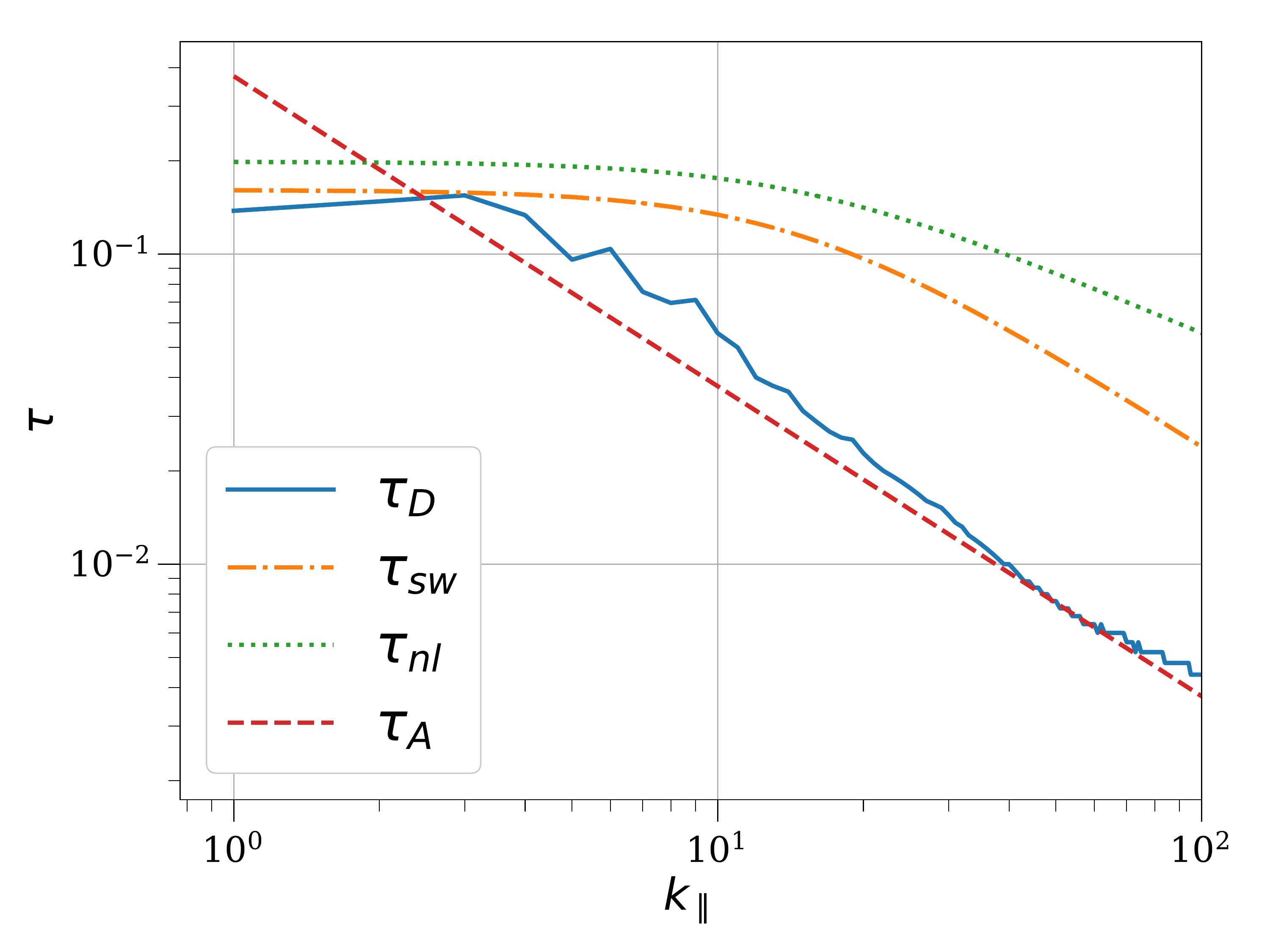}}
  \subfigure[$\vec{z}^+$, $B_0=8$, $\sigma_c=0.3$, $k_\perp=15$]{\includegraphics[width=0.9\columnwidth]{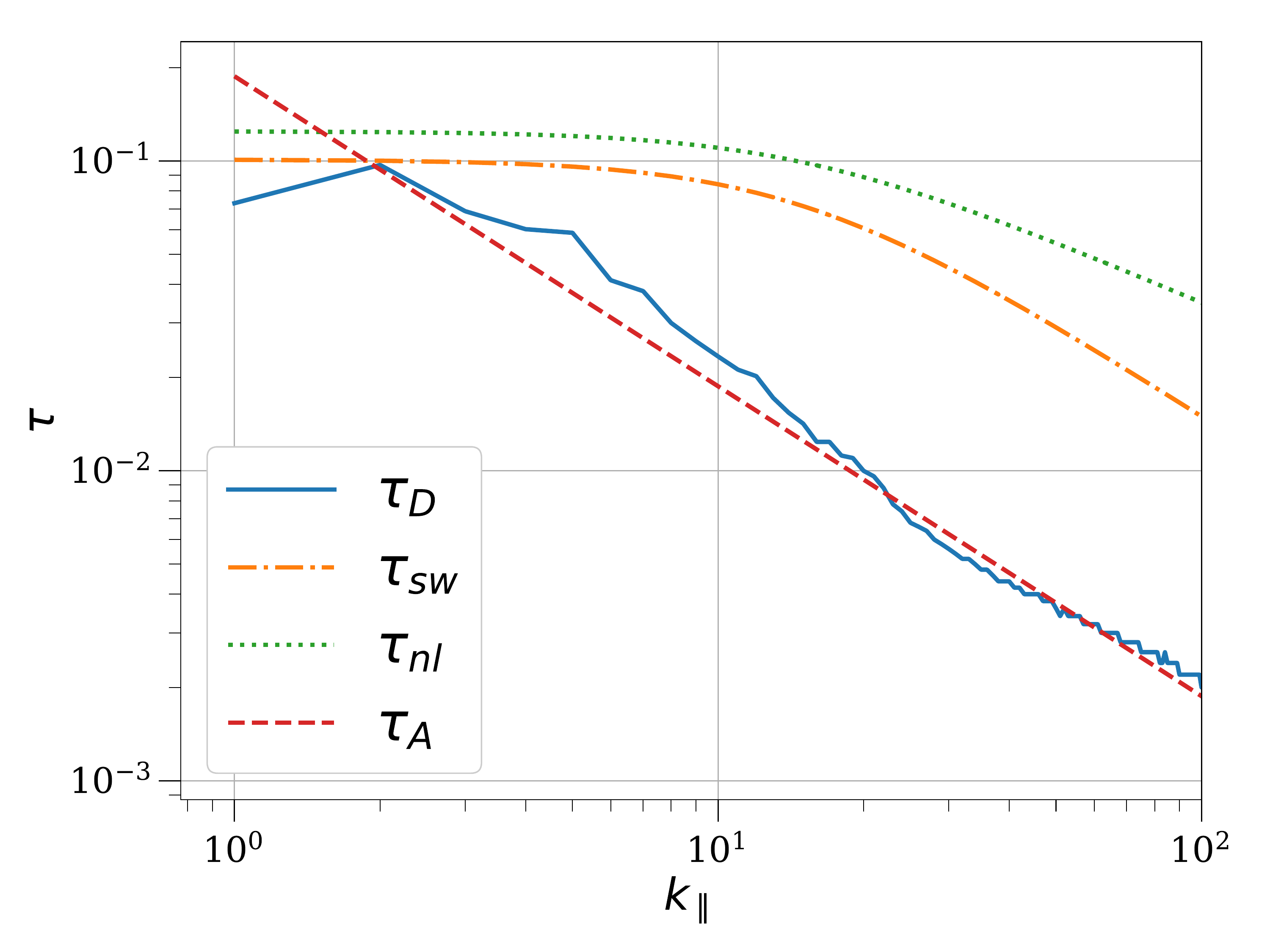}}
    \caption{Decorrelation times $\tau_D$ for the $\vec{z}^+$ field in
      simulations with $\sigma_c = 0.3$ and (a) $B_0=0.25$, (b) $1$,
      (c) $4$, and (d) $8$, for $k_\perp=15$ and as a function of
      $k_\parallel$. The theoretical prediction for the sweeping time
      $\tau_{sw}$, the non-linear time $\tau_{nl}$, and the Alfv\'en
      time $\tau_A$ are indicated as references.}
  \label{fig5:tD_vs_B0_2}
\end{figure*}

\begin{figure}
  \centering
  \subfigure[$\sigma_c = 0$]{\includegraphics[width=0.95\columnwidth]{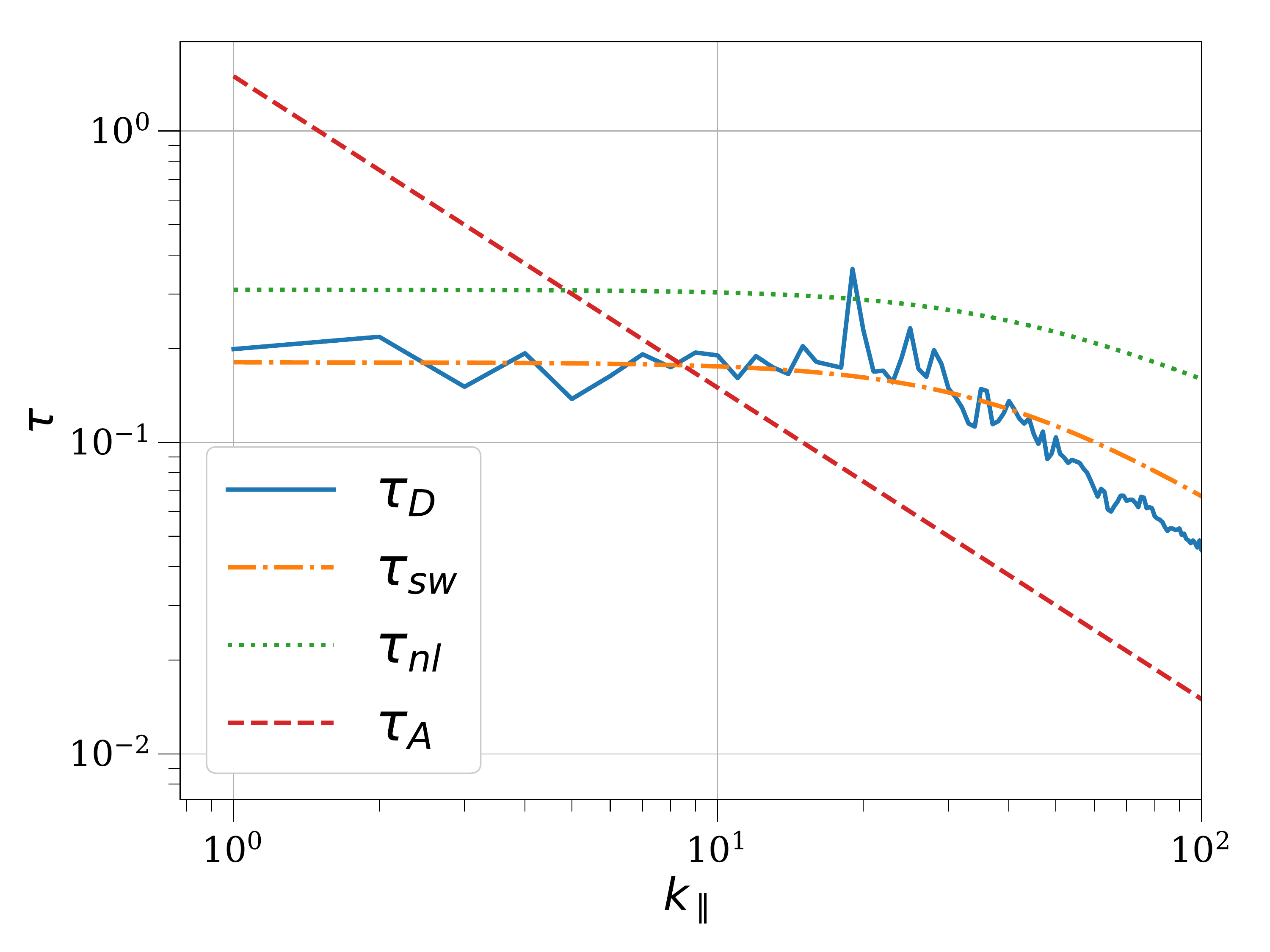}}
  \subfigure[$\sigma_c = 0.3$]{\includegraphics[width=0.95\columnwidth]{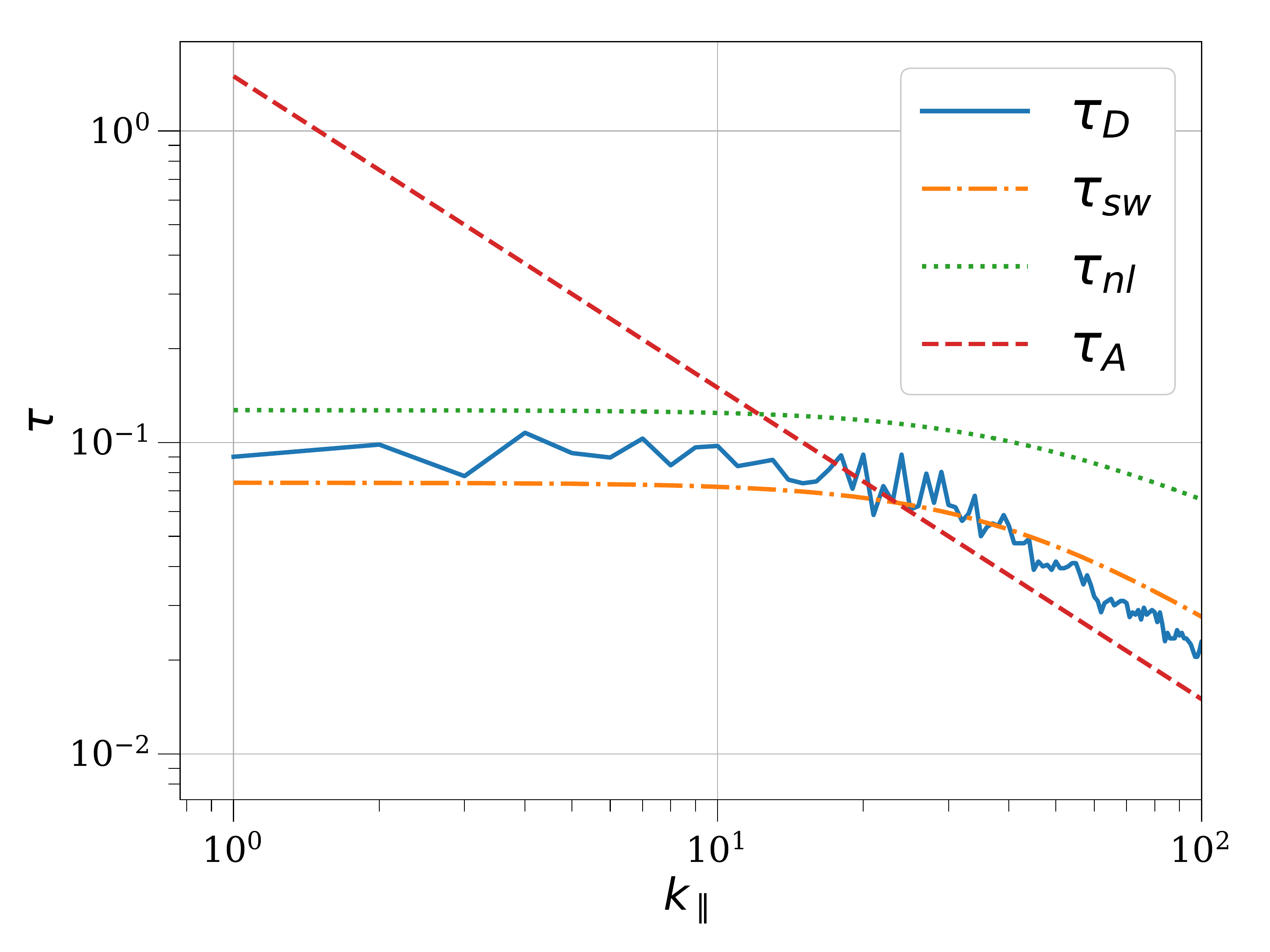}}
  \subfigure[$\sigma_c = 0.9$]{\includegraphics[width=0.95\columnwidth]{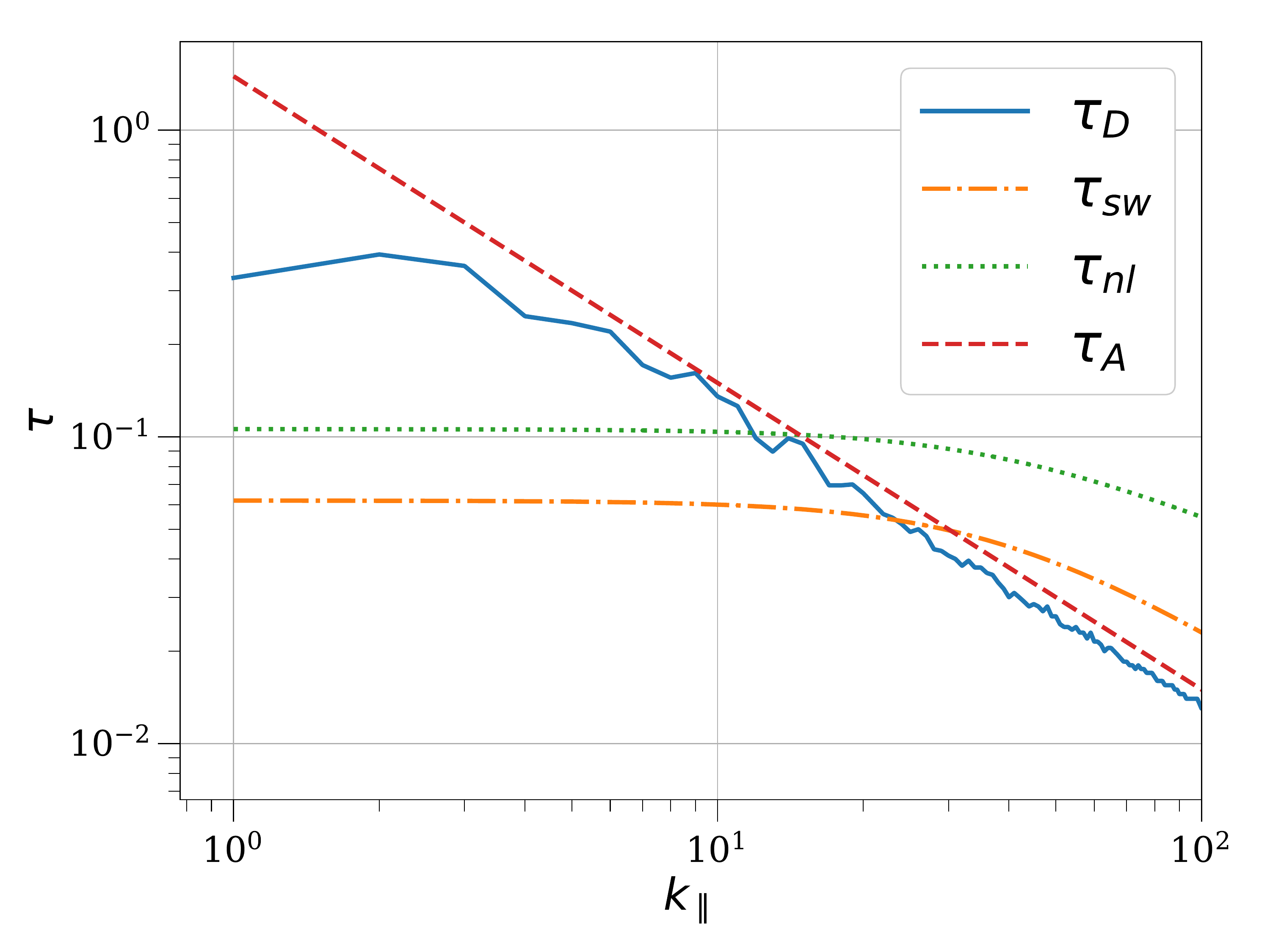}}
    \caption{Decorrelation times $\tau_D$ for the runs with $B_0=1$,
      for $k_\perp=40$ and as a function of $k_\parallel$. The panels
      correspond to (a) $\sigma_c = 0$, (b) $\sigma_c = 0.3$, and (c)
      $\sigma_c = 0.9$. The theoretical prediction for the sweeping
      time $\tau_{sw}$, the non-linear time $\tau_{nl}$, and the
      Alfv\'en time $\tau_A$ are indicated as references.}
  \label{fig5:tD_vs_Hc}
\end{figure}

As $B_0$ is increased, this effect becomes more evident. In
Fig.~\ref{fig3:B1_spectrum_Hc} we show the spatio-temporal spectra for
simulations with $B_0=1$. Now energy tends to concentrate near the
dispersion relation of the Alfv\'en waves for all values of
$\sigma_c$, i.e., as we increase the value of $B_0$ the relevance of
random sweeping decreases and Alfv\'en waves become more
important. For $\sigma_c=0$ we observe waves propagating in both
directions: $\vec{z}^+$ fluctuations propagate anti-parallel to the
guide field, and $\vec{z}^-$ fluctuations propagate parallel to this
field. Also, for values of $k_\parallel$ larger than $\approx 20$, the
dispersion in the excitation of modes increases and energy starts to
populate the funnel in spectral space associated with sweeping,
indicating random sweeping plays a role at sufficiently small vertical
scales. Instead, for $\sigma_c =0.3$ and $0.9$ energy accumulates only
near the wave dispersion relation, and we recover counter-propagation
of one of the wave motions: both $\vec{z}^+$ and $\vec{z}^-$ fields
propagate in the same direction, anti-parallel to the guide
field. Increasing $B_0$ further reduces this effect (see the cases
with $B_0=8$ in Fig.~\ref{fig3:B8_spectrum_Hc}), resulting in the
expected propagation for each excitation, or in very little or no
propagation of $\vec{z}^-$ when $\sigma_c$ is sufficiently small.

What is the origin of the observed $\vec{z}^-$ fluctuations
propagating in the same direction as the $\vec{z}^+$ fluctuations?
Based on the results of Hollweg \cite{hollweg_1990_wkb}, and on
Eq.~(\ref{eq:MHD_zpzm}), they must be caused by reflections in large
scale inhomogeneities of the mean magnetic field (note there is no
mean background flow in our simulations, nor density fluctuations).
Although our background guide field $B_0$ is uniform (i.e., constant
in space as well as in time), the total mean field a fluctuation sees
includes a slowly varying component (e.g., from magnetic field
fluctuations at large scales, such as those in $k=1$ modes, which
evolve on a slower time scale than fast waves and small-scale
fluctuations).  As a result, the flow has an effective Alfv\'en
velocity that depends on the spatial coordinates. We can then write
for either type of Els\"asser fluctuations the ideal linearized
Eq.~(\ref{eq:MHD_zpzm}) for constant density and for $\vec{U}=0$ (no
mean background flow) as
\begin{equation}
\partial_t \vec{z}^\pm = \pm \vec{V}_\textrm{A} \cdot \nabla \vec{z}^\pm 
    \mp \vec{z}^\mp \cdot \frac{\nabla \vec{B}'}{\sqrt{4\pi \rho}} ,
\end{equation}
where $\vec{V}_\textrm{A}$ can now include large-scale fluctuations of
the magnetic field, and $\vec{B}'$ as before is the total magnetic
field in Gaussian units. If the normalized cross-helicity $\sigma_c$ is close to 1,
that is, if $|\vec{z}^+| \gg |\vec{z}^-|$, we have for $\vec{z}^+$
\begin{equation}
\partial_t \vec{z}^+  \approx \vec{V}_\textrm{A} \cdot \nabla \vec{z}^+ ,
\end{equation}
and using $\vec{z}^\pm = \vec{z}_0^\pm e^{i(\vec{k}\cdot
  \vec{x}+\omega^\pm t)}$ we recover the usual dispersion relation for
waves propagating anti-parallel to the mean field $\omega^+ =
+\vec{V}_\textrm{A} \cdot \vec{k}$ (where now $\vec{V}_\textrm{A}$ can
fluctuate slowly in space and time). However, for $\vec{z}^-$ we
obtain
\begin{equation}
\partial_t \vec{z}^- \approx - \vec{V}_\textrm{A} \cdot \nabla \vec{z}^- + \vec{z}^+ 
    \cdot \frac{\nabla \vec{B}'}{\sqrt{4\pi \rho}} .
\label{eq:zmlinear}
\end{equation}
This equation indicates that the propagation of $\vec{z}^-$
perturbations (which are smaller in amplitude than $\vec{z}^+$) can be
strongly affected by the $\vec{z}^+$ field and by spatial variations
of the large-scale magnetic field.

From Eq.~(\ref{eq:zmlinear}) we can also extract some phenomenological
conditions for the behavior seen in Figs.~\ref{fig3:B0_spectrum_Hc} to
\ref{fig3:B8_spectrum_Hc} (and in particular, for the
counter-propagation of waves) to take place. Using again $\vec{z}^\pm
= \vec{z}_0^\pm e^{i(\vec{k}\cdot \vec{x}+\omega^\pm t)}$, and
assuming $\vec{B}' = \vec{B}_0' + \vec{b}_0'$ where $\vec{b}_0' =
\vec{\tilde{b}}_0'e^{i\vec{K} \cdot \vec{x}}$ is the slowly varying
large-scale magnetic field with wavenumber $K \ll k$,
Eq.~(\ref{eq:zmlinear}) reduces to
\begin{equation}
\left( \omega^- +\vec{V}_\textrm{A} \cdot \vec{k} \right) 
    \vec{z}_0^- e^{i \omega^- t} = 
    \frac{\left(\vec{K} \cdot \vec{z}_0^+\right) \vec{b}_0'}{\sqrt{4\pi \rho}} 
    e^{i \omega^+ t} .
\end{equation}
Taking the dot product with $\vec{z}_0^-$, defining Els\"asser energy
densities $e^\pm = |\vec{z}_0^\pm|^2/4$, and defining the fluctuations
in the Alfv\'en velocity (associated to the large-scale magnetic field
fluctuations) as ${\bf v}_\textrm{A} = \vec{b}_0'/\sqrt{4\pi \rho}$,
we finally get
\begin{equation}
\left( \omega^- + \vec{V}_\textrm{A} \cdot \vec{k} \right)
    e^{i \omega^- t} = 
    \frac{\left(\vec{K} \cdot \vec{z}_0^+\right)
    \left(\vec{v}_\textrm{A} \cdot \vec{z}_0^-\right)}
    {4e^-} 
    e^{i \omega^+ t} .
\end{equation}
This equation admits solutions
\begin{eqnarray}
    \omega^- &=& \omega^+ =
    + \vec{V}_\textrm{A} \cdot \vec{k}, 
    \label{eq:cond1} \\
    2 \vec{V}_\textrm{A} \cdot \vec{k} &=& 
    \left(\vec{K} \cdot \vec{z}_0^+\right)
    \left(\vec{v}_\textrm{A} \cdot \vec{z}_0^-\right) /
    (4e^-), \label{eq:cond2}
\end{eqnarray}
which correspond to both waves traveling in the same direction as long
as the second condition, given by Eq.~(\ref{eq:cond2}), can be
fulfilled. From dimensional analysis, this condition requires that
\begin{equation}
    2 \frac{V_\textrm{A}}{v_\textrm{A}} \frac{k}{K}
    \sim \sqrt{\frac{e^+}{e^-}},
\end{equation}
which (as $V_\textrm{A}\gtrsim v_\textrm{A}$ and $k\gg K$) cannot be
satisfied when $\sigma_c \approx 0$ (as observed in
Figs.~\ref{fig3:B0_spectrum_Hc} to \ref{fig3:B8_spectrum_Hc}), or when
the guide field becomes too strong for a fixed value of $\sigma_c$ (as
also observed in the spatio-temporal spectra). Thus, this last
qualitative argument indicates (in agreement with the simulations)
that $\vec{z}^-$ fluctuations can propagate with the same phase speed
and direction as the $\vec{z}^+$ fluctuations as long as $\sigma_c
\neq 0$ and $B_0$ is not too strong for a fixed value of the
normalized cross-helicity.

In other words, if $|\vec{z}^+|$ at large scales is comparable to
$|\vec{V}_\textrm{A}|$ and $\sigma_c \approx 1$, we can see
$\vec{z}^-$ fluctuations propagate in the same direction as
$\vec{z}^+$ fluctuations as the result of reflections in
inhomogeneities of the large-scale magnetic field. A similar behavior
can result, for example, from mass density fluctuations when the flow
is compressible, as is the case in some regions of the solar wind and
the interplanetary medium \cite{zhou_1989_nonWKBevolution}, and this
argument does not preclude other effects such as strong non-linear
interactions from resulting in reflection and counter-propagation of
excitations. Moreover, when the intensity of the background magnetic
field $\vec{B}_0$ is further increased, the arguments used above are
not valid anymore and the relevance of the reflections reduces. This
is compatible with the behavior seen in Fig.~\ref{fig3:B8_spectrum_Hc}
for the simulations with $B_0=8$, which show similar amounts of power
in both type of fluctuations when $\sigma_c=0$, less power in
$\vec{z}^-$ fluctuations when $\sigma_c=0.3$ (and propagating opposite
to the $\vec{z}^+$ field), and no appreciable power for $\vec{z}^-$
fluctuations when compared to $\vec{z}^+$ in the case with
$\sigma_c=0.9$.

\subsection{Decorrelation times}

From the discussions in Sec.~\ref{sec_Wfspectrum_and_Gamma}, another
way to identify dominant time scales for individual modes is to study
the decorrelation time $\tau_D$, i.e., the time it takes for each
Fourier mode with wave vector ${\bf k}$ to be decorrelated from its
previous history either by non-linear eddy interactions (if $\tau_D
\sim \tau_{nl}$), by the cross-over of waves (if $\tau_D \sim
\tau_A$), or by the sweeping by the large-scale flow (when $\tau_D
\sim \tau_{sw}$). Again, as $\tau_D$ depends on the wave vector ${\bf
  k}$, in the following we show it for fixed values of $k_\parallel$
or $k_\perp$, and as a function of the remaining wavenumber. In all
cases, the decorrelation time $\tau_D$ is obtained from the numerical
data by computing the correlation function
$\Gamma(k_{\perp},k_{\parallel},\tau)$, and looking at the value of
the time lag $\tau$ for which the correlation function decays to $1/e$
from its value for $\tau=0$. Note the choice of $1/e$ as a reference
value is arbitrary, but similar results are obtained if instead 
the decorrelation time is defined as the half width of $\Gamma$, or as the time when the correlation function crosses the zero 
\cite{lugones_2016_spatiotemporal, 
clark_di_leoni_quantification_2014}. With any of these choices,
$\tau_D$ is a measure of the characteristic time for the decay of the 
correlation.

Figure \ref{fig5:z+_vs_z-} shows the different decorrelation times for
a fixed value of $k_{\parallel}=10$ and as a function of $k_{\perp}$,
for the simulation with $B_0=1$ and $\sigma_c=0.3$. The theoretical
predictions for the different decorrelation times are also indicated
as a reference. Since the Alfv\'enic time is independent of
$k_{\perp}$ it shows as a constant vale in this figure. The
decorrelation time $\tau_D$ obtained from the numerical data is very
close to the Alfv\'enic time for small values of $k_{\perp}$ (up to
$k_\perp \approx 10$), but it deviates and becomes closer to the
sweeping time for large values of $k_{\perp}$ (i.e, for small
perpendicular lengthscales). This is more clear for $\vec{z}^-$
fluctuations than for $\vec{z}^+$ fluctuations, for which the
decorrelation time $\tau_D$ for $k_\perp > 10$ is in between the
scaling of $\tau_{sw}$ and of $\tau_{nl}$.

Figure \ref{fig5:tD_vs_B0} shows the decorrelation times $\tau_D$ for
the $\vec{z}^+$ field for cases with $\sigma_c = 0.3$, with a guide
field of $B_0=0.25$, $1$, $4$, and $8$, and for fixed $k_\parallel =
15$ as a function of $k_\perp$. Again, for low values of $B_0$,
$\tau_D$ is mostly dominated for the sweeping, either for all values
of $k_\perp$ (for $B_0=0.25$) or down to $k_\perp \approx 20$ (for
$B_0=1$). However, for larger values of $B_0$ (or for small values of
$k_\perp$ when $B_0=1$) Alfv\'enic effects become dominant, with
$\tau_D$ taking values close to $\tau_{A}$. Overall, the fastest time
scale at any given $k_\perp$ seems to be the dominant one. These
results are consistent with the previous ones we obtained
\cite{lugones_2016_spatiotemporal} for the case of strong
incompressible MHD turbulence with no cross-helicity, although the
presence of some cross-helicity in the flow seems to favor a
transition towards a flow more dominated by Alfv\'en waves as also
seen in the spatio-temporal spectra in Sec.~\ref{sec:wk}. This can be
also associated with the fact that under certain conditions the
nonlinear time of the dominant Els\"asser fluctuations becomes too
long, and the decorrelation time scale is then determined by the
so-called ``minority species'' as reported before in closure
calculations by Grappin \textit{et al.}
\cite{grappin_1983_dependence}.

This behavior can also be seen when $k_\perp$ is fixed, and $\tau_D$
is studied as a function of $k_\parallel$ (see
Fig.~\ref{fig5:tD_vs_B0_2}). For simulations with $\sigma_c=0.3$ and
with increasing $B_0$, we see that $\tau_D$ varies with $k_\parallel$
as $\tau_{sw}$ when $B_0$ is small or moderate and when $k_\parallel$
is small, and varies as $\tau_A$ when $B_0$ or $k_\parallel$ are
sufficiently large. In other words, modes with wave vectors
sufficiently aligned with the guide field are dominated by the
Alfv\'en time. And again, the fastest time scale in this figure is the
one that dominates the dynamics.

However, and as mentioned before, this picture changes when $\sigma_c$
is sufficiently large. This can be seen in Fig.~\ref{fig5:tD_vs_Hc},
where the decorrelation time $\tau_D$ is plotted for the simulations
with $B_0=1$, for fixed $k_\perp=40$, and as a functions of
$k_\parallel$ for $\sigma_c = 0$, $0.3$ and $0.9$. While for small
values of $\sigma_c$ we observe the same behavior as before, for large
values of $\sigma_c$ the Alfv\'en time becomes dominant, even when it
is slower than all the other time scales, as in the case of the
simulation with $\sigma_c=0.9$ and small values of $k_\parallel$.

Thus, while for small values of $\sigma_c$ the analysis of the
decorrelation time confirms the tendency observed in our previous
study \cite{lugones_2016_spatiotemporal} that the sweeping time
dominates the decorrelations except for the cases with medium and
large values of $B_0$ where the Alfv\'enic time is dominant for small
values of $k_{\perp}$ or large values of $k_{\parallel}$ (see also
studies of MHD turbulence in the weak regime in
Refs.~\cite{galtier_2000_weak,meyrand_weak_2015}, or of the transition
from weak to strong MHD turbulence in
Refs.~\cite{lugones_2016_spatiotemporal, meyrand_direct_2016}),
increasing the cross-helicity content of the flow has interesting
consequences. The appearance of the Alfv\'enic time as dominant
becomes more clear for large values of $\sigma_c$, even when it is not
the fastest time scale, and consistent with a linear (or weakly
non-linear) picture in which most of the fluctuations have a single
direction of propagation. However, as evidenced in the spatio-temporal
analysis of the energy spectrum of each Els\"asser field as a function
of $\vec{k}$ and $\omega$, inhomogeneities of the large scale magnetic
field can induce reflections, and turn on non-linear interactions
dominated by the Alfv\'en cross-over time between waves for modes with
wave vectors sufficiently aligned with the guide field, or by the
sweeping or non-linear time for other modes.

\section{Conclusions}\label{sec_Conclusions}

We analyzed the spatio-temporal behavior of MHD fluctuations
considering their polarizations in terms of the Els\"asser variables,
using direct numerical simulations of three-dimensional incompressible
MHD turbulence. We considered cases with relatively small,
intermediate, and large values of a mean background magnetic field,
and with null, small, and high cross-helicity. The correlation
function as a function of the wavenumber (decomposed in perpendicular
and parallel directions to the mean magnetic field) and of the time
lag was directly computed for all the different simulations
considered, as well as the spatio-temporal spectra. From the
correlation functions, we computed the decorrelation time for each
Fourier mode, and we compared it with different theoretical
predictions for relevant time scales in the system: the local
non-linear time, the random sweeping time and the Alfv\'enic time. It
was observed that time decorrelations are dominated by sweeping
effects for low values of the mean magnetic field and of the
cross-helicity, while for large values of the mean magnetic field or
of the cross-helicity, time decorrelations are controlled by
Alfv\'enic effects even when the Alfv\'en time is not the fastest
time, a new feature when compared with previous studies of
spatio-temporal behavior of strong MHD turbulence with zero
cross-helicity. In principle, this behavior could be interpreted as a
transition towards a regime with weaker non-linearities as the
cross-helicity is increased, as often argued on theoretical grounds
and apparently indicated by our numerical simulations.

However, it should be noted that the spatio-temporal spectra indicate
that even in this regime non-linear interactions are relevant: The
other main result obtained from our analysis is the finding of a
regime in which opposite polarizations $\vec{z}^-$ and $\vec{z}^+$
fluctuations are generated, and propagate in the same direction due to
wave reflections caused by inhomogeneities of the large-scale magnetic
field. This is more evident in the spatio-temporal spectra of the
Els\"asser fields for intermediate values of the background magnetic
field (that is, when the uniform and constant component of the
large-scale magnetic field is not too strong). A phenomenological
analysis based on previous ideas in Zhou and Matthaeus
\cite{zhou1990remarks} confirms the conclusions of Hollweg 
\cite{hollweg_1990_wkb}, that indicates that Alfv\'enic fluctuations with
opposite polarizations can indeed propagate in the same direction and
even with the same speed. If the background magnetic field becomes too
strong (or if the cross-helicity is close to zero), this effect is no
longer observed. Thus, the spatio-temporal analysis of the turbulent
flows provides direct evidence of a phenomena that was predicted before 
using WKB theory, and which can play a relevant role modifying wave 
propagation and nonlinear interactions in the interplanetary medium.

The results analyzed in this paper show in detail that, at least in
the strong turbulence regime, the wave picture is not complete enough
to describe the system of incompressible MHD. A broad band of
fluctuations appear in this system coming from local and non-local
(sweeping) effects, which bring in dispersion and non-linear
effects. It is important to recall, of course, that much of the
present study has concentrated on the study of the Eulerian decorrelation 
time, decomposed into a scale-dependent decorrelation time of individual 
Fourier modes. This decorrelation is generally interpreted as a 
competition between sweeping decorrelation by large scale fluctuations, 
and decorrelation originating from wave propagation. However, neither 
of these effects are in principle responsible for the spectral transfer
that gives rise to the turbulence cascade. In fact, the main effect of 
Alfv\'en propagation, from the perspective of the strong turbulent 
energy cascade,  is not to cause spectral transfer but to suppress it 
\cite{shebalin_1983_anisotropy}.  Understanding the cascade itself
requires examination of the strength of the nonlinearities. In this case, 
the appropriate characteristic time becomes the nonlinear time, whose 
isolation requires analysis of timescales in the Lagrangian frame
\cite{kraichnan_1964_kolmogorov} (note that only in a few particular cases 
in our analisys, the nonlinear time was positively identified as a 
candidate for the decorrelation time). Nevertheless, we have shown that 
physically relevant phenomena such as reflection and ``anomalous
propagation'' of reflected fluctuations can produce observable
effects in the flow energetics, and these phenomena have been recognized
in a variety of configurations of the different controlling parameters 
of the system, with potential applications.

For example, interesting effects associated with reflection add to the
complexity of the dynamics, even in the simplest case of
incompressible MHD considered here. This has important implications
for applications such as coronal heating, solar wind acceleration, and
particle energization in the interplanetary space
\cite{velli_1993_propagation, matthaeus_1999_coronal}. As a further example,
fluctuations observed in the solar wind, which tend to have the
magnetic and the velocity field aligned or anti-aligned (i.e., with
different Alfv\'enic polarizations), cannot always be trivially
interpreted as travelling ``downstream'' or ``upstream'' the mean
magnetic field. Extensions of this study to compressible MHD
\cite{andres_2017_interplay}, considering the dependence with the
cross-helicity in the flow and its interplay with compressible
effects, as well as a study considering other helicities such as 
the kinetic helicity $H_v$ or the magnetic helicity $H_b$ and the 
hybrid helicity for Hall-MHD, would be an
interesting follow up of the present study, and a
first step towards a deeper understanding of the role of non-linear
effects in the propagation of waves in plasma turbulence. 

\acknowledgments RL, PD and PM acknowledge support from PICT Grant
No.~2015-3530, PIP Grant No.~11220150100324CO, and UBACyT Grant
No.~20020170100508BA. AP is is thankful to LASP, and in particular to
Bob Ergun, for support. WHM partially supported by NASA Grant NNX17AB79G.

\bibliography{Bibliografia}
\bibliographystyle{unsrt}

\end{document}